\documentclass{article}
\usepackage{latexsym}
\usepackage{gastex}
\usepackage[usenames]{color}

\usepackage{amssymb}

\newtheorem{theorem}{Theorem}[section]

\newtheorem{corollary}[theorem]{Corollary}

\newtheorem{lemma}[theorem]{Lemma}

\newtheorem{proposition}[theorem]{Proposition}

\newtheorem{pro-example}[theorem]{Example}
\newenvironment{example}{\begin{pro-example}\rm}{\cqfd\end{pro-example}}

\newtheorem{pro-remark}[theorem]{Remark}
\newenvironment{remark}{\begin{pro-remark}\rm}{\cqfd\end{pro-remark}}

\newenvironment{Proof}{\rm \trivlist \item[\hskip \labelsep{\bf
Proof.}]}{\cqfd\endtrivlist}

\def\cqfd{\skip10=\parfillskip\parfillskip=0pt
\enspace\hfill\symbolecqfd\par\parfillskip=\skip10\par\medskip}
\def\symbolecqfd{\rlap{$\sqcap$}$\sqcup$}
\def\preuve{\begin{Proof}}
\def\proof{\begin{Proof}}
\def\eop{\end{Proof}}

\def\calB{\mathcal{B}}
\def\calC{\mathcal{C}}
\def\calF{\mathcal{F}}
\def\calG{\mathcal{G}}
\def\calH{\mathcal{H}}

\def\calP{\mathcal{P}}
\def\calS{\mathcal{S}}

\def\NLC{\form{NLC}}

\def\form#1{\mathsf{#1}}
\let\formrm\form

\def\StS{\calS t\calS}
\def\GP{\calG\calP}
\def\GS{\calG\calS}
\def\Grph{\form{Graph}}
\def\add{\formrm{add}}

\def\mark{\formrm{mark}}
\def\mdf{\formrm{mdf}}

\def\p{\form{p}}
\def\q{\form{q}}
\def\v{\form{v}}
\def\vloop{\v^{\form{loop}}}
\def\ren{\form{ren}}
\def\fg{\form{fg}}
\def\srcren{\form{srcren}}
\def\srcfg{\form{srcfg}}
\def\fus{\form{fus}}
\let\fuse\fus
\def\edg{\mathrel{\form{edge}}}

\def\inc{\form{inc}}
\def\mfus{\form{mfus}}

\def\del{\form{del}}

\def\card{\form{card}}

\def\HR{\form{HR}}
\def\VR{\form{VR}}

\def\sep{\form{sep}}

\def\inv{^{-1}}
\let\phi\varphi

\font\Bb=msbm10
\def\N{\hbox{\Bb N}}

\def\false{{\sf false}}
\def\true{{\sf true}}

\def\Knn{\overrightarrow{K}_{n,n}}
\def\Kmm{\overrightarrow{K}_{m,m}}

\def\sqbox{\mathbin{\Box}}

\begin{document}

\centerline{\LARGE The recognizability of sets of graphs}

\vskip.5cm

\centerline{\LARGE is a robust property\protect\footnote{This paper was
prepared in part while the second author was an invited Professor at
the University of Nebraska-Lincoln. The second author acknowledges
partial support from the \textit{AS 93} of the D\'epartement STIC of
CNRS, and the \textit{ACI S\'ecurit\'e Informatique} of the Minist\`ere
de la Recherche.}}

\vskip .5cm

\centerline{\Large Bruno Courcelle\protect\footnote{%
LaBRI, Universit\'e Bordeaux-1 -- 351 cours de la Lib\'eration -- 33405
Talence Cedex -- France.\enspace {\tt bruno.courcelle@labri.fr}},
\qquad
Pascal Weil\protect\footnote{%
LaBRI, CNRS -- 351 cours de la Lib\'eration -- 33405 Talence Cedex --
France.\\
{\tt pascal.weil@labri.fr}} }

\vskip 1cm

\begin{abstract}
    Once the set of finite graphs is equipped with an algebra structure
    (arising from the definition of operations that generalize the
    concatenation of words), one can define the notion of a
    recognizable set of graphs in terms of finite congruences.
    Applications to the construction of efficient algorithms and to the
    theory of context-free sets of graphs follow naturally. The class
    of recognizable sets depends on the signature of graph operations.
    We consider three signatures related respectively to Hyperedge
    Replacement ($\HR$) context-free graph grammars, to Vertex
    Replacement ($\VR$) context-free graph grammars, and to modular
    decompositions of graphs. We compare the corresponding classes of
    recognizable sets. We show that they are robust in the sense that
    many variants of each signature (where in particular operations are
    defined by quantifier-free formulas, a quite flexible framework)
    yield the same notions of recognizability. We prove that for graphs
    without large complete bipartite subgraphs, $\HR$-recognizability
    and $\VR$-recognizability coincide. The same combinatorial
    condition equates $\HR$-context-free and $\VR$-context-free sets of
    graphs. Inasmuch as possible, results are formulated in the more
    general framework of relational structures.
\end{abstract}

\section{Introduction}

The notion of a recognizable language is a fundamental concept in
Formal Language Theory, which has been clearly identified since the
1950's. It is important because of its numerous applications, in
particular for the construction of compilers, and also for the
development of the Theory: indeed, these languages can be specified in
several very different ways, by means of \textit{automata},
\textit{congruences}, \textit{regular expressions} and \textit{logical
formulas}. This multiplicity of quite different definitions is a clear
indication that the notion is central since one arrives at it in a
natural way from different approaches. The equivalence of definitions
is proved in fundamental results by Kleene, Myhill and Nerode, Elgot
and B\"uchi.

The notion of a recognizable set has been extended in the 1960's to
trees (actually to trees representing finite algebraic terms), to
infinite words and to infinite trees. In the present article we discuss
its extension to sets of finite graphs.

The recognizability of a set of finite words or trees can be defined in
several ways, as mentioned above, and in particular by finite
\textit{deterministic} automata. This definition (together with the
related effective translations from other definitions) provides
linear-time recognition algorithms, which are essential for compiler
construction, coding, text processing, and in other situations.
Recognizable sets of words can also be defined in an algebraic way by
finite saturating congruences relative to the monoid structure. These
definitions, by automata and congruences, extend smoothly to the case
of finite trees (i.e., algebraic terms), using the natural algebra
structure. The notion of recognizability in a general algebra is due to
Mezei and Wright \cite{MW}. We will not discuss here the extensions to
infinite words and trees, which raise specific problems surveyed by
Thomas \cite{ThomasHdbook} and Perrin and Pin \cite{PerrinPin}. Our aim
will be to consider sets of finite graphs.

For finite graphs, there is no automaton model, except in very special
cases, and in particular in the case of graphs representing certain
\textit{labelled partially ordered sets} and \textit{traces} (a trace
is a directed acyclic graph, representing the equivalence class of a
word w.r.t. a partial commutation relation), see the volume edited by
Diekert \cite{Diekert} and the papers by Lodaya and Weil
\cite{LWTCS,LWIC} and \'Esik and N\'emeth \cite{EsikNemeth}. Algebraic
definitions via finite congruences can be given because the set of
finite graphs can be equipped with an algebraic structure, based on
graph operations like the concatenation of words. However, many
operations on graphs can be defined, and there is no prominent choice
for a standard algebraic structure like in the case of words where a
unique associative binary operation is sufficient. Several algebraic
structures on graphs can be defined, and distinct notions of
recognizability follow from these possible choices. It appears
nevertheless that two graph algebras, called the \textit{$\HR$-algebra}
and the \textit{$\VR$-algebra} for reasons explained below, emerge and
provide robust notions of recognizability. The main purpose of this
paper is to demonstrate the robustness of these notions. By robustness,
we mean that taking variants of the basic definitions does not modify
the corresponding classes of recognizable sets of graphs.

In any algebra, one can define two family of sets, the recognizable
sets and the equational sets. The equational sets are defined as the
components of the least solutions of certain systems of recursive set
equations, written with set union and the operations of the algebra,
extended to sets in the standard way. Equational sets can be considered
as the natural extension of context-free languages in a general
algebraic framework (Mezei and Wright \cite{MW}, Courcelle
\cite{CourcelleAlgUniv} for a thorough development). The two graph
algebras introduced above, the $\HR$- and the $\VR$\textit{-algebra}, are
familiar to readers interested in graph grammars, because their
equational sets are the (context-free) \textit{Hyperedge Replacement}
($\HR$) sets of graphs on the one hand, and the (context-free)
\textit{Vertex Replacement} ($\VR$) sets on the other. Both classes of
context-free sets of graphs can be defined in alternative, more
complicated ways in terms of graph rewritings, and are robust in the
sense that they are closed under certain transformations expressible in
Monadic Second-Order Logic (Courcelle \cite{HbGraGraRozenberg97}).

The main results of this paper, described below in more detail, are:

1) the robustness of the classes of $\VR$- and $\HR$-recognizable sets of
graphs,

2) the robustness of the class of recognizable sets of finite relational
structures (equivalently of simple directed ranked hypergraphs), which
extends the two previous classes,

3) the exhibition of structural conditions on sets of graphs implying
that $\HR$-recognizability and $\VR$-recognizability coincide,

4) the comparison of the recognizable sets of the $\VR$-algebra and
those of a closely related algebra representing \textit{modular
decompositions} (modular decomposition is another useful notion for
graph algorithms).

\medskip

The notion of recognizability of a set of finite graphs is important
for several reasons.
First, because recognizability yields linear-time algorithms for the
verification of a wide class of graph properties on graphs belonging to
certain finitely generated graph algebras. These classes consist of
graphs of bounded tree-width and of bounded clique-width. These two
notions of graph complexity are important for constructions of
polynomial graph algorithms, see Downey and Fellows \cite{DF} and
Courcelle et al. \cite{CourcelleMakowskyRotics2000}. Furthermore, these
graph properties are not very difficult to identify because Monadic
second-order (MS) logic can specify them in a formalized and uniform
way. (In many cases, an MS formula can be obtained from the graph
theoretical expression of a property). More precisely, a central result
\cite{BCI,BCVII,HbGraGraRozenberg97,CourcelleMakowskyRotics2000} says
that every set of graphs (or graph property) definable by an MS formula
is recognizable (respectively admits such algorithms), for appropriate
graph algebras. This general statement covers actually several distinct
situations.

Another reason comes from the theory of Graph Grammars. The
intersection of a context-free set of graphs and of a recognizable set
is context-free (in the appropriate algebraic framework). This gives
immediately many closure properties for context-free sets of graphs,
via the use of MS logic as a specification language for graph
properties. Recognizability also makes it possible to construct
terminating and (in a certain sense) confluent graph rewriting rules by
which one can recognize sets of graphs of bounded tree-width by graph
reduction in linear time, see Arnborg \textit{et al.} \cite{ACPS}.

Finally, recognizability is a basic notion for dealing with languages
and sets of terms, and on this ground, its extension to sets of graphs
is worth investigating. Logical characterizations of recognizability
can be given using MS logic, extending many results in language theory
\cite{BCXIV,Hoogeboom,EsikNemeth,KuskeSP,KuskeMSC}. Several questions
remain open in this research field.

We have noted above that defining recognizability for sets of graphs
cannot be done in terms of finite automata, so that the algebraic
definition in terms of finite congruences has no alternative. Another
advantage of the algebraic definition is that it is given at the level
of universal algebra (Mezei and Wright \cite{MW}), and thus applies to
objects other than graphs. However, even in the case of graphs, the
algebraic setting is useful because it hides (temporarily) the
complexities of operations on graphs and makes it possible to
understand what is going on at a structural level.

We now present the main results of this article more in detail. The two
main algebraic structures on graphs called $\VR$ and $\HR$, originate
from algebraic descriptions of context-free graph grammars. Definitions
will be given in the body of the text. It is enough for this
introduction to retain that the operations of $\VR$ are more powerful
than those of $\HR$. Hence every $\HR$-context-free set of graphs
(i.e., defined by a grammar based on the operations of $\HR$) is
$\VR$-context-free, but not vice-versa. For recognizability, the
inclusion goes in the opposite direction : every $\VR$-recognizable set
is $\HR$-recognizable but the converse is not true. However, if the
graphs of a set $L$ have no subgraph of the form $K_{n,n}$ (the
complete bipartite graph on $n+n$ vertices) for some $n$, then $L$ is
$\HR$-recognizable if and only if it is $\VR$-recognizable (this is the
main theorem of Section~\ref{sec Knn}). A similar statement is known to
hold under the same hypothesis for context-free sets: if $L$ is without
$K_{n,n}$ (\textit{i.e.}, no graph in $L$ contains a subgraph
isomorphic to $K_{n,n}$), then it is $\HR$-context-free if and only if
it is $\VR$-context-free (Courcelle, \cite{BC:VR/HR}). The proofs of
the two statements are however different (and both difficult).

Up to now we have only discussed graphs, but our approach, which
extends the approach developped by Courcelle in \cite{BCVII}, also
works for hypergraphs and for relational structures.

The operations on graphs, hypergraphs and structures are basically of
three types defined in Section~\ref{sec: AlgRelStr}: we use only one
binary operation, the disjoint union; we use unary operations defined
by quantifier-free first-order formulas; and basic graphs and structures
corresponding to nullary operations. In this way we can generate graphs
and structures by finite algebraic terms. The quantifier-free definable
operations can modify vertex and edge labels, add or delete edges. This
notion is thus quite flexible. What is remarkable is that these
numerous operations can be added without altering the notion of
recognizability.

The main result of Section \ref{sec GP} states that the same
recognizable sets of graphs are obtained if one uses the basic
$\VR$-algebra (closely connected to the definition of
\textit{clique-width}), the same algebra enriched with quantifier-free
definable operations, and even the larger algebra dealing with
relational structures. Variants of the $\VR$-algebra which are useful,
in particular for algorithmic applications, are also considered, and
they are proved to yield the same class of recognizable sets.

In Section \ref{sec GS}, we discuss similarly the $\HR$-algebra which is
very important because of its relation with \textit{tree-width} and
with context-free graph grammars. We prove a robustness result relative
to the subclass such that the distinguished vertices denoted by
distinct labels (nullary operations) are different. The $\HR$-operations
are appropriate to handle graphs and hypergraphs with multiple edges
and hyperedges (whereas the $\VR$-operations are not). The original
definitions (see Courcelle \cite{BCI}) were given for graphs with
multiple edges and hyperedges. In Section \ref{sec simple multiple}, we
prove that for a set of simple graphs, $\HR$-recognizability is the same
in the $\HR$-algebra of simple graphs and in the larger $\HR$-algebra of
graphs with multiple edges. Without being extremely difficult, the
proof is not just a routine verification.

In Section \ref{sec: modular}, we consider an algebra arising from the
theory of modular decomposition of graphs. We show that under a natural
finiteness condition, the corresponding class of recognizable sets is
equal to that of $\VR$-recognizable ones.

In an appendix, we clarify the definitions of certain equivalences of
logical formulas, focusing on cases where they are decidable, and we
give upper bounds to the cardinalities of the quotient sets for these
equivalences. These results yield upper bounds to the number of
equivalence classes in logically based congruences. They are thus
useful for the investigation of recognizability in view of the cases
where the sets under consideration are defined by logical formulas.
They also provide elements to appreciate (an upper bound of) the
complexity of the algorithms underlying a number of the effective
proofs in the main body of the paper.

This work has been presented in invited lectures by B. Courcelle
\cite{DLT04} and P. Weil \cite{MFCS04}.

\section{Recognizability}

The notion of a recognizable set is due to Mezei and Wright \cite{MW}.
It was originally defined for one-sort structures, and we adapt it to
many-sorted ones with infinitely many sorts. We begin with definitions
concerning many-sorted algebras.

\subsection{Algebras}

We follow essentially the notation and definitions from \cite{Wechler},
see also \cite{CourcelleAlgUniv}. Let $\form{S}$ be a set called the
set of \emph{sorts}. An $\form{S}$-\emph{signature} is a set $\calF$
given with two mappings $\alpha \colon \calF\longrightarrow
seq(\form{S)}$ (the set of finite sequences of elements of $\form{S}$),
called the \emph{arity} mapping, and
$\sigma\colon\calF\longrightarrow\form{S}$, called the \emph{sort}
mapping. We denote by $\rho(f)$ the length of the sequence $\alpha(f)$,
which we call also \textit{arity}. The \emph{type} of $f$ in $\calF$ is
the pair $(\alpha(f),\sigma(f))$ that we shall rather write $\alpha(f)
\rightarrow \sigma(f)$, or $(\form{s_{1}}, \form{s_{2}}, \ldots,
\form{s_{n}}) \longrightarrow \form s$ if $\alpha(f) = (\form{s_{1}},
\cdots, \form{s_{n}})$ and $\sigma(f) = \form{s}$. The sequence
$\alpha(f)$ may be empty (that is, $n=0$), in which case $f$ is called
a \emph{constant of type $\sigma(f) = \form{s}$}.

An \emph{$\calF$-algebra} is an object $M = \langle(M_{\form s})_{\form
s\in\form{S}}, (f_{M})_{f\in \calF}\rangle$, where for each $\form s
\in \form{S}$, $M_{\form s}$ is a non-empty set, called the
\emph{domain of sort $\form s$} of $M$. For a nonempty sequence of
sorts $\mu=(\form{s_{1}}, \cdots, \form{s_{n}})$, we denote by
$M_{\mu}$ the product $M_{\form{s_{1}}}\times M_{\form{s_{2}}} \times
\cdots \times M_{\form{s_{n}}}$. If $\rho(f) > 0$, then $f_{M}$ is a
total mapping from $M_{\alpha(f)}$ to $M_{\sigma(f)}$. If $f$ is a
constant of type $\form s$, then $f_{M}$ is an element of $M_{\form
s}$. The objects $f_{M}$ are called the \emph{operations} of $M$. We
assume that $M_{\form s}\cap M_{\form s^{\prime}}=\emptyset$ for $\form
s\neq \form s^{\prime}$. We also let $M$ denote the union of the
$M_{\form s}$ ($\form s\in\form{S}$). For $d\in M$, we let $\sigma(d)$
denote the unique $\form s\in\form S$ such that $d\in M_{\form s}$.

A mapping $h\colon M\rightarrow M^{\prime}$ between $\calF$-algebras is
a \textit{homomorphism} (or \emph{$\calF$-homomorphism} if it is useful
to specify the signature) if it maps $M_{\form s}$ into $M_{\form
s}^{\prime}$ for each sort $\form s$ and it commutes with the
operations of $\calF$.

We denote by $T(\cal F)$ the set of finite well-formed terms built with
$\cal F$ (we will call them \emph{$\calF$-terms}), and by
$T(\calF)_{\form s}$ the set of those terms of sort $\form s$ (the sort
of a term is that of its leading symbol). If $\calF$ has no constant
the set $T(\calF)$ is empty.

There is a standard structure of $\calF$-algebra on $T(\calF)$. Its
domain of sort $\form s$ is $T(\calF)_{\form s}$, and $T(\calF)$ can be
characterized as the \emph{initial $\calF$-algebra}.\ This means that
for every $\calF$-algebra $M$, there is a unique homomorphism
$val_{M}\colon T(\calF)\longrightarrow M$. If $t\in T(\calF)_{\form s}$,
the image of $t$ under $val_{M}$ is an element of $M_{\form s}$, also
denoted by $t_{M}$. It is nothing but the evaluation of $t$ in $M$,
where the function symbols are interpreted by the corresponding
functions of $M$. One can consider $t$ as a term \emph{denoting}
$t_{M}$, and $t_{M}$ as the \emph{value} of $t$ in $M$. The set of
values in $M$ of the terms in $T(F)$ is called the \textit{subset
generated} by $\calF$. We say that a subset of $M$ is \emph{finitely
generated} if it is the set of values of terms in $T(\calF^{\prime })$
for some finite subset $\calF^{\prime}$ of $\calF$.

Let $\calF$ be an $\form S$-signature, $\calF^\prime$ be an $\form
S'$-signature where $\form S'\subseteq \form S$. We say that $\calF'$
is a \emph{subsignature} of $\calF$, written $\calF'\subseteq \calF$ ,
if $\calF'$ is a subset of $\calF$ and the types of every $f$ in
$\calF'$ are the same with respect to $\calF$ and to $\calF'$. We say
then that an $\calF'$-algebra $M'$ is a \emph{subalgebra} of an
$\calF$-algebra $M$ if $M'_{\form s} \subseteq M_{\form s}$ for every
$\form s \in \form S'$, and every operation of $M'$ coincides with the
restriction to the domains of $M'$ of the corresponding operation of
$M$.

We will often encounter the case where an $\calF$-algebra $M$ is also
the carrier of a $\calG$-algebra, and the $\calG$-operations of $M$ can
be expressed as $\calF$-terms: in that case, we say that the
$\calG$-operations of $M$ are \textit{$\calF$-derived}, and the
$\calG$-algebra $M$ is an \textit{$\calF$-derived algebra} (or it is
\textit{derived from} $M$).

More formally, an $\form{S}$-\emph{sorted set of variables} is a pair
$(X,\sigma)$ consisting of a set $X$ and a \emph{sort mapping}
$\sigma\colon X\longrightarrow \form{S}$ (usually denoted simply by
$X$). We let $T(\calF,X)$ be the set of $(\calF\cup X)$-terms written
with $\calF\cup X$, where it is understood that the variables are among
the nullary symbols (constants) of $\calF\cup X$. $T(\calF,X)_{\form
s}$ denotes the subset of those terms of sort $\form s$.
Now if $\mathcal{X}$ is a finite sequence of pairwise distinct
variables from $X$ and $t\in T(\calF,X)_{\form s}$, we denote by
$t_{M,\mathcal{X}}$ the mapping from $M_{\sigma(\mathcal{X)}}$ to
$M_{\form s}$ associated with $t$ in the obvious way
($\sigma(\mathcal{X)}$ denotes the sequence of sorts of the elements of
$\mathcal{X}$). We call $t_{M,\mathcal{X}}$ a \emph{derived operation
of the algebra} $M$. If $\mathcal{X}$ is known from the context, we
write $t_{M}$ instead of $t_{M,\mathcal{X}}$. This is the case in
particular if $t$ is defined as a member of
$T(\calF,\{x_{1},\cdots,x_{k}\})$ : the sequence $\mathcal{X}$ is
implicitly $(x_{1},\cdots,x_{k})$.

\subsection{Recognizable subsets}

Let $\calF$ be an $\form{S}$-signature. An $\calF$-algebra $M$ is
\emph{locally finite} if each domain $M_{\form s}$ is finite. If $M$ is
an $\calF$-algebra and $\form s\in\form{S}$ is a sort, a subset $L$ of
$M_{\form s}$ is $M$-\emph{recognizable} if there exists a locally
finite $\cal F$-algebra $A$, a homomorphism $h\colon M\longrightarrow
A$, and a (finite) subset $C$ of $A_{\form s}$ such that $L =
h^{-1}(C)$.

We denote by $Rec(M)_{\form s}$ the family of $M$-recognizable subsets
of $M_{\form s}$. In some cases it will be useful to stress the
relevant signature and we will talk of $\calF$-recognizable sets
instead of $M$-recognizable sets.

An equivalent definition can be given in terms of finite congruences. A
\emph{congruence} on $M$ is an equivalence relation $\approx$ on $M =
\bigcup_{\form s\in\form S}M_{\form s}$, such that each set $M_{\form
s}$ is a union of equivalence classes, and which is stable under the
operations of $M$. It is \emph{locally finite} if for each sort $\form
s$, the restriction $\approx_{\form s}$ of $\approx$ to $M_{\form s}$
has finite index. A congruence \emph{saturates a set} if this set is a
union of classes. A subset $L$ of $M_{\form s}$ is $M$-recognizable if
and only if it is saturated by a locally finite congruence on $M$.

The following facts are easily verified from the definition of
recognizability or its characterization in terms of congruences (see
\cite{CourcelleAlgUniv}), and will be used freely in the sequel.

\begin{proposition}\label{easy recog facts}
    Let $M$ be an $\calF$-algebra.
    \begin{itemize}
	\item For each sort $\form s$, the family $Rec(M)_{\form s}$
	contains $M_{\form s}$ and the empty set, and it is closed
	under union, intersection and difference.
	\item If $h$ is a unary derived operation of $M$ or a
	homomorphism of $M'$ into $M$, (where $M'$ is another
	$\calF$-algebra), then the inverse image under $h$ of an
	$M$-recognizable set is recognizable.
	\item If $N$ is a $\calG$-algebra with the same domain as $M$,
	and if every $\calG$-congruence of $N$ is an $\calF$-congruence
	of $M$ (e.g. $N$ is derived from $M$, or $\calG$ is obtained
	from $\calF$ by adding constants), then every $M$-recognizable
	set is $N$-recognizable. If in addition $\calG$ contains
	$\calF$, then $M$ and $N$ have the same recognizable subsets.
	\item If $M'$ is a subalgebra of $M$ and $L$ is an
	$M$-recognizable set, then $L\cap M'$ is $M'$-recognizable.
	This includes the case where $M'$ has the same domain as $M$,
	and is an $\calF'$-algebra for some subsignature $\calF'$ of
	$\calF$.
	\item Suppose that $M$ is generated by $\calF$ and let
	$val_{M}$ be the evaluation homomorphism from $T(\calF)$ onto
	$M$. A subset $L$ of $M_{\form s}$ is $\calF$-recognizable if
	and only if $val_{M}^{-1}(L)$ is a recognizable subset of
	$T(\calF)$. If in addition $\calF$ is finite, then this is
	equivalent to the existence of a finite tree-automaton
	recognizing $val_{M}^{-1}(L)$.
    \end{itemize}
\end{proposition}

\begin{example}
    On the set of all words over a finite alphabet $A$, let us consider
    the binary operation of the concatenation product, and the unary
    operation $u\mapsto u^2$, which is derived from the concatenation
    product. Then the 3rd statement in Proposition~\ref{easy recog
    facts} shows that we have the same recognizable subsets as if we
    considered only the concatenation product. It is interesting to
    note that, in contrast, adding the operation $u\mapsto u^2$ to the
    signature adds new equational languages, \textit{e.g.} the set of
    all squares.
\end{example}    

We will see more technical conditions that guarantee the transfer of 
recognizability between algebras in Section~\ref{sec technical lemmas}
below.

\subsection{Remarks on the notion of
recognizability}\label{significance}

We gather here some observations on the significance of
recognizability.

First, we note that if $f$ is an operation of an $\calF$-algebra $M$,
with arity $k$, and if $B_{1},\ldots,B_{k}$ are $M$-recognizable, then
$f(B_{1},\ldots,B_{k})$ is not necessarily recognizable. This is
discussed for instance in \cite{BC-MSCS}, where sufficient conditions
are given to ensure that $f(B_{1},\ldots,B_{k})$ is recognizable. It is
well-known for instance that the product of two recognizable subsets of
the free monoid (word languages) or of the trace monoid is
recognizable; a similar result holds for recognizable sets of trees.

Now, let $M$ be an $\calF$-algebra and let $\calF'$ be a signature
which differs from $\calF$ only by the choice of constants and their
values. In particular, $\calF'$ may be obtained from $\calF$ by the
addition of countably many new constants. Then the congruences on $M$
are the same with respect to $\calF$ and to $\calF'$ and it follows
that a subset of $M$ is $\calF$-recognizable if and only if it is
$\calF'$-recognizable.

It is customary to assume that the $\calF$-algebra $M$ is generated by
the signature $\calF$. If $M$ is a countable $\calF$-algebra that is
not generated by $\calF$, we can enrich $\calF$ to $\calF'$ by adding
to $\calF$ one constant of the appropriate sort for each element of
$M$. Then $\calF'$ generates $M$ (in a trivial way). As noted above,
$M$ has the same $\calF$- and $\calF'$-recognizable subsets. If $L$ is
one of these subsets, the set $val_{M}^{-1}(L)$ of $\calF'$-terms is
recognizable but we cannot do much with it, because we lack the notion
of a finite tree-automaton. See the conclusion of the paper for a
further discussion of this point.

Finally, we can question the interest of the notion of a recognizable
set. Is it interesting in every algebra? The answer is clearly no. Let
us explain why.

If the algebraic structure over the considered set $M$ is poor, for
example in the absence of non-nullary functions, then every set $L$ is
recognizable, by a congruence with two classes, namely $L$ and its
complement. The notion of recognizability becomes void.

Another extreme case is when the algebraic structure is so rich that
there are very few recognizable sets. For an example, consider the set
$\mathbb{N}$ of natural integers equipped with the successor and the
predecessor functions (predecessor is defined by $\textsf{pred}(0)=0$,
$\textsf{pred}(n+1) = n$). The only recognizable sets are $\mathbb{N}$
and the empty set. Indeed, if $\sim$ is a congruence and if $n \sim
n+p$ for some $n\ge 0$, $p>0$, then by using the function \textsf{pred}
$n+p-1$ times, we find that $0 \sim 1$. It follows (using the successor
function repeatedly) that any two integers are equivalent.

Intuitively, if one enriches an algebraic structure by adding new
operations, one gets fewer recognizable sets.

For another example, let us consider the monoid $\{a,b\}^{\ast}$ of
words over two letters. Let us add a unary operation, the
\emph{circular shift}, defined by : $sh(1) = 1$ and $sh(au)=ua$,
$sh(bu)=ub$, for every word $u$. The language $a^{\ast}b$ is no longer
recognizable w.r.t. this new structure, however recognizability does
not degenerate completely since every commutative language that is
recognizable in the usual sense remains recognizable in the enriched
algebraic structure.

It is not completely clear yet which algebraic condition makes
recognizability ``interesting''.

\subsection{Technical results on recognizability}\label{sec technical
lemmas}

The statements in this section explain how to transfer a locally finite
congruence from one algebra to another, possibly with a different
signature, and hence how to transfer recognizability properties between
algebras. Proposition~\ref{easy recog facts} above contains examples of
such results.


The statements that follow will be used in the proof of some of our
main results, in Section~\ref{sec GP}. They are, unfortunately, heavily
technical in their statements (but not in their proofs\dots)

\begin{lemma}\label{FGH-congruence}
    Let $\calF$ be an $\form S$-signature and let $\calG$ be a $\form
    T$-signature. Let $S$ be an $\calF$-algebra and let $T$ be a
    $\calG$-algebra. Let also $\calH$ be a collection
    $(\calH_{\form{t},\form{s}})$ such that, for each $\form{t}\in\form{T}$
    and $\form{s}\in\form{S}$, $\calH_{\form{t},\form{s}}$ consists of
    mappings from $T_{\form{t}}$ into $S_{\form{s}}$ with the following
    property:
    
	{\narrower\noindent for each operation $g\in\calG$ of type
	$(\form{t}_{1},\ldots,\form{t}_{r})\mapsto\form{t}$ and for
	each $h\in\calH_{\form{t},\form{s}}$, there exist sorts
	$\form{s}_{1},\ldots,\form{s}_{r}\in\form{S}$, mappings
	$h_{i}\in\calH_{\form{t}_{i},\form{s}_{i}}$ ($1\le i\le r$) and
	an $\calF$-derived operation $f$ of type
	$(\form{s}_{1},\ldots,\form{s}_{r})\mapsto\form{s}$ such that,
	for every $x_{1}\in T_{1}$, \dots, $x_{r}\in T_{r}$,
	$h(g(x_{1},\ldots,x_{r})) =
	f(h_{1}(x_{1}),\ldots,h_{r}(x_{r}))$.\par}
    \noindent Finally, let $\equiv$ be an $\calF$-congruence on $S$ and
    let $\approx$ be the equivalence relation defined, on each
    $T_{\form{t}}$, by
    $$x\approx y\enspace\hbox{if and only if}\enspace \hbox{$h(x)\equiv 
    h(y)$ for every $h\in\calH_{\form{t},\form{s}}$,
    $\form{s}\in\form{S}$.}$$
    Then $\approx$ is a $\calG$-congruence on $T$.
\end{lemma}

\proof
Let $g$ be an operation in $\calG$, of type
$(\form{t}_{1},\ldots,\form{t}_{r})\mapsto\form{t}$, and let
$x_{1},y_{1}\in T_{\form{t}_{1}},\ldots,x_{r},y_{r} \in
T_{\form{t}_{r}}$ such that $x_{i}\approx y_{i}$ for each $i =
1,\ldots,r$. Let also $h\in \calH_{\form{t},\form{s}}$ with
$\form{s}\in\form{S}$.

By hypothesis, there exist sorts
$\form{s}_{1},\ldots,\form{s}_{r}\in\form{S}$, mappings
$h_{i}\in\calH_{\form{t}_{i},\form{s}_{i}}$ (for $i = 1,\ldots,r$) and
an $\calF$-derived operation $f$ of type
$(\form{s}_{1},\ldots,\form{s}_{r})\mapsto\form{s}$ such that
\begin{eqnarray*}
   h(g(x_{1},\ldots,x_{r})) &=& f(h_{1}(x_{1}),\ldots,h_{r}(x_{r}))\cr
   h(g(y_{1},\ldots,y_{r})) &=& f(h_{1}(y_{1}),\ldots,h_{r}(y_{r})).
\end{eqnarray*} 
Since $x_{i}\approx y_{i}$ for each $i$, we have $h_{i}(x_{i})\equiv
h_{i}(y_{i})$; and since $\equiv$ is an $\calF$-congruence, it
follows that $h(g(x_{1},\ldots,x_{r})) \equiv h(g(y_{1},\ldots,y_{r}))$. 
Thus we have $g(x_{1},\ldots,x_{r}) \approx g(y_{1},\ldots,y_{r})$,
which concludes the proof.
\eop

With the notation of Lemma~\ref{FGH-congruence}, for each sort
$\form{t}\in\form{T}$, let $\le_{\form{t}}$ be the quasi-order relation
defined on $\calH_{\form{t}} =
\bigcup_{\form{s}\in\form{S}}\calH_{\form{t},\form{s}}$ by
$$h\le_{\form{t}} h'\hbox{ if there exists an $\calF$-derived
    unary operation $f$ such that $h' = f\circ h$.}$$

\begin{lemma}\label{FGH-finiteness}
    With the notation of Lemma~\ref{FGH-congruence}, if for each
    $\form{t}$ the order relation associated with $\le_{\form{t}}$ has
    a finite number of minimal elements, and if the $\calF$-congruence
    $\equiv$ on $S$ is locally finite, then the $\calG$-congruence
    $\approx$ on $T$ is locally finite.
\end{lemma}

\proof
Let $\textsf{t}\in\form{T}$. We want to show that there are only
finitely many $\approx$-classes in $T(\textsf{t})$. By assumption,
there exist elements $h_{1},\ldots,h_{k}\in\calH_{\form{t}}$ such that
every mapping of $\calH_{\form{t}}$ is of the form $f\circ h_{i}$ for
some $1\le i\le k$ and some $\calF$-derived operation $f$.

For each $i$, let $S_{\form{s}_{i}}$ be the range of $h_{i}$ and let
$n_{i}$ be the number of $\equiv$-classes in $S_{\form{s}_{i}}$. It is
immediately verified from the definition of $\le_{\form{t}}$ that if
$x,y\in T_{\form{t}}$, then $x\approx y$ if and only if $h_{i}(x)\equiv
h_{i}(y)$ for each $1\le i\le k$. In particular, $T_{\form{t}}$ has at
most $n_{1}\cdots n_{k}$ $\approx$-classes, which concludes the proof.
\eop

We will actually need even more technical versions of these
lemmas.

\begin{lemma}\label{FGH-congruence+zeta(T)}
    Let $S$, $T$, $\calF$, $\calG$ and $\calH$ be as in
    Lemma~\ref{FGH-congruence}, and let $\zeta$ be a $\calG$-congruence
    on $T$ such that:

    {\narrower\noindent for each operation $g\in\calG$ of type
    $(\form{t}_{1},\ldots,\form{t}_{r})\mapsto\form{t}$, for each
    $h\in\calH_{\form{t},\form{s}}$ and for each $\vec z =
    (z_{1},\ldots,z_{r})$ where each $z_{i}$ is a $\zeta$-class of
    $T_{\form{t}_{i}}$, there exist sorts $\form{s}_{1,\vec
    z},\ldots,\form{s}_{r,\vec z}\in\form{S}$, mappings $h_{i,\vec
    z}\in\calH_{\form{t}_{i},\form{s}_{i,\vec z}}$ ($1\le i\le r$) and
    an $\calF$-derived operation $f_{\vec z}$ of type
    $(\form{s}_{1,\vec z},\ldots,\form{s}_{r,\vec z})\mapsto\form{s}$
    such that, in $T$, $h(g(x_{1},\ldots,x_{r})) = f_{\vec z}(h_{1,\vec
    z}(x_{1}),\ldots,h_{r,\vec z}(x_{r}))$ if each $x_{i}$ is in
    $z_{i}$.\par}
    \noindent Finally, let $\equiv$ be an $\calF$-congruence on $S$ and
    let $\approx$ be the equivalence relation defined, on each
    $T_{\form{t}}$, by
    $$x\approx y\enspace\hbox{if and only if}\enspace
    \hbox{$x\mathrel{\zeta} y$ and $h(x)\equiv h(y)$ for every
    $h\in\calH_{\form{t},\form{s}}$, $\form{s}\in\form{S}$.}$$
    \noindent Then $\approx$ is a $\calG$-congruence on $T$. Moreover,
    if $\calH$ satisfies the hypothesis of Lemma~\ref{FGH-finiteness}
    and $\equiv$ and $\zeta$ are locally finite, then $\approx$ is
    locally finite as well.
\end{lemma}

\proof
The proof is the same as for Lemmas~\ref{FGH-congruence} and
\ref{FGH-finiteness}.
\eop

\section{Algebras of relational structures}\label{sec: AlgRelStr}

Even though we are ultimately interested in studying sets of graphs, it
will be convenient to handle the more general case of relational
structures. Furthermore, relational structures can be identified with
simple directed hypergraphs. Such hypergraphs form a natural
representation of terms. See for instance the chapter on hypergraphs in
\cite{HbGraGraRozenberg97} for applications.

In this paper, all graphs and structures are finite or countable. Our
proofs will not usually depend on cardinality assumptions on the graphs
or structures, and hence our results will hold for finite as well as
for infinite graphs or structures. However, recognizability in the
algebraic sense we defined, is really interesting only for dealing with
finitely generated objects, and hence for finite graphs and structures.
For dealing with infinite words, trees and graphs, other tools are
necessary, see for instance \cite{PerrinPin,ThomasHdbook,KuskeSP,KuskeMSC}.

\subsection{Relational structures}\label{sec rel structures}

Let $R$ be a finite set of relation symbols, and $C$ be a finite set of
nullary symbols. Each symbol $r\in R$ has an associated positive
integer called its \emph{rank}, denoted by $\rho(r)$. An
$(R,C)$-structure is a tuple $S=\langle D_{S},(r_{S})_{r\in
R},(c_{S})_{c\in C}\rangle$ such that $D_{S}$ is a (possibly empty) set
called the \emph{domain} of $S$, each $r_{S}$ is a $\rho(r)$-ary
relation on $D_{S}$, i.e., a subset of $D_{S}^{\rho(r)}$, and each
$c_{S}$ is an element of $D_{S}$, called the \emph{$c$-source} of $S$.

We denote by $\StS(R,C)$ the class of (finite or countable)
$(R,C)$-structures, and we sometimes write $\StS(R)$ for
$\StS(R,\emptyset)$. By convention, isomorphic structures will be
considered as equal. In the notation $\StS$, $\mathcal St$ stands for
\textit{structures}, while the second $\mathcal S$ stands for
\textit{sources}.

A structure $S\in\StS(R,C)$ is \emph{source-separated} if $c_{S} \neq
c'_{S}$ for $c\neq\ c'$. We will denote by $\StS_{\sep}(R,C)$ the class
of source-separated structures in $\StS(R,C)$. See
Corollary~\ref{StSsep recognizable} and Section~\ref{sec source
separation} below.

In order to handle graphs, we will consider particular kinds of
structures in the sequel. We let $E = \{\edg\}$ be the set of relation
symbols consisting of a single binary relation $\edg$, intended to
represent directed edges. Thus graphs can be seen as the elements of
$\StS(E)$, also written $\Grph$. Clearly these graphs are directed,
simple (we cannot represent multiple edges) and they may have loops.
For a discussion of graphs with multiple edges, see Section~\ref{sec
simple multiple}.

We let $\GS(C)$ denote the set $\StS(E,C)$. These structures are called
\textit{graphs with sources}. We let $\GS_{\sep}(C)$ denote the
intersection $\GS(C) \cap \StS_{\sep}(R,C)$.

We will discuss also \textit{graphs with ports} (Section~\ref{sec GP}): if
$P$ is a finite set of unary relation symbols called \emph{port
labels}, then we denote by $E_{P}$ the set of relational symbols $E\cup
P$ and by $\GP(P)$ the class $\StS(E_{P})$. Port labels are useful for
studying the clique-width of graphs, see \cite{CER93,BC-Olariu} and
Remark~\ref{rk clique-width} below.

\subsection{The algebra $\StS$}\label{sec algebra StS}

We first define some operations on structures.

\paragraph{Disjoint union}

Let $C$ and $C'$ be disjoint sets of constants and let $S\in\StS(R,C)$
and $S'\in\StS(R',C')$. Let us also assume that $S$ and $S'$ have
disjoint domains. We denote by $S\oplus S'$ the union of $S$ and $S'$,
which is naturally a structure in $\StS(R\cup R',C\cup C')$.

If $S$ and $S'$ are not disjoint, we replace $S'$ by a disjoint copy.
We need not be very precise on how to choose this copy because
different choices will yield isomorphic $\oplus$-sums, and we are
interested in structures up to isomorphism.

\begin{remark}
   It is also possible to define a similar operation, without the
   restriction that $C$ and $C'$ are disjoint (as in, say,
   \cite{BCVII,BCX}). See Section~\ref{sec StS-parallel} below for a
   discussion.
\end{remark}

\paragraph{Quantifier-free definable operations}

Our purpose is now to define functions from $\StS(R,C)$ to
$\StS(R',C')$ by quantifier-free formulas. We denote by
$QF(R,C,\{x_{1},...,x_{n}\})$ the set of quantifier-free formulas on
$(R,C)$-structures with variables in $\{x_{1},...,x_{n}\}$.

A \textit{qfd operation scheme} from $\StS(R,C)$ to $\StS(R',C')$ is a
tuple
$$(\delta, (\phi_{r})_{r\in R'}, (\kappa_{c,d})_{c\in C,d\in C'}),$$
where $\delta\in QF(R,C,\{x\})$, $\phi_{r}\in
QF(R,C,\{x_{1},\ldots,x_{\rho(r)}\})$ if $r$ is a $\rho(r)$-ary
relation symbol, $\kappa_{c,d}\in QF(R,C,\emptyset)$, such that the
following formulas are valid in every structure in $\StS(R,C)$, for all
$c, c'\in C$, $d\in C'$ and $r\in R'$ of arity $\rho(r)$:
\begin{itemize}
    \item $\kappa_{c,d} \land \kappa_{c',d} \Longrightarrow c = c'$;
    
    \item $\bigvee_{e\in C}\kappa_{e,d}$;
    
    \item $\kappa_{c,d}\Longrightarrow \delta(c)$;
  
    \item $\forall x_{1},\ldots,x_{\rho(r)}\
    \Big(\phi_{r}(x_{1},\ldots,x_{\rho(r)})\Longrightarrow
    \bigwedge_{i=1}^{\rho(r)}\delta(x_{i})\Big)$.
\end{itemize}
The reason for these conditions becomes apparent with the following
definition of the \textit{qfd operation}
$g\colon\StS(R,C)\to\StS(R',C')$ defined by such a scheme. Let
$S\in\StS(R,C)$. The domain of the structure $g(S)$ is the subset of
the domain of $S$ defined by formula $\delta$ and the relation $r$
($r\in R'$) on $g(S)$ is described by formula $\phi_{r}$. Finally, if
$d\in C'$, then $d_{g(S)} = c_{S}$ if $c\in C$ and $S$ satisfies
$\kappa_{c,d}$. The first two conditions imposed above assert that
relative to $S$, $c$ is uniquely defined for each $d$, the third
condition asserts that $d_{g(S)}$ always lies in the domain of $g(S)$,
and the fourth condition asserts that the relation $\phi_{r}$ ($r\in
R'$) can only relate elements of the domain of $g(S)$.


\begin{remark}
   Note that in the first condition, $c=c'$ does not mean that $c$ and
   $c'$ are the same constant, but that they have the same value in the
   considered structure.
\end{remark}

\begin{remark}
   The conditions to be verified by a qfd operation scheme are
   decidable. It follows that the notion of a qfd operation scheme is
   effective. See the appendix (Remark \ref{remark A4} in particular)
   for a discussion of this decidability result.
\end{remark}

\begin{example}\label{def HR ops}
    Let $R$ be a finite set of relational symbols, $C$ be a
    finite set of source labels and let $a,b$ be source labels. We
    define the following operations.

    \begin{itemize}
	\item if $a\in C$ and $b\not\in C$, $\srcren_{a\rightarrow b}$
	is the unary operation of type $(R,C)\rightarrow
	(R,C\setminus\{a\}\cup\{b\})$ which renames the $a$-source of
	a structure to a $b$-source;
	\item if $a\in C$, $\srcfg_{a}$ is the unary operation of type
	$(R,C)\rightarrow (R, C\setminus\{a\})$ which forgets the
	$a$-source of a structure;
	
	\item if $a\ne b\in C$, $\fus_{a,b}$ is the unary operation of
	type $(R,C)\rightarrow (R,C)$ which identifies the $a$-source
	and the $b$-source of a structure (so the resulting domain
	element is both the $a$-source and the $b$-source), and
	reorganizes the tuples of the relational structure accordingly.
    \end{itemize}
    
    Note that the operation names $\srcren_{a\rightarrow b}$,
    $\srcfg_{a}$ and $\fus_{a,b}$ are overloaded: they denote different
    operations when the sets $R$ and $C$ are allowed to vary. A
    completely formal definition would use operation names
    such as $\srcren_{a\rightarrow b, R, C}$, which would be
    inconvenient.

    It is immediately verified that the operations of the form
    $\srcren_{a\rightarrow b}$ and $\srcfg_{a}$ are qfd. It is probably
    worth showing explicitly a qfd operation scheme defining the
    operation $\fus_{a,b}$.

    Let $\delta(x)$ be the formula $(a = b) \lor ((a \ne b) \land (x
    \ne a))$. If $r\in R$ has arity $\rho(r) = n$, let
    $\phi_{r}(x_{1},\ldots,x_{n})$ be the formula
    \begin{eqnarray*}
	&&\Big((a=b) \land r(x_{1},\ldots,x_{n})\Big) \lor\\
	&&\Big((a\ne b)\land
	\bigvee_{I\subseteq\{1,\ldots,n\}}(\bigwedge_{i\in I}(x_{i} = b) \land
	    \bigwedge_{i\not\in I}(x_{i}\ne b)\land
	    r(y_{1},\ldots,y_{n}))\Big),
    \end{eqnarray*}
    where for each $I$, $y_{i} = a$ if $i\in I$ and $y_{i} = x_{i}$
    otherwise. For each $d\in C$ such that $d\ne a$ and for each $c\in
    C$, let $\kappa_{c,d}$ be the formula $c = d$; let $\kappa_{b,a}$
    be the formula $\true$, and let $\kappa_{c,a}$ be the formula $c =
    a$ for each $c\ne b$. It is now routine to verify that the scheme
    $(\delta, (\phi_{r})_{r\in R}, (\kappa_{c,d})_{c,d\in C})$ defines
    $\fus_{a,b}$.
\end{example}

\begin{remark}
    There is no qfd operation from $\StS(R)$ into $\StS(R',C)'$ if
    $C'\ne\emptyset$, because in the absence of constants in the
    input structure, we cannot define constants in the output
    structure. 
\end{remark}

\begin{example}\label{example silent}
   The natural inclusion of $\StS(R,C)$ into $\StS(R',C)$ when $R'$
   contains $R$ is a qfd operation in natural way: the formulas
   intended to define relations in $R'\setminus R$ are taken to be
   identically $\false$.
\end{example}

\paragraph{The signature $\calS$}

We define the algebra $\StS$ of \textit{structures with sources} as
follows. First, let us fix once and for all a countable set of relation
symbols containing $\edg$ and countably many relation symbols of each
arity, and a countable set of constants. In the sequel, finite sets of
relation symbols $R$ and finite sets of constants $C$ will be taken in
these fixed sets. The set of sorts consists of all such pairs $(R,C)$. 
The set of elements of $\StS$ of sort $(R,C)$ is $\StS(R,C)$.

The signature $\calS$ consists of the following operations (interpreted
in $\StS$). First, for each pair of sorts $(R,C)$ and $(R',C')$ such
that $C\cap C' = \emptyset$, the disjoint union $\oplus$ is an
operation of type $((R,C), (R',C'))\rightarrow (R\cup R',C\cup C')$.
Note that we overload the symbol $\oplus$, that is, we denote in the
same way an infinite number of operations on $\StS$. Next, every qfd
operation is a (unary) operation in $\calS$.

Finally, we observe that the signature $\calS$ contains the natural
inclusions of $\StS(R,C)$ into $\StS(R',C)$ when $R'$ contains $R$,
which are qfd (Example~\ref{example silent}).

As for constants in $\calS$, one can pick a single source label $a$,
and consider a single constant $\form a$, denoting the structure with a
single element, which is an $a$-source, and no relations. Together with
the operations in $\calS$, this constant suffices to generate all
finite relational structures. As noted in Section~\ref{significance},
the choice of constants does not affect recognizability. It only
affects the generating power of the signature, but this is not our
point in this paper.

\subsection{Elementary properties of $\StS$}\label{sec elementary
properties}

We first consider the composition of qfd operations.

\begin{proposition}\label{composition of qfd}
    Qfd operations in $\StS$ are closed under composition (whenever
    types fit for defining meaningful composition).
\end{proposition}

\proof
Let $g\colon\StS(R,C)\longrightarrow \StS(R',C')$ and
$g'\colon\StS(R',C')\longrightarrow \StS(R'',C'')$ be qfd operations,
given respectively by the schemes $(\delta,(\psi_{r})_{r\in R'},
(\kappa_{c,d})_{c\in C,d\in C'})$ and $(\delta',(\psi'_{r})_{r\in R''},
(\kappa'_{c,d})_{c\in C',d\in C''})$.

The composite $g'\circ g$ turns an $(R,C)$-structure into an 
$(R'',C'')$-structure.

Let $\delta^0$, $\psi^0_{r}$ ($r\in R''$) and $\kappa^0_{c,d}$ ($c\in
C'$, $d\in C''$) be obtained from $\delta'$, $\psi'_{r}$ and
$\kappa'_{c,d}$ by 
replacing every occurrence of $r(y_{1},\ldots,y_{\rho(r)})$ ($r\in R'$)
by $\psi_{r}(y_{1},\ldots,y_{\rho(r)})$; our formulas are now in the
language of $(R,C')$-structures and we need to ``translate'' the
constants $d\in C'$ into elements of $C$. However, this translation, 
a mapping from $C'$ to $C$, depends on the structure in which we
operate.

To reflect this observation, for each mapping $h\colon C'\rightarrow C$,
we let $h(\delta^{0})$ be the conjunction of the formulas
$\kappa_{h(d),d}$ ($d\in C'$) and the formula obtained from
$\delta^0$ by replacing each occurrence of $d$ ($d\in C'$) by $h(d)$.
Finally, we let $\delta''$ be the disjunction of the $h(\delta^0)$
when $h$ runs over all mappings from $C'$ to $C$.
    
We proceed in the same fashion to define $\psi''_{r}$ and
$\kappa''_{c,d}$ for each $r\in R''$ and each $c\in C'$, $d\in C''$. 
Finally, if $b\in C$ and $d\in C''$, we let $\lambda_{b,d} =
\bigvee_{c\in C'}(\kappa'_{b,c}\wedge\kappa''_{c,d})$.

It is a routine verification that $(\delta'',(\psi''_{r})_{r\in
R''},(\lambda_{b,d})_{b\in C,d\in C''})$ is a qfd operation scheme,
which defines the composite operation $g'\circ g$. This completes the
proof.
\eop

For each $S\in\StS(R,C)$, we define the \textit{type of $S$}, written
$\zeta(S)$, to be the restriction of $S$ to its set of sources. That
is: the domain of $\zeta(S)$ is the set of $C$-sources of $S$, and the
relations of $\zeta(S)$ are those tuples of $C$-sources that are
relations in $S$. In order to simplify notation, we also denote by
$\zeta$ the equivalence relation on $\StS$ given by
$$S\ \zeta\ T\quad\hbox{if and only if}\quad\hbox{$\zeta(S)$ and
$\zeta(T)$ are isomorphic.}$$

\begin{lemma}\label{type equivalence}
    Let $S,T\in\StS(R,C)$. Then $S\ \zeta\ T$ if and only if $S$ and 
    $T$ satisfy the same formulas in $QF(R,C,\emptyset)$.
\end{lemma}

\proof
A formula in $QF(R,C,\emptyset)$ is a Boolean combination of atoms of
the form $c = d$ where $c,d\in C$, or $r(x_{1},\ldots,x_{n})$ where
$r\in R$ has arity $n$ and the $x_{i}$ are in $C$. It is immediate that
such an atom is true in $S$ if and only if it is true in $\zeta(S)$.
Thus $S$ and $\zeta(S)$ satisfy the same formulas in
$QF(R,C,\emptyset)$: in particular, $\zeta$-equivalent structures
satisfy the same formulas in $QF(R,C,\emptyset)$. Thus, if we denote by
$Th^{FO}_{0,R,C}(S)$ the set of formulas in $QF(R,C,\emptyset)$ that
are satisfied by $S$ (see Section~\ref{sec FFVV}), we find that
$Th^{FO}_{0,R,C}(S) = Th^{FO}_{0,R,C}(\zeta(S))$.

Conversely, we observe that if $S$ is a structure in $\StS(R,C)$, which
consists only of its $C$-sources (that is, $S = \zeta(S)$), then $S$ is
entirely described by some formula in $QF(R,C,\emptyset)$. Thus, if
$\zeta(S) \ne \zeta(T)$, then $Th^{FO}_{0,R,C}(S) \ne
Th^{FO}_{0,R,C}(T)$. This suffices to conclude the proof.
\eop

The type relation $\zeta$ has the following important property.

\begin{proposition}\label{FFVV light}
    The type relation $\zeta$ is a locally finite congruence on $\StS$.
\end{proposition}

\proof
The verification that $\zeta(S\oplus S') = \zeta(S)\oplus \zeta(S')$
($S\in\StS(R,C)$, $S'\in\StS(R',C')$ and $C\cap C' = \emptyset$) is
immediate. Let us now consider a qfd operation
$g\colon\StS(R,C)\longrightarrow \StS(R',C')$, specified by the qfd
operation scheme $(\delta,(\psi_{r})_{r\in R'}, (\kappa_{c,d})_{c\in
C,d\in C'})$. By Lemma \ref{type equivalence}, $S$ and $\zeta(S)$
satisfy the same formulas of $QF(R,C,\emptyset)$. In particular, for
each $c\in C$ and $d\in C'$, $S$ and $\zeta(S)$ both satisfy
$\kappa_{c,d}$, or both satisfy its negation. Thus $g(S)$ and
$g(\zeta(S))$ have the same sources, and hence $\zeta(g(S)) =
\zeta(g(\zeta(S)))$.

We have just shown that the type relation is a congruence. To complete
the proof, it suffices to show that for each sort $(R,C)$, the set of
types of sort $(R,C)$, that is, the set $\zeta(\StS(R,C))$ is finite.
Note that if $S\in\StS(R,C)$, then $\zeta(S)$ has cardinality at most
$\card(C)$ (and also at most $\card(S)$). It follows that
$\card(\zeta(\StS(R,C))) \le\card(C)!\ \prod_{r\in
R}2^{\card(C)^{\rho(r)}}$.
\eop

\begin{remark}
    Proposition~\ref{FFVV light} can be seen as a particular case of a
    result of Feferman and Vaught \cite{FefermanVaught},
    Theorem~\ref{FFVV} below, which will be used in Section \ref{sec
    Knn}. The simple formulation above will be very useful.
\end{remark}

Note that the knowledge of $\zeta(S)$ is sufficient to determine
whether $S$ is a source-separated structure. This observation is used
to prove the following corollary.

\begin{corollary}\label{StSsep recognizable}
    Let $(R,C)$ be a sort in $\StS$. Then $\StS_{\sep}(R,C)$ is
    a recognizable subset of $\StS(R,C)$.
\end{corollary}

\proof
Whether a structure $S$ is source-separated depends only on its type
$\zeta(S)$: in particular, the type congruence $\zeta$ saturates
$\StS_{\sep}(R,C)$. By Proposition~\ref{FFVV light}, this relation is a
locally finite congruence, and hence $\StS_{\sep}(R,C)$ is recognizable.
\eop

\subsection{A result of Feferman and Vaught}\label{sec FFVV}

If $(R, C)$ is a sort of $\StS$, we denote by $FO(R,C)$ the set of
closed first-order formulas over $R$ and $C$. For each integer $d$, we
denote by $FO_{d}(R,C)$ the set of those formulas of quantifier-depth
at most $d$. Up to a decidable syntactic equivalence (taking into
account Boolean laws, properties of equality, renaming of quantified
variables, see Appendix~\ref{appendix}), there are only finitely many
formulas in each set $FO_{d}(R,C)$. Thus, we can reason as if
$FO_{d}(R,C)$ was actually finite.

For an $(R,C)$-structure $S$, we let its \emph{$FO_{d}$-theory} be the
set $Th_{d,R,C}^{FO}(S)$ of formulas in $FO_{d}(R,C)$ that are valid in
$S$. It is finite since it is a subset of the finite set $FO_{d}(R,C)$.

\begin{theorem}\label{FFVV}
    Let $d\ge 0$.
    \begin{itemize}
	\item[(1)] For every qfd operation $f$ of type $(R,C)
	\rightarrow (R',C')$, there exists a
	mapping $f_{d}^{\#}$ such that, for every $(R,C)$-structure
	$S$
	$$Th_{d,R',C'}^{FO}(f(S))=f_{d}^{\#}(Th_{d,R,C}^{FO}(S)).$$
	
	\item[(2)] For every $(R,C)$ and $(R',C')$, where
	$C$ and $C'$ are disjoint, there exists a
	binary function $\oplus_{d}^{\#}$ such that, for every
	$(R,C)$-structure $S$, and every $(R',C')$-structure
	$S'$,
	$$Th_{d,R\cup R',C\cup C'}^{FO}(S\oplus S') = Th_{d,R,C}^{FO}(S)
	\oplus_{d}^{\#} Th_{d,R',C'}^{FO}(S').$$
    \end{itemize}    
\end{theorem}

\begin{remark}
    The second assertion was proved in \cite{FefermanVaught} for
    first-order logic, and extended by Shelah to monadic second-order
    logic \cite{Shelah}. The importance of this result is discussed by
    Makowsky in \cite{Makowsky}.
\end{remark}

\begin{remark}
    The functions $f_{d}^{\#}$ and $\oplus_{d}^{\#}$ have finite
    domains and codomains. However these sets are quite large. These
    functions can be (at least in principle) effectively determined for
    given $(R,C)$, $(R',C')$, and $d$.
\end{remark}

\subsection{Variants of the algebra of relational structures}

In the literature on recognizable and equational graph languages,
several variants of the signature $\calS$ and the algebra $\StS$ are
considered, notably a variant where the definition of the disjoint
union is replaced by a more general parallel product, and a variant
where all structures are assumed to be source-separated. We verify in
this section that these variants do not yield different notions of
recognizability.

\subsubsection{Parallel composition vs. disjoint union}\label{sec
StS-parallel}

In the literature (e.g. \cite{BCVII,BCX}), the operation of disjoint
union $\oplus$ is sometimes replaced by the so-called
\textit{parallel composition} (or \textit{product}), written
$\parallel$, an operation of type $((R,C),(R',C'))\rightarrow (R\cup
R', C\cup C')$ for which we do not assume that $C$ and $C'$ are
disjoint. If $S\in \StS(R,C)$ and $S'\in\StS(R',C')$, the parallel
composition $S\parallel S'$ is obtained by taking the (set-theoretic)
disjoint union of $S$ and $S'$ and then identifying the $c$-sources of
$S$ and $S'$ for each $c\in C\cap C'$. Let $\calS_{\parallel}$ denote
the signature obtained from $\calS$ by substituting $\parallel$ for
$\oplus$.

\begin{proposition}\label{prop StS-parallel}
    Let $L$ be a subset of $\StS$. Then $L$ is $\calS$-recognizable
    if and only if it is $\calS_{\parallel}$-recognizable.
\end{proposition}

\proof
We first observe that the operation $\oplus$ is a particular case of
$\parallel$. Therefore $\calS$ is a sub-signature of
$\calS_{\parallel}$ and hence, every $\calS_{\parallel}$-recognizable
set is $\calS$-recognizable.

To prove the converse, it suffices to verify that $\parallel$ is an
$\calS$-derived operation by Proposition~\ref{easy recog facts}.
Indeed, if $S\in\StS(R,C)$ and $S'\in\StS(R',C')$, the parallel
composition $S\parallel S'$ can be obtained by the following sequence
of $\calS$-operations (see Example~\ref{def HR ops} for their
definition):

- for each $c\in C\cap C'$, apply the qfd operation
$\srcren_{c\rightarrow \bar c}$ which renames the $c$-source in $S'$
with a new source label, say $\bar c$, not in $C$; let $\bar S'$ be the
resulting structure;

- take the disjoint union $S\oplus \bar S'$;

- for each $c\in C\cap C'$, apply the operation $\fus_{c,\bar c}$ which
identifies the $c$-source and the $\bar c$-source in $S\oplus\bar S'$;

- apply the source-forgetting operation $\srcfg_{\bar c}$ for each
$c\in C\cap C'$.
\eop

\subsubsection{Source-separated structures}\label{sec source separation}

The property that $c_{S} \ne c'_{S}$ for $c\ne c'$ is called
\textit{source separation}. This property makes it easier to work with
operations on structures and graphs, and hence we discuss a variant of
the $\calS$-algebra $\StS$, which handles source-separated structures.
We will also use it in Section~\ref{sec Knn}.

Recall that $\StS_{\sep}(R,C)$ denotes the set of source-separated
structures in $\StS(R,C)$. We now define a subsignature $\calS_{\sep}$
of $\calS$ such that $\StS_{\sep}$ is a sub-algebra of $\StS$.

Disjoint union $\oplus$ clearly preserves source separation, and is
part of $\calS_{\sep}$. Next we include in $\calS_{\sep}$ the operations
specified by qfd operation schemes such that, for each $c\in C$ and
$d\ne d'\in C'$ (see the notation in Section~\ref{sec algebra StS}),
\begin{equation}\label{eq source separation}
    \kappa_{c,d} \Longrightarrow \neg\kappa_{c,d'},
\end{equation}
which guarantees that the operation preserves source separation.

\begin{example}\label{exemple HRsep}
    The operations $\srcren_{a\rightarrow b}$ and $\srcfg_{a}$
    defined in Example~\ref{def HR ops} are in $\calS_{\sep}$. The
    operation $\fus_{a,b}$ defined in the same example is not.
    
    In contrast, the operation written $\fus_{a\rightarrow b}$, which
    identifies the $a$-source and the $b$-source of a structure as in
    $\fus_{a,b}$, and makes the resulting element of the domain a
    $b$-source but not an $a$-source, preserves source separation. It
    can be written as $\fus_{a\rightarrow b} = \srcfg_{a}\circ
    \fus_{a,b}$.
    
    The operation which, given a graph with source labels $a$ and $b$,
    exchanges the source labels $a$ and $b$ if the corresponding
    vertices are linked by an edge and does nothing otherwise, is
    another example of a qfd operation in $\calS_{\sep}$.
\end{example}

Regarding the effectiveness of the definition of $\calS_{\sep}$, we
observe the following.

\begin{proposition}
    Given a qfd operation scheme, one can decide whether the
    corresponding qfd operation preserves source separation.
\end{proposition}

\proof
Let $g$ be the qfd operation specified by the given qfd operation
scheme, and let $\StS(R,C)$ be the domain of $g$. One can effectively
construct the images under $g$ of every type in $\StS(R,C)$, since
there are only finitely many of them, and they can all be enumerated.
One can then verify whether the operation preserves souce-separation on
types.

Now it follows from the proof of Proposition~\ref{FFVV light} that for
each $S\in\StS(R,C)$, we have $\zeta(g(\zeta(S))) = \zeta(g(S))$. In
particular, $g$ preserves source separation if and only if it preserves
it for the structures of the form $\zeta(S)$. Thus one can effectively
decide whether $g\in\calS_{\sep}$.
\eop


We now show that the restriction to source-separated structures does
not change the notion of recognizability.

\begin{theorem}\label{thm Ssep-recognizable}
    Let $L$ be a subset of $\StS_{\sep}$. Then $L$ is
    $\calS$-recognizable if and only if it is
    $\calS_{\sep}$-recognizable.
\end{theorem}

\proof
By definition, $\calS_{\sep}$ is a subsignature of $\calS$, so every
$\calS$-recognizable set is $\calS_{\sep}$-recognizable.

To prove the converse, we first define a mapping $h$, which maps a
structure $S\in \StS(R,C)$ to a source-separated structure $h(S) \in
\StS_{\sep}(R,C)$ by splitting sources that were identified in $S$.

We assume that the countable set of constant symbols (from which $C$ is
taken, see Section~\ref{sec algebra StS}) is linearly ordered. Let
$h_{0}^S\colon C\rightarrow C$ be given by
$$h_{0}^S(c) = \min\{d\in C\mid c_{S} = d_{S}\}.$$
We let $C_{0}^S = h_{0}^S(C)$ and $C_{1}^S = C\setminus C_{0}^S$. The
structure $h(S)$ has domain set the disjoint union of $S$ and
$C_{1}^S$. For each $c\in C_{0}^S$, the $c$-source of $h(S)$ is the
$c$-source of $S$, and for each $c\in C_{1}^S$, the $c$-source of
$h(S)$ is the element $c\in C_{1}^S$. Finally, for each $r\in R$, the
relation $r_{h(S)}$ equals the relation $r_{S}$ (so it does not involve
the elements of $C_{1}^S$). Observe that $h$ is not a qfd operation,
and that $h_{0}^S$, $C_{0}^S$ and $C_{1}^S$ depend only on $\zeta(S)$.

Now let $L$ be an $\calS_{\sep}$-recognizable subset of $\StS_{\sep}$ and
let $\equiv$ be a locally finite $\calS_{\sep}$-congruence recognizing
it. We need to construct a locally finite $\calS$-congruence $\sim$ on
$\StS$ which recognizes $L$.

The relation $\sim$ on $\StS$ is defined as follows. If
$S,T\in\StS(R,C)$, we say that $S\sim T$ if $\zeta(S) = \zeta(T)$ and
$h(S)\equiv h(T)$. It is immediately verified that $\sim$ is an
equivalence relation. Moreover, the $\sim$-class of a structure $S$
is determined by its $\zeta$-class, and by the $\equiv$-class of
$h(S)$. Since both $\zeta$ and $\equiv$ are locally finite, $\sim$
also is locally finite.

Let us now prove that $\sim$ is an $\calS$-congruence. Let $S\sim
T\in\StS(R,C)$ and $S'\sim T'\in\StS(R',C')$, with $C\cap C' =
\emptyset$. By Proposition~\ref{FFVV light}, $\zeta(S\oplus S') =
\zeta(T\oplus T')$. It is not difficult to verify that
$$h(S\oplus S') = h(S) \oplus h(S').$$
It follows that $h(S\oplus S') \equiv h(T\oplus T')$ since $\oplus$
is an operation in $\calS_{\sep}$. Thus $S\oplus S' \sim T\oplus T'$.

Next let $g$ be a qfd operation from $\StS(R,C)$ to $\StS(Q,B)$, given
by the qfd operation scheme $(\delta,(\psi_{q})_{q\in Q},
(\kappa_{c,b})_{c\in C, b\in B})$. Let $S$ and $T$ be $\sim$-equivalent
elements of $\StS(R,C)$, which will remain fixed for the rest of this
proof. We need to show that $g(S) \sim g(T)$. We already know from
Proposition~\ref{FFVV light} that if $S\sim T\in \StS(R,C)$, then
$\zeta(g(S)) = \zeta(g(T))$, and we want to show that $h(g(S)) \equiv
h(g(T))$.

Since $\zeta(g(S)) = \zeta(g(T))$, the mappings $h_{0}^{g(S)}$ and
$h_{0}^{g(T)}$, from $B$ to $B$, coincide. Let $B_{0} =
h_{0}^{g(S)}(B)$ and $B_{1} = B \setminus B_{0}$. Without loss of
generality, we may assume that $B_{1}\cap C = \emptyset$. The domain
set of $h(g(S))$ (resp. $h(g(T))$) is the disjoint union of the domain
of $g(S)$ (resp. $g(T)$) and $B_{1}$.

It suffices to show that there exists a qfd operation
$k\in\calS_{\sep}$, depending on $g$ and $\zeta(S)$, such that $h(g(S))
= k(h(S)\oplus B_{1})$ and $h(g(T)) = k(h(T)\oplus B_{1})$ (where
$B_{1}$ is the source-only element of $\StS_{\sep}(\emptyset,B_{1})$).
Indeed, the fact that $\equiv$ is an $\calS_{\sep}$-congruence will then
imply that $h(g(S)) \equiv h(g(T))$.

Let $\delta'$ be obtained from $\delta$ by replacing every occurrence
of $c\in C$ by $h_{0}^S(c)$. For each $q\in Q$, $c\in C$ and $b\in B$,
let $\psi'_{q}$ be obtained from $\psi_{q}$ and $\kappa'_{c,b}$ be
obtained from $\kappa_{c,b}$ in the same fashion.

Let now $k'\colon \StS(R,C\cup B_{1})\rightarrow \StS(Q,B)$ be defined
by the scheme
\begin{eqnarray*}
    &&(\gamma', (\chi'_{q})_{q\in Q},
    (\lambda'_{c,b})_{c\in C\cup B_{1}, b\in B})\hbox{ defined as
    follows:} \\
    &&\gamma'(x) = \Big(\delta'(x) \land \bigwedge_{c\in
    C_{1}^S}\neg(x = c)\Big) \vee \bigvee_{b\in B_{1}}(x = b) \\
    &&\chi'_{q} = \psi'_{q}\hbox{ for each $q\in Q$}\\
    &&\lambda'_{b,b} = \true\hbox{ if $b\in B_{1}$} \\
    &&\lambda'_{c,b} = \false\hbox{ if $b\in B_{1}$ and $c\ne b$} \\
    &&\lambda'_{c,b} = \false\hbox{ if $b\in B_{0}$ and $c\in C_{1}^S$} \\
    &&\lambda'_{c,b} = \bigvee_{h_{0}^{g(S)}(a) =
    b,\ h_{0}^{S}(d) = c}
    \kappa'_{d,a}\hbox{ if
    $b\in B_{0}$ and $c\in C_{0}^S$.}
\end{eqnarray*}

It is now a routine verification that (for our fixed structure $S$)
$k'(h(S)\oplus B_{1}) = h(g(S))$. Since all our definitions depend only
on $\zeta(S)$, we also have $k'(h(T)\oplus B_{1}) = h(g(T))$.

One last step is required in this proof as the qfd operation $k'$ may
not preserve source separation for all structures, that is, $k'$ may
not lie in $\calS_{\sep}$. It does for the particular structures
$h(S)\oplus B_{1}$ and $h(T)\oplus B_{1}$, but perhaps not for others.
Actually, structures $U$ such that $\zeta(U) \ne \zeta(h(S)\oplus
B_{1}) = \zeta(h(T)\oplus B_{1})$ do not matter in this context, so we
can replace $k'$ by the operation $k$, with the same domain and range
as $k'$, which maps a structure $U$ to $k'(U)$ if $\zeta(U) =
\zeta(h(S)\oplus B_{1})$, and to the source-only source-separated
structure $B\in \StS(Q,B)$ where all relations are empty. This new
operation $k$ preserves source separation by construction, and it is
easily verified to be qfd. This completes (at last) the proof.
\eop

\section{The algebra $\GP$ of graphs with ports}\label{sec GP}

Graphs with ports were introduced in Section~\ref{sec rel structures}.
Recall that if $P$ is a set of unary relation symbols, then $E_{P}$
denotes the set $E_{P} = \{\edg\}\cup P$ and the class of graphs with
ports in $P$, written $\GP(P)$ can be identified with $\StS(E_{P})$. We
observe that a vertex of a graph with ports in $P$ can be a $p$-port
for one or several port labels $p\in P$, or for none at all.

For convenience, we will consider that $P$ is a finite subset of the
set \N\ of natural integers.

\subsection{The signature $\VR$ on graphs with ports}

We define the set of sorts of the algebra $\GP$ to be the set of
finite subsets of \N. For each such subset $P$, the set of elements
of $\GP$ of sort $P$ is the set $\GP(P)$ of graphs with ports in $P$.

The signature $\VR$ consists of constants, unary operations and binary
operations. These operations (interpreted in $\GP$) are as follows.

First, if $P,Q$ are finite subsets of \N, then $\oplus$ is as in
$\StS$, and is thus a binary operation of type $(E_{P},E_{Q})
\rightarrow E_{P\cup Q}$. In $\GP$, we consider $\oplus$ as an
operation of type $(P,Q)\rightarrow P\cup Q$.

Next, the unary operations of $\VR$ are the following (clearly qfd)
operations:
\begin{itemize}
    \item if $p,q$ are distinct integers, $\add_{p,q}$ is an operation
    of type $P\rightarrow P$ for each sort $P$ such that $p,q\in P$: it
    modifies neither the domain (the set of vertices) nor the unary
    relations $p$ ($p\in P$); the new edge relation has the existing
    edges, \textit{plus} every edge from a $p$-port to a $q$-port: it
    is given by
    $$\edg(x,y) \lor (p(x) \land q(y));$$
    
    \item if $D$ is a finite subset of $\N\times\N$, $\mdf_{D}$ is an
    operation of type $P\rightarrow Q$ where $P$ is any finite set
    containing the domain of the relation $D$ and $Q$ is any finite set
    containing the range of $D$; it modifies neither the domain (set of
    vertices) nor the edge relation; for each $q\in Q$, the $q$-ports
    of the output structure are the vertices of the input structure
    that are $p$-ports for some $p$ such that $(p,q)\in D$; that is,
    $q(x)$ is given by $\bigvee_{(p,q)\in D}p(x)$.
\end{itemize}

Finally, for each integer $p$, we let $\p$ be the constant of type
$\{p\}$ denoting the graph with a single vertex, no edges, and whose
vertex is a $p$-port. We also let $\p^{\form{loop}}$ be the same graph,
with a single loop.

\begin{remark}\label{rename, forget}
    The following operations on graphs with ports occur in the
    literature, and are particular cases of $\VR$-operations.
    
    Let $p\ne q$ be integers, $P$ be a subset of \N\ containing $p$ and
    $Q = P\setminus\{p\}\cup\{q\}$. The operation $\ren_{p\rightarrow
    q}$, of type $P\rightarrow Q$ which \textit{renames} every $p$-port
    to a $q$-port, is an operation of $\VR$: it is equal to $\mdf_{D}$
    where $D = \{(r,r)\mid r\in P\setminus\{p\}\} \cup \{(p,q)\}$.
    Observe that this operation fuses the sets of vertices defined by
    $p$ and $q$.
    
    Let $p$ be an integer, and let $P$ be a subset of \N\ containing
    $p$. The operation $\fg_{p}$, of type $P\rightarrow
    P\setminus\{p\}$, which \textit{forgets} $p$-ports is an operation
    of $\VR$: it is equal to $\mdf_{D}$ where $D = \{(r,r)\mid r\in
    P\setminus\{p\}\}$.
\end{remark}

\begin{remark}
    In our definition of graph with ports, an element of $\GP(Q)$ does
    not need to have $q$-ports for each $q\in Q$. Thus, if
    $P\subseteq Q$, every graph with ports in $P$ can also be viewed
    as a graph with ports in $Q$. The natural inclusion of $\GP(P)$
    into $\GP(Q)$ is part of the signature $\VR$: it is equal to
    $\mdf_{D}$ where $D = \{(p,p)\mid p\in P\}$.
\end{remark}

\begin{remark}
    Again (as in Example \ref{def HR ops}), the operations introduced
    in this section are denoted by overloaded symbols. A formal
    definition should specify the type of the operation, and would read
    something like $\add_{p,q,P}$ or $\mdf_{D,P,Q}$. We prefer the more
    concise notation introduced here.
\end{remark}

\subsection{A technical result}

The following result describes the action of a qfd operation on a
disjoint union of structures. It is the key to the main results of this
section, described in Section \ref{VR equivalence} below.

\begin{proposition}\label{unary-parallel}
    Let $\zeta$ be the type congruence (see Section~\ref{sec elementary
    properties}). Let $h$ be a unary qfd operation on $\StS$, from
    $\StS(R,C)$ to $\StS(E_{Q},\emptyset) = \GP(Q)$, let
    $(R_{1},C_{1})$ and $(R_{2},C_{2})$ be sorts of $\StS$ such that $R
    = R_{1}\cup R_{2}$, $C_{1}\cap C_{2} = \emptyset$ and $C =
    C_{1}\cup C_{2}$, and let $\vec z = (z_{1},z_{2})$ with $z_{1}$ a
    $\zeta$-class in $\StS(R_{1},C_{1})$ and $z_{2}$ a $\zeta$-class in
    $\StS(R_{2},C_{2})$.
    
    Then there exist quantifier-free definable operations
    $g_{1,\vec z}\colon \StS(R_{1},C_{1})\to \GP(Q_{1,\vec z})$,
    $g_{2,\vec z}\colon\StS(R_{2},C_{2})\to\GP(Q_{2,\vec z})$, and
    $f_{\vec z}\colon\GP(Q_{1,\vec z}\cup Q_{2,\vec z})\to\GP(Q)$, such
    that
    \begin{itemize}
	\item $f_{\vec z}$ is a composition of unary operations in
	$\VR$;
	
	\item for each $x_{1}\in\StS(R_{1},C_{1})$ in class $z_{1}$ and
	each $x_{2}\in\StS(R_{2},C_{2})$ in class $z_{2}$,
	$h(x_{1}\oplus x_{2}) = f_{\vec z}(g_{1,\vec z}(x_{1})\oplus
	g_{2,\vec z}(x_{2}))$.
    \end{itemize}
\end{proposition}    

\proof
Let $(\delta,\psi_{\edg},(\psi_{q})_{q\in Q})$ be the qfd operation
scheme defining the operation $h$: here $\psi_{\edg}$ defines the
$\edg$ relation, $\psi_{q}$ defines the $q$-ports ($q\in Q$), and there
is no formula of the form $\kappa_{c,d}$ since the range of $h$ is in
$\GP(Q) = StS(E_{Q},\emptyset)$. The formulas $\delta$, $\psi_{\edg}$
and $\psi_{q}$, for $q\in Q$, are in the language of
$(R,C)$-structures.

The atoms of $\delta(v)$ are either of the form
$r(y_{1},\ldots,y_{\rho(r)})$ ($r\in R$), or $v = c$, or $c_{1} =
c_{2}$ ($c,c_{1},c_{2}\in C$). Let $\delta^{1}$ be the formula obtained
from $\delta(v)$ by substituting the Boolean value $0$ (false) for the
following atoms, which are certainly false in a disjoint sum
$x_{1}\oplus x_{2}$, with $x_{1}\in\StS(R_{1},C_{1})$,
$x_{2}\in\StS(R_{2},C_{2})$ and the variable $v$ interpreted in
$x_{1}$:
\begin{itemize}
    \item each $r$-atom such that $r\not\in R_{1}$ and an argument of
    $r$ is $v$ or a constant in $C_{1}$;
    
    \item each $r$-atom such that $r\not\in R_{2}$ and an argument of
    $r$ is a constant in $C_{2}$;
	
    \item each $r$-atom such that $r\in R_{1}\cap R_{2}$, an argument
    of $r$ is a constant in $C_{2}$, and another argument of $r$ is $v$
    or a constant in $C_{1}$;
	
    \item each atom of the form $y = c$ such that $c\in
    C_{2}$ and $y$ is equal to $v$ or to a constant in $C_{1}$.
\end{itemize}
The remaining atoms in $\delta^1$ are either in $QF(R_{1},C_{1},\{v\})$
or in $QF(R_{2},C_{2},\emptyset)$. Note that the $\zeta$-class of an
element of $\StS(R_{2},C_{2})$ determines entirely which formulas in
$QF(R_{2},C_{2},\emptyset)$ it satisfies. For each $\vec z$ as in the
statement of the proposition, we let $\delta^{1,\vec z}$ be the formula
in $QF(R_{1},C_{1},\{v\})$ obtained from $\delta^{1}$ by replacing each
atom in $QF(R_{2},C_{2},\emptyset)$ by the Boolean value 0 or 1
according to the $\zeta$-class $z_{2}$. We observe that if $v$ is a
vertex of $x_{1}\oplus x_{2}$ which happens to be in $x_{1}$, then
$$\delta(v) \iff \delta^{1,\vec z}(v)\qquad\hbox{whenever the
$\zeta$-class of $x_{2}$ is $z_{2}$.}$$
For each $q\in Q$, let  $\psi^{1,\vec z}_{q}$ be defined similarly.
Then we also have, if $v$ is a vertex of $x_{1}\oplus x_{2}$ in
$x_{1}$,
$$\psi_{q}(v) \iff \psi^{1,\vec z}_{q}(v)\qquad\hbox{whenever the
$\zeta$-class of $x_{2}$ is $z_{2}$.}$$
Let also $\delta^{2,\vec z}$ and $\psi^{2,\vec z}_{q}$ be defined
dually. And again, if $i,j\in\{1,2\}$, we let $\psi_{\edg}^{i,j}(v,w)$ be
the formula obtained from $\psi_{\edg}$ by substituting the Boolean value
$0$ for the atoms that are certainly false in a disjoint sum
$x_{1}\oplus x_{2}$ for the variable $v$ interpreted in $x_{i}$ and the
variable $w$ interpreted in $x_{j}$:
\begin{itemize}
    \item each $r$-atom such that $r\not\in R_{i}$ and $v$ is an
    argument of $r$;

    \item each $r$-atom such that $r\not\in R_{j}$ and $w$ is an
    argument of $r$;

    \item each $r$-atom such that $r\not\in R_{1}$ and a constant in
    $C_{1}$ is an argument of $r$;

    \item each $r$-atom such that $r\not\in R_{2}$ and a constant in
    $C_{2}$ is an argument of $r$;

    \item each $r$-atom such that $r\in R_{1}\cap R_{2}$, an argument
    of $r$ is a constant in $C_{2}$, and another argument of $r$ is a
    constant in $C_{1}$;

    \item each $r$-atom such that $r\in R_{1}\cap R_{2}$, an argument
    of $r$ is $v$ (resp. $w$) and another argument of $r$ is a
    constant in $C_{3-i}$ (resp. $C_{3-j}$);

    \item each atom of the form $v = c$ with $c\in C_{3-i}$, $w = c$
    with $c\in C_{3-j}$, or $c_{1} = c_{2}$ with $c_{1}\in C_{1}$ and
    $c_{2}\in C_{2}$;

    \item if $i\ne j$, each $r$-atom such that $r\in R_{1}\cap R_{2}$,
    and $v$ and $w$ are  arguments of $r$.
\end{itemize}
As above, the remaining atoms in $\psi_{\edg}^{1,1}$ are in
$QF(R_{1},C_{1},\{v,w\}) \cup QF(R_{2},C_{2},\emptyset)$, and for
each $\vec z$, we let $\psi_{\edg}^{1,1,\vec z}$ be obtained from
$\psi_{\edg}^{1,1}$ by substituting the Boolean values 0 or 1 for the
atoms in $QF(R_{2},C_{2},\emptyset)$ according to the $\zeta$-class
$z_{2}$. If $v,w$ are vertices of $x_{1}\oplus x_{2}$ in $x_{1}$, and
if the $\zeta$-class of $x_{2}$ is $z_{2}$, then
$$\psi_{\edg}(v,w) \iff \psi^{1,1,\vec z}_{\edg}(v,w).$$
We define $\psi_{\edg}^{2,2,\vec z}$ similarly, and get the analogous
equivalence.

If $i\ne j$, the atoms of $\psi_{\edg}^{i,j}$ are in
$QF(R_{i},C_{i},\{v\})$ and in $QF(R_{j},C_{j},\{w\})$ -- which may
include atoms in $QF(R_{1},C_{1},\emptyset)$ and in
$QF(R_{2},C_{2},\emptyset)$. Again, we let $\psi_{\edg}^{i,j,\vec z}$ be
obtained from $\psi_{\edg}^{i,j}$ by substituting the Boolean values 0 or
1 for the atoms without free variables according to the $\zeta$-classes
$z_{1}$ and $z_{2}$. And we observe that if $v,w$ are vertices of
$x_{1}\oplus x_{2}$, $v$ is in $x_{i}$ and in the $\zeta$-class
$z_{i}$, $w$ is in $x_{j}$ and in the $\zeta$-class $z_{j}$, then
$$\psi_{\edg}(v,w) \iff \psi^{i,j,\vec z}_{\edg}(v,w).$$

Now let $k = 1+\max(Q)$, let $X_{k+1},\ldots,X_{\ell}$ be an
enumeration of the subsets of $QF(R_{1},C_{1},\{y\})$, and let
$Y_{\ell+1},\ldots,Y_{m}$ be an enumeration of the subsets of
$QF(R_{2},C_{2},\{y\})$. Let us denote by $Q_{1}$ the set
$Q\cup\{k+1,\ldots,\ell\}$ and by $Q_{2}$ the set
$Q\cup\{\ell+1,\ldots,m\}$.

We define the qfd operation $g_{1,\vec z}\colon \StS(R_{1},C_{1})\to
\GP(Q_{1})$ defined by the following operation scheme:
$$\delta^{1,\vec z},\qquad \psi_{\edg}^{1,1,\vec z},\qquad
\psi_{q}^{1,\vec z}\hbox{ ($q\in
Q$)},\qquad \theta_{n}\hbox{ ($k+1\le n\le \ell$)}$$
where for each $k+1\le n\le \ell$, $\theta_{n}(v)$ holds if the set of
quantifier-free formulas in $QF(R_{1},C_{1},\{y\})$ satisfied by $v$ is
exactly $X_{n}$.

Similarly, the qfd operation $g_{2,\vec z}\colon \StS(R_{2},C_{2})\to
\GP(Q_{2})$ is defined by the operation scheme
$$\delta^{2,\vec z},\qquad \psi_{\edg}^{2,2,\vec z},\qquad
\psi_{q}^{2,\vec z}\hbox{ ($q\in
Q$)},\qquad \theta_{n}\hbox{ ($\ell+1\le n\le m$)}$$
where for each $\ell+1\le n\le m$, $\theta_{n}(v)$ holds if the set of
quantifier-free formulas in $QF(R_{2},C_{2},\{y\})$ satisfied by $v$ is
exactly $X_{n}$.

Finally, we consider structures $x_{1}\in\StS(R_{1},C_{1})$ and
$x_{2}\in\StS(R_{2},C_{2})$, with $\zeta$-classes respectively $z_{1}$
and $z_{2}$, and we compare the graphs with ports $g_{1,\vec
z}(x_{1})\oplus g_{2,\vec z}(x_{2})$ and $h(x_{1}\oplus x_{2})$. The
above remarks show that these two graphs have the same set of vertices,
the same $q$-ports ($q\in Q$), and the same edges between two vertices
of $x_{1}$ or two vertices of $x_{2}$. On the other hand, $g_{1,\vec
z}(x_{1})\oplus g_{2,\vec z}(x_{2})$ misses the edges of $h(x_{1}\oplus
x_{2})$ that connect a vertex of $x_{1}$ with a vertex of $x_{2}$.

These edges are captured by the formulas $\psi_{\edg}^{1,2,\vec z}$ and
$\psi_{\edg}^{2,1,\vec z}$. Now, if $v$ is a vertex of $x_{1}$ and $w$ is
a vertex of $x_{2}$, we already observed that the truth values of
$\psi_{\edg}^{1,2,\vec z}(v,w)$ and $\psi_{\edg}^{2,1,\vec z}(w,v)$ are
entirely determined by the quantifier-free formulas with one free
variable satisfied by $v$ in $x_{1}$ and by $w$ in $x_{2}$: that is,
they are entirely determined by the (unique) index $k+1\le n\le\ell$
such that $\theta_{n}(v)$ and by the (unique) index $\ell+1\le n\le m$
such that $\theta_{n}(w)$. In other words, $\psi_{\edg}^{1,2,\vec z}(a,b)$
and $\psi_{\edg}^{2,1,\vec z}(b,a)$ are equivalent to disjunctions of
conjunctions of the form
$$\theta_{n}(a)\land \theta_{u}(b)\qquad\hbox{for some $k+1\le
n\le\ell$ and $\ell+1\le u\le m$.}$$

Thus the edges in $h(x_{1}\oplus x_{2})$ from a vertex of $x_{1}$ to
a vertex of $x_{2}$ can be created from $g_{1,\vec z}(x_{1})\oplus
g_{2,\vec z}(x_{2})$ by applying repeatedly the operations (in $\VR$) of
the form $\add_{n,u}$ such that $n\in [k+1,\ell]$,
$\theta_{n}\land\theta_{u}$ is a disjunct of $\psi_{\edg}^{1,2,\vec z}$.

Similarly, the edges in $h(x_{1}\oplus x_{2})$ from a vertex of $x_{2}$
to a vertex of $x_{1}$ can be created from $g_{1,\vec z}(x_{1})\oplus
g_{2,\vec z}(x_{2})$ by applying the appropriate operations of the form
$\add_{u,n}$. The last operation consists in forgetting the auxiliary
ports numbered $k+1$ to $m$, that is, in applying the operation
$\mdf_{D}$, with $D = \{(q,q)\mid q\in Q\}$.
\eop

\subsection{Recognizable sets of graphs with ports}\label{VR
equivalence}

In this section, we consider different notions of recognizability that
can be used for sets of graphs with ports. Let $L\subseteq \GP(P)$.
Then $L$ can be $\VR$-recognizable, as a subset of the $\VR$-algebra
$\GP$. It can also be $\calS$-recognizable, as a subset of the
$\calS$-algebra $\StS$ since $\GP(P) = \StS(E_{P})$. Finally, we
introduce another signature, written $\VR^+$, on $\GP$: it is obtained
from $\VR$ by adding all the qfd operations between the sorts of $\GP$.

\begin{theorem}\label{thm VR equivalence}
    Let $P$ be a finite subset of $\N$ and let $L$ be a subset of
    $\GP(P)$. The following properties are equivalent:
    \begin{description}
	\item[1] $L$ is $\calS$-recognizable;
	
	\item[2] $L$ is $\VR^+$-recognizable;
	
	\item[3] $L$ is $\VR$-recognizable;
    \end{description}
\end{theorem}

\proof
Since the operations of $\VR$ are operations of $\VR^+$, and the
operations of $\VR^+$ are operations of $\calS$, it follows from
Proposition~\ref{easy recog facts} that (1) implies (2), and (2)
implies (3). Thus, we only need to verify that (3) implies (1).

We use Lemma~\ref{FGH-congruence+zeta(T)}, with $\calF = \VR$, $S =
\GP$, $\calG = \calS$, $T = \StS$, and $\zeta$ the type congruence (see
Section~\ref{sec elementary properties}), which relates structures with
sources of the same sort, provided they satisfy the same
quantifier-free formulas. We use the collection $\calH$ of sets
$\calH_{(R,C),P}$ of unary qfd operations from $\StS(R,C)$ to $\GP(P)$.

Let $L$ be a $\VR$-recognizable subset of $\GP(P)$ and let $\equiv$ be
a locally finite $\VR$-congruence on $\GP$ such that $L$ is a union of
$\equiv$-classes. Since $\zeta$ is a locally finite $\calS$-congruence
on $\StS$ (Proposition~\ref{FFVV light}), its restriction to $\GP$ is also a
locally finite $\VR$-congruence; and the intersection of $\equiv$ and
$\zeta$ is a locally finite $\VR$-congruence on $\GP$ which saturates
$L$. Thus we can assume, without loss of generality, that
$\equiv$-equivalent elements of $\GP$ are also $\zeta$-equivalent.

Next we consider the equivalence relation $\approx$ on $\StS$ defined
as in Lemma~\ref{FGH-congruence+zeta(T)}. Note that the identity of
$\GP(P)$ belongs to $\calH_{(E_{P},\emptyset),P}$, so that
$\approx$-equivalent elements of $\GP(P) = \StS(E_{P},\emptyset)$ are
also $\equiv$-equivalent. In particular, $\approx$ saturates $L$ and it
suffices to show that $\approx$ is locally finite and is a
$\calS$-congruence. In view of Lemma~\ref{FGH-congruence+zeta(T)}, it
is enough to verify that $\calH$ satisfies the assumptions of
Lemma~\ref{FGH-congruence} and \ref{FGH-finiteness}.

We first verify the hypothesis of Lemma~\ref{FGH-congruence}.
Let $g$ be an operation of $\calS$: either $g$ is a unary qfd operation
or $g = \oplus$. In the latter case, Proposition~\ref{unary-parallel}
states precisely that the required property holds.

If $g$ is a qfd operation of type $(R_{1},C_{1})\to(R,C)$, and
$h\in\calH_{(R,C),P}$, then $h\circ g$ is a qfd operation
(Lemma~\ref{composition of qfd}) and hence, $h_{1} = h\circ
g\in\calH_{(R_{1},C_{1}),P}$. Now letting $f$ be the identity mapping
of $\GP(P)$, we find that $h(g(x)) = f(h_{1}(x))$ as required. In this
case, $h_{1}$ and $f$ do not depend on the $\zeta$-class of $x$.

Next, we turn to the verification of the hypothesis of
Lemma~\ref{FGH-finiteness}. Let $\phi_{1}$, \dots, $\phi_{k}$ be an
enumeration of the elements of $QF(R,C,\{x\})$ and let
$\chi_{1}$, \dots, $\chi_{\ell}$ be an enumeration of the elements of
$QF(R,C,\{x,y\})$.

Thus, a qfd operation scheme from $\StS(R,C)$ into $\GP(Q)$ consists in
the choice of a formula $\delta = \phi_{i_{0}}$ ($1\le i_{0}\le k$), a
formula $\psi_{\edg} = \chi_{j}$ ($1\le j\le \ell$), a sequence of
formulas $\phi_{i_{1}},\ldots,\phi_{i_{r}}$ ($1\le i_{1} < \ldots <
i_{r} \le k$), and a partition of $Q$ as $Q = Q_{1} \cup \cdots \cup
Q_{r}$: if $q\in Q_{j}$, then $\psi_{q} = \phi_{i_{j}}$. (If $Q =
\emptyset$, then $r=0$.)

Let us now consider two unary qfd operations $g\colon
\StS(R,C)\rightarrow \GP(Q)$ and $g'\colon \StS(R,C)\rightarrow
\GP(Q')$, associated with the same choice of values $i_{0}$, $j$ and
$i_{1} < \ldots < i_{r}$. Let $Q = Q_{1} \cup \cdots \cup Q_{r}$ and
$Q' = Q'_{1} \cup \cdots \cup Q'_{r}$ be the corresponding partitions
of $Q$ and $Q'$. Finally let $\pi, \pi_{0}, \pi_{1},\ldots,\pi_{r}$ be
the following operations in the signature $\VR$. These operations have
the common particularity to not alter the graph structure, and to
modify only the port predicates.

The mapping $\pi_{0}$ shifts every port index of an element of $\GP(Q)$
by $m = \max(Q')$, to yield a graph with ports in $Q+m$, whose port
names do not intersect $Q'$. We let $R_{h} = Q_{h}+m$ for $1\le h \le
r$.

For $1\le h\le r$, $\pi_{h} = \mdf_{D_{h}}$ where
$$D_{h} = \{(a,a)\mid a\in \bigcup_{i<h}Q'_{i} \cup 
\bigcup_{i>h}R_{i}\} \cup (R_{h}\times Q'_{h}).$$
Thus $\pi_{h}$ turns a graph with ports in $Q'_{1}+\cdots+ Q'_{h-1} +
R_{h} + R_{h+1} + \cdots R_{r}$ into a graph with ports in
$Q'_{1}+\cdots+ Q'_{h-1} + Q'_{h} + R_{h+1} + \cdots R_{r}$, with the
same vertex set, the same edge relation, the same $q$-ports for each
$q\in \bigcup_{i<h}Q'_{i} \cup \bigcup_{i>h}R_{i}$, and with each
$r$-port ($r\in R_{h}$) turned into a $q$-port for each $q\in Q'_{h}$.

It is now an easy verification that, if $\pi =
\pi_{r}\circ\cdots\circ\pi_{1}\circ\pi_{0}$, then $g'(x) = \pi(g(x))$
for each $x\in\StS(R,C)$. Thus the quasi-order $\le_{(R,C)}$ defined in
Lemma~\ref{FGH-finiteness} is in fact a finite index equivalence
relation, and this concludes the proof.
\eop

\begin{remark}
    This actually proves also that we get the same recognizable sets of
    graphs with ports, if we consider $\GP(Q)$ as a domain of sort $Q$
    in the algebra of structures \textit{without} sources --- which
    consists of the domains $\StS(R,\emptyset)$ equipped with the
    operations of $\calS$ between them. If we were only interested in
    the equivalence of this recognizability with $\VR$- and
    $\VR^+$-recognizability (or just the equivalence between $\VR$- and
    $\VR^+$-recognizability), we could do with
    Lemmas~\ref{FGH-congruence} and \ref{FGH-finiteness} instead of
    Lemma~\ref{FGH-congruence+zeta(T)}, and with a simpler version of
    Proposition~\ref{unary-parallel}, making no reference to $\zeta$.
\end{remark}

\subsection{Variants of the algebra of graphs with ports}

The first variant considered here replaces the signature $\VR$ by a
smaller signature, which we will see is equivalent to $\VR$ in terms
of recognizability. The second one concerns a certain class of graphs 
with ports, and is central in the definition of the clique-width of a
finite graph.

\subsubsection{A variant of $\VR$ on $\GP$}

In Section~\ref{VR equivalence}, we exhibited signatures larger than
$\VR$, for which all the $\VR$-recognizable sets of graphs with ports
are recognizable: namely the signature $\VR^+$ on $\GP$ and the
signature $\calS$ on the wider algebra $\StS$. In contrast, we exhibit
in this section a smaller signature (in fact, a signature consisting of
$\VR$-derived operations) which does not create new recognizable
subsets.

The basic idea behind the definition of this new signature is the
following: when we evaluate a $\VR$-term $t$ of the form
$\add_{p,q}(t')$, then we add edges from each $p$-port of $G'$, the
value of $t'$, to each of its $q$-ports. It may happen that some edges
from a $p$-port to a $q$-port already exist in $G'$. In this case, we
do not add a parallel edge since we are dealing with simple graphs.
Thus the term $t$ presents a form of redundancy, since some of its
edges may be, in some sense, defined twice.

For disjoint sets of port labels $P$ and $Q$, we denote by $J(P,Q)$ the
set of $\VR$-derived unary operations defined by terms of the form
$f_{1}(f_{2}(\ldots(f_{n}(x))\ldots))$, where the $f_{i}$ are of the
forms $\add_{p,q}$ or $\add_{q,p}$ for $p$ in $P$ and $q$ in $Q$. Since
the operations $\add_{p,q}$ are idempotent and commute with one
another, an operation in $J(P,Q)$ is completely described by a subset
of $(P\times Q)\cup(Q\times P)$. Thus $J(P,Q)$ is finite, although one
can write infinitely many terms specifying its elements. For each
element $J\in J(P,Q)$, we let $\otimes_{J}$ denote the binary operation
defined, for $G\in \GP(P)$ and $H\in\GP(Q)$, by $G\otimes_{J} H =
J(G\oplus H)$.

We observe that in the evaluation of a term of the form
$t\otimes_{J}t'$, the application of $\otimes_{J}$ does not recreate
edges that already exist in $G$, the value of $t$, or in $G'$, the
value of $t'$ since the $\add_{p,q}$ operations forming $\otimes_{J}$
add edges between the disjoint graphs $G$ and $G'$ (because $p$ and $q$
are not port labels of the same argument graphs).

Now the signature $\NLC$ consists of the operations $\otimes_{J}$ as
above, the unary qfd operations of the form $\fg_{p}$ and
$\ren_{p\rightarrow q}$ as defined in Remark~\ref{rename, forget}, and
the constants $\p$ and $\p^{\form{loop}}$ as in $\VR$. We denote by
$\GP^{\NLC}$ the $\NLC$-algebra of graphs with ports.

\begin{remark}
    The notation $\NLC$ refers to a very similar algebra used by Wanke
    \cite{Wanke94}.
\end{remark}

\begin{example}\label{important example NLC}
    We have in fact already encountered $\NLC$-operations and
    $\NLC$-derived operations.

    The $\VR$-derived operation $f_{\vec z}$ whose existence is proved
    in Proposition~\ref{unary-parallel} is actually $\NLC$-derived.
    Consider indeed the last paragraphs of the proof of that
    proposition: the operation $f_{\vec z}$ is obtained by first
    composing operations of the form $\add_{n,u}$ and $\add_{u,n}$,
    where the pairs $(n,u)$ lie in a certain subset of $[k+1,\ell]
    \times [\ell+1,m]$ and the pairs $(u,n)$ lie in another subset of
    $[\ell+1,m] \times [k+1,\ell]$, and then composing operations of
    the form $\fg_{p}$.
    
    One can also check that the operations $\pi_{0},\ldots,\pi_{r}$ at
    the end of the proof of Theorem~\ref{thm VR equivalence} are
    $\NLC$-derived.
\end{example}

\begin{proposition}\label{prop NLC}
    Let $P$ be a finite subset of $\N$ and let $L$ be a subset of
    $\GP(P)$. Then $L$ is $\VR$-recognizable if and only if $L$ is
    $\NLC$-recognizable.
\end{proposition}

\proof
The proof is a simple extension of the proof of Theorem~\ref{thm VR
equivalence}.

Since the operations of $\NLC$ are $\VR$-derived, every
$\VR$-recognizable subset of $\GP$ is $\NLC$-recognizable. For the
converse, we observe that the proof that (1) implies (3) in
Theorem~\ref{thm VR equivalence} can be modified to show that an
$\NLC$-recognizable set of $\GP$ is $\calS$-recognizable.

Again, we rely on Lemma~\ref{FGH-congruence+zeta(T)}, but now with
$\calF = \NLC$, $S = \GP$, and $\calG$, $T$, $\zeta$ and $\calH$ as
in Theorem~\ref{thm VR equivalence}.

In order to justify the fact that the arguments used in the proof of
Theorem~\ref{thm VR equivalence} are also valid with these assumptions, we
refer to Example~\ref{important example NLC}. Indeed this example shows
two things: on one hand, the operation $f_{\vec z}$ in
Proposition~\ref{unary-parallel} is in fact $\NLC$-derived, so that the
first hypothesis of Lemma~\ref{FGH-congruence+zeta(T)} is satisfied by
this new choice of $\calF$ and $S$. On the other hand the finiteness
hypothesis of Lemma~\ref{FGH-finiteness} is also satisfied with this
new value of $\calF = \NLC$. This completes the proof.
\eop

\subsubsection{Graphs whose port labels partition the vertex set}\label{port partition}

In certain contexts, and in particular in the definition of the
clique-width of a graph (see Remark~\ref{rk clique-width} below), one
needs to consider graphs with ports where port labels partition the
vertex set. More precisely, for each set of port labels $P$, let
$\GP^\pi(P)$ be the set of elements of $\GP(P)$ such that each vertex
is a port, and no vertex is both a $p$-port and a $q$-port for $p\ne
q$. Let also $\GP^\pi = (\GP^\pi(P))$.

Note that $\GP^\pi$ is preserved by the operations of the form
$\oplus$, $\add_{p,q}$ and $\ren_{p\rightarrow q}$. These operations
form the signature $\VR^\pi$, and $\GP^\pi$ is a $\VR^\pi$-algebra.

\begin{remark}
    The operation $\add_{p,q}$ is written $\alpha_{p,q}$ in
    \cite{BC-Olariu}.
\end{remark}

\begin{remark}\label{rk clique-width}
    The \emph{clique-width} of a finite graph $G$, denoted by $cwd(G)$,
    is defined as the smallest cardinality of a set $P$ such that $G$
    is the value of a (finite) $\VR^\pi$-term using a set $P$ of port
    labels, see \cite{BC-Olariu,ar:Corneil-etal99}.
    
    For algorithmic applications \cite{CourcelleMakowskyRotics2000}, it
    is useful to have efficient recognition algorithms for classes of
    graphs of clique-width at most $k$. At the moment we only know that
    this problem is $NP$. It is polynomial for $k\leq3$, see
    \cite{ar:Corneil-etal99}.
\end{remark}

\begin{proposition}
    Let $L$ be a subset of $\GP^\pi(P)$. Then $L$ is
    $\VR^\pi$-recognizable if and only if $L$ is $\VR$-recognizable.
\end{proposition}

\proof
Since $\VR^\pi$ consists of operations in $\VR$, every locally finite
$\VR$-cong\-ruence on $\GP$ induces a locally finite $\VR^\pi$-congruence
on $\GP^\pi$. In particular, if $L$ is $\VR$-recognizable, and hence is
saturated by a locally finite $\VR$-congruence on $\GP$, then $L$ is
saturated by a locally finite $\VR^\pi$-congruence on $\GP^\pi$, and
hence $L$ is $\VR^\pi$-recognizable.

To prove the converse, we first introduce the mapping $\sigma\colon
\GP\rightarrow \GP^\pi$ defined as follows. If $G\in \GP(P)$, then
$\sigma(G)$ is the graph in $\GP^\pi(2^P)$ with the same set of
vertices and the same edge relation as $G$, and such that for each
vertex $v$ and each $X\subseteq P$, $v$ is an $X$-port in $\sigma(G)$
if and only if $X$ is the set of $p\in P$ such that $v$ is a $p$-port
in $G$. We say that a port label $p$ is \textit{void in $G$} if there
are no $p$-ports in $G$.

Now let us assume that $L$ is $\VR^\pi$-recognizable, and let $\equiv$
be a locally finite congruence on $\GP^\pi$ saturating it. If $G, H\in
\GP(P)$, we let $G \sim H$ if $\sigma(G)$ and $\sigma(H)$ have the same
non-void port labels, and $\sigma(G) \equiv \sigma(H)$. It is
immediately verified that $\sim$ is a locally finite equivalence
relation.

We now verify that $\sim$ is a $\VR$-congruence. If $G\in\GP(P)$ and
$H\in\GP(Q)$, it is easily seen that $\sigma(G\oplus H) = \sigma(G)
\oplus \sigma(H)$. If $p,q\in P$, then $\sigma(\add_{p,q}(G)) =
f(\sigma(G))$ where $f$ is the composition of the operations
$\add_{X,Y}$ for each $X,Y\subseteq P$ such that $p\in X$ and $q\in Y$.
Finally, one can verify that if $D\subseteq P\times Q$, then
$\sigma(\mdf_{D}(G)) = g(\sigma(G))$ where $g$ is the composition of
the operations $\ren_{X\rightarrow Y}$, where $X\subseteq P$,
$Y\subseteq Q$ and $Y = D\inv(X) = \{q\in Q\mid (p,q)\in D\hbox{ for
some $p\in P$}\}$.

It is a routine task to derive from these observations the fact that
$\sim$ is a $\VR$-congruence. We now need to verify that $\sim$
saturates $L$. Let $G\in L$ and $G \sim H$. In particular,
$G\in\GP^\pi$, so that the non-void port labels of $\sigma(G)$ are
exactly the sets $\{p\}$ where $p$ is a non-void port label of $G$.
Since $\sigma(G)$ and $\sigma(H)$ have the same non-void port labels,
$H$ is also in $\GP^\pi$. Moreover, if $h$ is the composition of the
operations $\ren_{\{p\}\rightarrow p}$ ($p$ non-void in $G$), then $G =
h(\sigma(G))$ and $H = h(\sigma(H))$. Since $h$ is $\VR^\pi$-derived,
it follows that $G \equiv H$, and hence $H\in L$.
This concludes the proof.
\eop

\section{The algebra of graphs with sources}\label{sec GS}

Recall that we call \textit{graphs with sources} the elements of $\StS$
of sort $(E,C)$, where $E = \{\edg\}$ and $C$ is some finite set of
source labels, and that we write $\GS(C)$ for $\StS(E,C)$ (see
Section~\ref{sec rel structures}).

\subsection{The signature $\HR$}\label{sec signature HR}

The disjoint union $\oplus$ and the operations of the form
$\srcren_{a\rightarrow b}$, $\srcfg_{a}$ and $\fus_{a,b}$ (defined in
Example~\ref{def HR ops}) preserve graphs with sources. We denote by
$\HR$ the signature consisting of all these operations, so $\GS$ is an
$\HR$-algebra.

We note the following properties of $\HR$-recognizability.

\begin{proposition}\label{S2HR}
    Let $C$ be a finite set of source labels. Every
    $\calS$-recognizable subset of $\StS(E,C)$ is $\HR$-recognizable.
\end{proposition}

\proof
This is a simple consequence of Proposition~\ref{easy recog facts} and
of the observation given above that the operations of $\HR$ are also
operations of $\calS$.
\eop

Note that the class $\Grph$ of graphs, defined in Section~\ref{sec rel
structures}, is equal to $\GP(\emptyset)$ as well as to $\GS(\emptyset)
= \StS(E)$. Thus $\VR$-recognizability and $\HR$-recognizability are
properties of subsets of $\Grph$.

\begin{corollary}\label{VR2HR}
    Let $L$ be a set of graphs (a subset of $\Grph$). If $L$ is
    $\VR$-recognizable, then it is $\HR$-recognizable.
\end{corollary}

\proof
This follows immediately from Proposition~\ref{S2HR} and
Theorem~\ref{thm VR equivalence}.
\eop

\begin{remark}
    Intuitively, the $\VR$-operations are more powerful than the
    $\HR$-operations (every $\HR$-context-free set of simple graphs is 
    $\VR$-context-free but the converse is not true, Courcelle
    \cite{BC:VR/HR}), but the $\HR$-operations are not among the
    $\VR$-operations, nor are they derived from them.
\end{remark}

We will see in Sections~\ref{proof no Kmm} and \ref{other finiteness}
sufficient conditions for $\HR$-recognizable sets to be
$\VR$-recognizable, and in Section~\ref{HR not VR}, examples of
$\HR$-recognizable sets which are not $\VR$-recognizable.

\subsection{Variants of the algebra of graphs with sources}

We find in the literature a number of variants of the signature $\HR$
or of the algebra $\GS$. We now discuss these different variants, to
verify that they do not introduce artefacts from the point of view of
recognizability.

\subsubsection{The signature $\HR_{\parallel}$}

Let $\HR_{\parallel}$ denote the signature on $\GS$ obtained by
substituting the parallel composition $\parallel$ for $\oplus$ (see
Section~\ref{sec StS-parallel}). With the same proof as Proposition
\ref{prop StS-parallel}, we get the following result.

\begin{proposition}\label{prop HR-parallel}
    Let $L$ be a subset of $\GS$. Then $L$ is $\HR$-recognizable
    if and only if it is $\HR_{\parallel}$-recognizable.
\end{proposition}

\subsubsection{Source-separated graphs}

As in Section~\ref{sec source separation}, we now discuss the class
$\GS_{\sep}$ of source separated graphs. The operations of $\HR$ all
preserve source separation, except for $\fus_{a,b}$, but we defined in
Example~\ref{exemple HRsep} the operation $\fus_{a\rightarrow b} =
\srcfg_{a}\circ\fus_{a,b}$ which does. Let $\HR_{\sep}$ be the signature
on $\GS_{\sep}$ consisting of $\oplus$ and the qfd unary operations of
the form $\srcren_{a\rightarrow b}$, $\srcfg_{a}$ and
$\fus_{a\rightarrow b}$.

\begin{proposition}\label{prop HRsep}
    Let $L$ be a subset of $\GS_{\sep}$. Then $L$ is $\HR$-recognizable
    if and only if it is $\HR_{\sep}$-recognizable.
\end{proposition}

\proof
Since $\HR_{\sep}$ consists only of $\HR$-derived operations, every
$\HR$-recog\-nizable set subset of $\GS_{\sep}$ is also
$\HR_{\sep}$-recognizable.

The proof of the converse is a variant of the proof of Theorem~\ref{thm
Ssep-recognizable}. First we note that the type relation $\zeta$ (see
Section~\ref{sec elementary properties}) is also an $\HR$-congruence on
$\GS$. We use the same mapping $h$ defined in the proof of
Theorem~\ref{thm Ssep-recognizable}, that maps a graph with sources
$S\in \GS(C)$ to a source-separated graph $h(S) \in \GS_{\sep}(C)$ by
splitting sources that were identified in $S$. We refer to that proof
for notation used here.

If $L$ is an $\HR_{\sep}$-recognizable subset of $\GS_{\sep}$ and $\equiv$
is a locally finite $\HR_{\sep}$-congruence recognizing it, we define a
relation $\sim$ on $\GS$ as follows. If $S,T\in\GS(C)$, we say that
$S\sim T$ if $\zeta(S) = \zeta(T)$ and $h(S)\equiv h(T)$. As in the
proof of Theorem~\ref{thm Ssep-recognizable}, $\sim$ is easily seen to
be a locally finite equivalence relation. It is also easily seen that
$\sim$ is preserved under the $\HR_{\sep}$-operation $\oplus$.

We now need to verify that if $S\sim T\in\GS(C)$ and $g$ is one of the
unary operations of $\HR_{\sep}$ defined on $GS(C)$, then $g(S) \sim
g(T)$. Again, Proposition~\ref{FFVV light} shows that
$\zeta(g(\zeta(S))) = \zeta(g(\zeta(T)))$ and we want to show that
$h(g(S)) \equiv h(g(T))$. The graphs $S$ and $T$ are fixed for the rest
of this proof. We write $h_{0}$, $C_{0}$ and $C_{1}$ for $h_{0}^S$,
$C_{0}^S$ and $C_{1}^S$.

As in the proof of Theorem~\ref{thm Ssep-recognizable}, it suffices to
construct an $\HR_{\sep}$-derived operation $k$, depending on $g$ and
$\zeta(S)$, such that $h(g(S)) = k(h(S))$ and $h(g(T)) = k(h(T))$.
There is no reason why the operation $k$ constructed in the proof of
Theorem~\ref{thm Ssep-recognizable} should be $\HR_{\sep}$-derived, but
the operations $g$ considered here, namely $\srcren_{a\rightarrow b}$,
$\srcfg_{a}$ and $\fus_{a\rightarrow b}$ are simple enough that we can
directly construct a suitable $k$ in each case.

\paragraph{If $g = \srcren_{a\rightarrow b}$} Then $g$ is defined on
$\GS(C)$ (where $a\in C$ and $b\not\in C$) and its range is
$\GS(C\setminus \{a\} \cup \{b\})$. One verifies that $h(\srcren_{a\rightarrow
b}(S))$ is equal to:
\begin{itemize}
    \item $\srcren_{a\rightarrow b}(h(S))$ if $a\in C_{1}$ and $b >
    h_{0}(a)$;
    \item $\srcren_{a\rightarrow h_{0}(a)}(\srcren_{h_{0}(a)\rightarrow
    b}(h(S)))$ if $a\in C_{1}$ and $b < h_{0}(a)$;
    \item $\srcren_{a\rightarrow b}(h(S))$ if $a\in C_{0}$ and $b < c$
    for every $c\in C_{1}$ such that $h_{0}(c) = a$;
    \item $\srcren_{c\rightarrow b}(\srcren_{b\rightarrow c}(h(S)))$ if
    $a\in C_{0}$ and $b > c = \min\{d\in C_{1} \mid h_{0}(d) = a$.
\end{itemize}

\paragraph{If $g = \srcfg_{a}$} Then $g$ is defined on $\GS(C)$ (where
$a\in C$) and its range is $\GS(C\setminus a)$. One verifies that
$h(\srcfg_{a}(S))$ is equal to:
\begin{itemize}
    \item $\fus_{a\rightarrow h_{0}(a)}(h(S))$ if $a\in C_{1}$;
    \item $\fus_{a\rightarrow c}$ if $a\in C_{0}$, ${h_{0}}\inv(a) \ne
    \emptyset$ and $c = \min\{{h_{0}}\inv(a)\}$;
    \item $\srcfg_{a}(h(S))$ if $a\in C_{0}$ and ${h_{0}}\inv(a) =
    \emptyset$.
\end{itemize}

\paragraph{If $g = \fus_{a\rightarrow b}$} Then $g$ is defined on
$\GS(C)$ (where $a\ne b\in C$) and its range is $\GS(C\setminus a)$. One
verifies that $h(\fus_{a\rightarrow b}(S))$ is equal to:
\begin{itemize}
    \item $\srcren_{a\rightarrow h_{0}(a)}(\fus_{h_{0}(a)\rightarrow
    h_{0}(b)}(h(S)))$ if $a\in C_{1}$ and $h_{0}(b) < h_{0}(a)$;
    \item $\srcren_{a\rightarrow h_{0}(b)}(\fus_{h_{0}(b)\rightarrow
    h_{0}(a)}(h(S)))$ if $a\in C_{1}$ and $h_{0}(b) > h_{0}(a)$;
    \item $\srcren_{a\rightarrow h_{0}(a)}(h(S))$ if $a\in C_{1}$ and
    $h_{0}(b) = h_{0}(a)$;
    \item $\fus_{a\rightarrow h_{0}(b)}(h(S))$ if $a\in C_{0}$ and $a >
    h_{0}(b)$;
    \item $\srcren_{a\rightarrow c}(\fus_{h_{0}(b)\rightarrow
    a}(h(S)))$ if $a\in C_{0}$, and $c =
    \min\{h_{0}(b),h_{0}\inv(a)\}$, and $a < h_{0}(b)$;
    \item $\srcren_{a\rightarrow c}(\fus_{a\rightarrow c}(h(S)))$ if
    $a\in C_{0}$, $a = h_{0}(b)$ and $c = \min\{d\in C_{1}\mid h_{0}(d)
    = a\}$.
\end{itemize}
This concludes the proof.
\eop

Again with the same proof as for Proposition \ref{prop StS-parallel},
we can show that the operation $\oplus$ can be replaced by $\parallel$
in the signature $\HR_{\sep}$ -- yielding the signature $\HR_{\sep,\parallel}$.

\begin{proposition}\label{prop HR-sep-parallel}
    Let $L$ be a subset of $\GS_{\sep}$. Then $L$ is $\HR_{\sep}$-recognizable
    if and only if it is $\HR_{\sep,\parallel}$-recognizable.
\end{proposition}

\subsubsection{Other variants}\label{sec: other variants}

The equivalence between $\HR_{\sep,\parallel}$- and
$\HR_{\parallel}$-recognizability for a set of source-separated graphs
-- a consequence of Propositions~\ref{prop HR-parallel}, \ref{prop
HRsep} and \ref{prop HR-sep-parallel} -- was already established by
Courcelle in \cite{BC-MSCS} for graphs with multi-edges (see
Section~\ref{sec simple multiple}). In the same paper, Courcelle
established the equivalence between $\HR_{\sep}$- and
$\calB$-recognizability for several variants $\calB$ of the signature
$\HR$, which we now describe. We refer to \cite{BC-MSCS} for the
proofs.

For each finite set $C$ of source labels, let $\srcfg_{all}$ be the
composition of the operations $\srcfg_{c}$ for each $c\in C$ (in any
order). Let also $\sqbox_{C}$ be the following binary operation on
$\GS_{\sep}$, of type $(C,C)\rightarrow \emptyset$: if
$G,H\in\GS_{\sep}(C)$, then $G\Box_{C}H = \srcfg_{all}(G\parallel H)$:
$G\sqbox_{C} H$ is obtained by first taking the parallel composition
$G\parallel H$, and then forgetting all source labels.

Let $\calC\calS$ be the signature on $\GS_{\sep}$, which consists only
of the $\sqbox_{C}$ operations.

Let $\HR^{\fg}$ be the derived signature of $\HR_{\parallel}$, which
consists of the operations $\srcfg_{all}$ and $\parallel$.

Let $\HR^{\ren}$ be the subsignature of $\HR_{\parallel}$, which consists
of the operations $\srcren_{p\rightarrow q}$ and $\parallel$.

Let $\HR_{\sep}^{\ren}$ be the subsignature of $\HR_{\sep,\parallel}$, which
consists of the operations $\parallel$ and those operations
$\srcren_{p\rightarrow q}$ which preserve source separation.

The following result is a compilation of \cite[Section 4]{BC-MSCS}.

\begin{proposition}\label{prop: other variants}
    If $L\subseteq \GS$, then $L$ is $\HR$-recognizable if and only if
    $L$ is $\HR^{\ren}$-recognizable.

    If $L\subseteq \GS_{\sep}$, then $L$ is $\HR_{\sep}$-recognizable if
    and only if $L$ is $\HR_{\sep}^{\ren}$-recognizable.

    If $L\subseteq \Grph$, the following are equivalent:
    \begin{itemize}
	\item $L$ is $\HR$-recognizable;
	
	\item $L$ is $\calC\calS$-recognizable;
	
	\item $L$ is $\HR^{\fg}$-recognizable.
    \end{itemize}
\end{proposition}

\begin{remark}
    The notation $\calC\calS$ refers to the notion of \textit{fully
    cutset-regular} sets of graphs, introduced by Abrahamson and
    Fellows \cite{Abr-Fel}. Full cutset-regularity is equivalent to
    $\calC\calS$-recognizability.
\end{remark}

In \cite{BC-MSCS}, Courcelle also shows a number of closure properties
of the class of $\HR_{\sep}$-recognizable sets of source-separated graphs
with sources. In particular, it is shown that this class contains all
singletons and it is closed under the operations of $\HR_{\sep}$
\cite[Section 6]{BC-MSCS}.

Finally Courcelle shows the following result \cite[Theorem
6.7]{BC-MSCS}.

\begin{proposition}
    Let $L\in\GS(C)$. Then $L$ is $\HR$-recognizable if and only if
    $\srcfg_{all}(L)$ is $\HR$-recognizable.
\end{proposition}

\section{Finiteness conditions ensuring that $\HR$- and $\VR$-recognizability coincide}\label{sec Knn}

We saw that a $\VR$-recognizable set of graphs is always
$\HR$-recognizable (Corollary~\ref{VR2HR}). The converse does not hold
in general, as we discuss in Section \ref{HR not VR}. We first explore
structural conditions on graphs, which are sufficient to guarantee that
an $\HR$-recognizable set of graphs is also $\VR$-recognizable.

Let $\Knn$ be the directed complete bipartite graph with $n+n$
vertices. A directed graph $G\in\Grph$ is \emph{without $\Knn$} if it
has no subgraph isomorphic to $\Knn$. The main result in this section
is the following.

\begin{theorem}\label{no Kmm}
    Let $n$ be an integer. An $\HR$-recognizable set of graphs without
    $\Knn$ is $\VR$-recognizable.
\end{theorem}

This theorem is proved in Section \ref{proof no Kmm}, and some of its
corollaries are discussed in Section~\ref{other finiteness}.

Note that results similar to Corollary~\ref{VR2HR} and
Theorem~\ref{no Kmm} hold for $\VR$- and $\HR$-equational sets of
graphs. As explained in the introduction, such sets are exactly the
context-free sets of graphs, formally specified in terms of recursive sets of
equations using the operations of $\VR$ and $\HR$ respectively.
Specifically, the following results are known to hold:

\begin{itemize}
    \item every $\HR$-equational set of simple directed graphs is
    $\VR$-equational (Courcelle \cite{HbGraGraRozenberg97});
    
    \item if a $\VR$-equational set of directed graphs is without
    $\overrightarrow K_{n,n}$ for some $n$, then it is $\HR$-equational
    (by the main theorem in Courcelle \cite{BC:VR/HR} and
    Lemma~\ref{no Kpp} below).
\end{itemize}

Thus the same combinatorial condition is sufficient to guarantee the
equivalence between $\VR$- and $\HR$-recognizability, as well as
between $\VR$- and $\HR$-equationality. A further similar result
concerning monadic second-order definability and using a stronger
combinatorial property will be discussed in Section~\ref{sparse MSO}.

\subsection{Proof of Theorem~\ref{no Kmm}}\label{proof no Kmm}

We first record the following observation.

\begin{lemma}\label{lemma YYY}
    Let $G$ be a directed graph and let $x,y$ be two vertices of $G$
    that are not adjacent, and such that there is no vertex $z$ such
    that both $(x,z)$ and $(y,z)$ (resp. both $(z,x)$ and $(z,y)$) are
    edges. Let $H$ be obtained from $G$ by identifying $x$ and $y$. If
    $G$ contains $\Kmm$ as a subgraph, then so does $H$.
\end{lemma}

\proof
    Let $K$ be a subgraph of $G$ isomorphic to $\Kmm$. From the
    hypothesis, the vertices $x$ and $y$ are not both in $K$. It
    follows that $K$ is still isomorphic to a subgraph of $H$.
\eop

The proof of Theorem~\ref{no Kmm} will proceed as follows. We consider
an $\HR$-recognizable set $L$ of finite graphs without $\Knn$ and we
denote by $m$ the largest integer such that $\Kmm$ is a subgraph of a
graph in $L$. Such an integer exists by hypothesis.

Since we are talking about source-less graphs, the set $L$ is
$\HR_{\sep}$-recog\-nizable by Proposition~\ref{prop HRsep}, and we
consider a locally finite $\HR_{\sep}$-congruence $\equiv$ saturating
$L$. We will define a locally finite $\NLC$-congruence $\sim$ on $\GP$
that also saturates $L$. By Proposition~\ref{prop NLC}, this suffices
to show that $L$ is $\VR$-recognizable. The definition of $\sim$ makes
use of the notion of expansion of a graph, defined below.

Note that the following definitions depend on the integer $m$, even
though terminology and notation do not make this dependence explicit.

\paragraph{Small and large port labels and formulas}
Let $G\in\GP(P)$ be a graph with ports. If $p\in P$, we denote by
$p_{G}$ the set of $p$-ports of $G$. We say that a port label $p$ is
\emph{void in $G$} if $p_{G}$ is empty, we say that $p$ is \emph{small
in $G$} if $1\leq \card(p_{G}) \leq m$ and that $p$ is \emph{large in
$G$} if $\card(p_{G}) > m$.

Observe that if the port labels $p$ and $q$ are both large in $G$, then
$\add_{p,q}(G)$ contains $\overrightarrow{K}_{m+1,m+1}$ as a subgraph.

Moreover, if $p$ is large in $G$, if $r_{1},\ldots,r_{k}$ are small in
$G$, let
$$H = \add_{p,r_{1}}\add_{p,r_{2}}\cdots\add_{p,r_{k}}(G).$$
For $i=1,\ldots,k$, let $n_{i} = \card({r_{i}}_{G})$. If $H$ does not
contain $\overrightarrow{K}_{m+1,m+1}$, then we must have $n_{1}+\cdots
+n_{k} \le m$. If $G$ already contains edges from the $p$-ports to
other vertices, then $n_{1}+\cdots +n_{k} < m$. The notion of expansion
below will make it possible to handle this sort of complicated
situation (see Example~\ref{ex: expansion} below).

Let us say that a closed first-order formula is \textit{small} if it
has quantifier-depth at most $2m+2$. Note that the existence of a
subgraph isomorphic to $\overrightarrow{K}_{m+1,m+1}$ can be expressed
by a first-order formula of quantifier-depth $2m+2$.

\paragraph{Expansions}
We will define supergraphs of $G\in\GP(P)$ called \emph{expansions},
that contain information relevant to the distribution of small and
large port labels, and where ports are represented by sources.
Furthermore, it will be possible to simulate an $\NLC$-operation on $G$
that does not create $\overrightarrow{K}_{m+1,m+1}$ subgraphs by
$\HR$-operations on expansions of $G$. These expansions will then be
used to transform the $\HR_{\sep}$-congruence $\equiv$ into an
$\NLC$-congruence $\sim$.

Furthermore, we will define $\sim$ in such a way that two equivalent
graphs satisfy the same small first-order formulas.

We now give formal definitions. For each port label $p$, we define a
set $C(p)$ of source labels,
$$C(p) = \{in(p,i),out(p,i),s(p,i) \mid 1\leq i \leq m\}.$$
If $P$ is a set of port labels, $C(P)$ denotes the union of the $C(p)$,
for $p$ in $P$.

Let $G\in \GP(P)$ be a graph with ports, let $C \subseteq C(P)$, and
let $\bar{G}$ be a graph in $\GS_{\sep}(C)$. We say that $\bar{G}$ is an
\emph{expansion} of $G$ if the following conditions hold:
\begin{itemize}
    \item[(1)] $\bar G$ has no subgraph isomorphic to
    $\overrightarrow{K}_{m+1,m+1}$.
    
    \item[(2)] Except for the labeling of ports and sources, $G$ is a
    subgraph of $\bar{G}$. The sources of $\bar G$, and its vertices 
    and edges not in $G$, are specified by Conditions (3) and (4).
    
    \item[(3)] If $p$ is small in $G$, then each $p$-port of $G$ is
    an $s(p,i)$-source of $\bar{G}$ for some integer $i\leq m$.
    Different $p$-ports are of course labelled by different source
    labels. There are no $in(p,j)$- or $out(p,j)$-sources.
    
    \item[(4)] If $p$ is large in $G$, then there may be vertices of 
    $\bar G$ that are not in $G$, with source labels of the form
    $in(p,i)$ or $out(p,i)$ for some $i\le m$. Moreover, there is
    an edge in $\bar G$ from each vertex of $p_{G}$ to each
    $in(p,i)$-source, and from each $out(p,i)$-source to each vertex
    in $p_{G}$. There are no $s(p,j)$-sources.
\end{itemize}

In particular, $G$ may have several different expansions, but it has
only a finite number of expansions (up to isomorphism). This number
is bounded by a function depending on $m$ and the cardinality of $P$.
Indeed, for each small port label $p$, there is only a bounded number
of ways to make $p$-ports into $s(p,i)$-sources (see (3)), and for
each large port label $p$, there is a bounded number of ways to create
$in(p,i)$- and $out(p,i)$-sources (see (4)).


\begin{example}\label{ex: expansion}
    Let $m=2$, and let $G$ be a graph with port labels $p,q,r$.
    Suppose that $G$ has 4 $p$-ports, 2 $q$-ports and 1 $r$-port, so 
    that $p$ is large, and $q$, $r$ are small in $G$, see
    Figure~\ref{fig: example expansion}. Then in any
    expansion of $G$, every $q$- and $r$-port will be a source, say
    labeled by $s(q,1)$, $s(q,2)$ and $s(r,2)$ (there is only one
    $s(r,i)$-source, but it is not required that these sources should
    be labeled with consecutive numbers starting at 1).
    \begin{figure}[htb]
	\centering
	\begin{picture}(70,55)(0,-50)
	\put(10,-47){\framebox(60,35){}}
	\put(2,-45){\hbox{$H$}}
	\put(12,-45){\hbox{$G$}}


	\node[Nfill=y,fillcolor=Black,Nw=0.6,Nh=0.6,Nmr=0.2,ExtNL=y,NLangle=-100.0,NLdist=2.0](n0)(12.0,-12.0){$p$}
	\node[Nfill=y,fillcolor=Black,Nw=0.6,Nh=0.6,Nmr=0.2,ExtNL=y,NLangle=-100.0,NLdist=2.0](n1)(22.0,-12.0){$p$}
	\node[Nfill=y,fillcolor=Black,Nw=0.6,Nh=0.6,Nmr=0.2,ExtNL=y,NLangle=-100.0,NLdist=2.0](n2)(32.0,-12.0){$p$}
	\node[Nfill=y,fillcolor=Black,Nw=0.6,Nh=0.6,Nmr=0.2,ExtNL=y,NLangle=-80.0,NLdist=2.0](n3)(42.0,-12.0){$p$}
	\node[Nfill=y,fillcolor=Black,Nw=0.6,Nh=0.6,Nmr=0.2,ExtNL=y,NLangle=-180.0,NLdist=1.0](n4)(70.0,-17.0){$q$}
	\node[Nfill=y,fillcolor=Black,Nw=0.6,Nh=0.6,Nmr=0.2,ExtNL=y,NLangle=-180.0,NLdist=1.0](n5)(70.0,-23.0){$q$}
	\node[Nfill=y,fillcolor=Black,Nw=0.6,Nh=0.6,Nmr=0.2,ExtNL=y,NLangle=-180.0,NLdist=1.0](n6)(70.0,-35.0){$r$}
	\node[Nfill=y,fillcolor=Black,Nw=0.6,Nh=0.6,Nmr=0.2,ExtNL=y,NLangle=0.0,NLdist=1.0](n4)(70.0,-17.0){$s(q,1)$}
	\node[Nfill=y,fillcolor=Black,Nw=0.6,Nh=0.6,Nmr=0.2,ExtNL=y,NLangle=0.0,NLdist=1.0](n5)(70.0,-23.0){$s(q,2)$}
	\node[Nfill=y,fillcolor=Black,Nw=0.6,Nh=0.6,Nmr=0.2,ExtNL=y,NLangle=0.0,NLdist=1.0](n6)(70.0,-35.0){$s(r,2)$}
	\node[Nfill=y,fillcolor=Black,Nw=0.6,Nh=0.6,Nmr=0.2,ExtNL=y,NLangle=10.0,NLdist=2.0](n7)(7.0,0.0){$out(p,2)$}
	\node[Nfill=y,fillcolor=Black,Nw=0.6,Nh=0.6,Nmr=0.2,ExtNL=y,NLangle=10.0,NLdist=1.0](n8)(27.0,0.0){$in(p,1)$}
	\node[Nfill=y,fillcolor=Black,Nw=0.6,Nh=0.6,Nmr=0.2,ExtNL=y,NLangle=10.0,NLdist=1.0](n9)(43.0,0.0){$in(p,2)$}
	\node[Nfill=y,fillcolor=Black,Nw=0.6,Nh=0.6,Nmr=0.2,ExtNL=y,NLangle=0.0,NLdist=2.0](n10)(37.0,-30.0){$x$}

	\drawedge(n10,n0){}
	\drawedge(n10,n1){}
	\drawedge(n10,n2){}
	\drawedge(n10,n3){}

	\drawedge(n7,n0){}
	\drawedge(n7,n1){}
	\drawedge(n7,n2){}
	\drawedge(n7,n3){}

	\drawedge(n0,n8){}
	\drawedge(n1,n8){}
	\drawedge(n2,n8){}
	\drawedge(n3,n8){}

	\drawedge(n0,n9){}
	\drawedge(n1,n9){}
	\drawedge(n2,n9){}
	\drawedge(n3,n9){}

	\end{picture}
	\caption{$H$ is an expansion of $G$}
	\label{fig: example expansion}
    \end{figure}
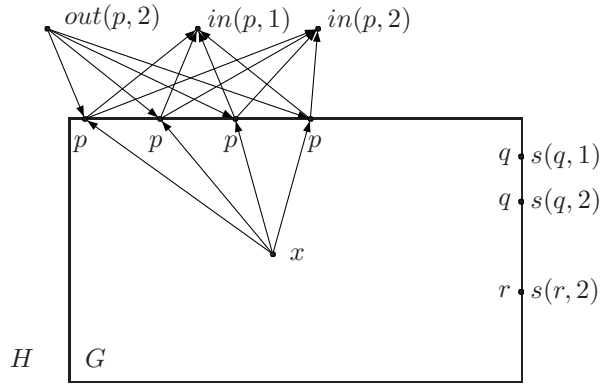
   
    Moreover, an expansion of $G$ may have up to two new vertices
    that are $in(p,j)$-sources, and at most one
    $out(p,j)$-source. Say, an expansion $H$ could have new
    vertices as $in(p,1)$- and $in(p,2)$-sources, with edges from
    each of the 4 $p$-ports to each $in(p,j)$-source; and it could
    have a new vertex as a, say, $out(p,2)$-source, with edges from
    that vertex to each of the $p$-ports.
    
    Note that if $G$ has a vertex $x$ with an edge from $x$ to at least
    3 $p$-sources, then an expansion cannot have 2 $out(p,j)$-sources:
    otherwise it would contain a copy of $\overrightarrow K_{3,3}$,
    which is not allowed for an expansion.
\end{example}

\begin{remark}
    It is not always the case that $G$ is determined by each of its
    expansions $\bar G$. If $p$ is large in $G$ but $\bar G$ has no
    $in(p,i)$- or $out(p,i)$-sources, then it is not possible to
    determine which of its vertices are $p$-ports.
\end{remark}

\paragraph{Construction of an $\NLC$-congruence from an
$\HR_{\sep}$-congruence}

Let $\equiv$ be a locally finite $\HR_{\sep}$-congruence saturating $L$.
We define a relation $\sim$ on $\GP$ as follows.
For $G$ and $G'$ in $\GP(P)$  we let $G \sim G'$ if and only if

\begin{itemize}
    \item[(a)] either $G$ and $G'$ both contain
    $\overrightarrow{K}_{m+1,m+1}$ as a subgraph, or neither does
    and in that case, the following two conditions hold:

    \item[(b)] $G$ and $G'$ satisfy the same small first-order formulas
    (\textit{i.e.}, with quantifier-depth at most $2m+2$) on graphs with
    ports.

    \item[(c)] for every expansion $\bar{G}$ of $G$, there exists an
    expansion $\bar{G}'$ of $G'$ such that $\bar{G} \equiv \bar G'$ and
    $\bar{G}$ and $\bar G'$ satisfy the same small first-order formulas
    on graphs with sources (we say that $\bar G$ and $\bar G'$ are
    \textit{equivalent expansions}); and conversely, for every
    expansion $\bar G'$ of $G'$ there exists an expansion $\bar{G}$ of
    $G$ equivalent to $\bar G'$.
\end{itemize}

Note that Condition (b) implies that $G$ and $G'$ have the same void,
small and large port labels, and Condition (c) implies that $\bar{G}$
and $\bar{G}'$ have the same source labels.

The relation $\sim$ is clearly an equivalence relation on each set
$\GP(P)$. It has finitely many classes on each $\GP(P)$ since a finite
graph has a uniformly bounded number of expansions (up to isomorphism),
the $\HR_{\sep}$-congruence $\equiv$ is locally finite, and there are
finitely many first-order formulas of each quantifier-depth on graphs
with sources in a subset of $C(P)$.

Now a graph without ports and without $\overrightarrow K_{m+1,m+1}$ has
a unique expansion: itself. It follows that, for graphs without ports
and without $\overrightarrow K_{m+1,m+1}$, the equivalences $\equiv$\
and $\sim$ coincide. In particular, $\sim$ saturates $L$ since $\equiv$
does.

It remains to prove that $\sim$ is an $\NLC$-congruence. Recall that 
the signature $\NLC$ consists of the operations of the form
$\fg_{p}$, $\ren_{p\rightarrow q}$ and $\otimes_{J}$.

\paragraph{The port forgetting operation}
We first consider the operation $\fg_{p}$. We consider
$G,G'$ with $G\sim G'$ and we want to prove that $H\sim H'$, where
$H = \fg_{p}(G)$ and $H' = \fg_{p}(G')$.

First of all, the underlying graphs of $G$ and $H$ (resp. $G'$ and
$H'$) are identical, so that $G$ and $G'$ contain
$\overrightarrow{K}_{m+1,m+1}$ if and only if so do $H$ and $H'$. If
this is the case, then $G\sim G'$ and $H\sim H'$. We now exclude this
case and assume that $G$ and $G'$ are without
$\overrightarrow{K}_{m+1,m+1}$. Note also that if $p$ is void in $G$,
then it is in $G'$ as well, and we have $H = G$, $H' = G'$, so that
$H\sim H'$. We now assume that $p$ is not void in $G$.

It is an immediate consequence of Theorem~\ref{FFVV} that $H$ and $H'$
satisfy the same small first-order formulas on graphs with ports, so
Condition (b) is verified.

We now consider Condition (c). Let $\bar H$ be an expansion of $H$. We
will show that there exists an expansion $\bar{G}$\ of $G$ and a unary
$\HR_{\sep}$-term $t$ such that $\bar{H}$ = $t(\bar{G})$. Since $G\sim
G'$, there exists an equivalent expansion $\bar{G'}$ of $G'$, and
$t(\bar{G'})$ will be the desired expansion of $H'$. Using the fact
that $\equiv$ is an $\HR_{\sep}$-congruence and Theorem~\ref{FFVV}, we
will have $H\sim H'$ as expected.

If $p$ is large in $G$, the situation is particularly simple: $\bar H$
is also an expansion of $G$, so we can choose $t$ to represent the
identity. If $\bar G'$ is an expansion of $G'$, equivalent to $\bar H$,
then $\bar G'$ does not use source labels of the form $s(p,i)$,
$in(p,i)$ or $out(p,i)$, so $\bar G'$ is also an expansion of $H'$.

If $p$ is small in $G$, let $\bar G$ be a graph with source obtained
from $\bar H$ by letting each $p$-port of $G$ be an $s(p,i)$-source
(where distinct source labels are used for distinct $p$-ports). Then
$\bar G$ is an expansion of $G$, and $\bar H = t(\bar G)$ where $t$ is
the composition of the operations $\srcfg_{s(p,i)}$ ($1\le i\le m$).
Using the definition of $\sim$, there exists an expansion $\bar G'$ of
$G'$ which is equivalent to $\bar G$, and we only need to verify that
$\bar H' = t(\bar G')$ is an expansion of $H'$. The only point to check
here is the fact that $H'$ is a subgraph of $\bar H'$: this follows
from the facts that $G$ is a subgraph of $\bar G$ and the operations
$t$ and $\fg_{p}$ do not change the underlying graph structures.

\paragraph{The renaming operation}
We now consider the operation $\ren_{p\rightarrow q}$. Let $G,G'$ with
$G\sim G'$: as with the port forgetting operation, we want to prove
that $H\sim H'$ where $H = \ren_{p\rightarrow q}(G)$ and $H' =
\ren_{p\rightarrow q}(G')$. As above, we can reduce the proof to the
case where neither $G$ nor $G'$ contains
$\overrightarrow{K}_{m+1,m+1}$, and where $p$ is not void in $G$ (if
$p$ is void in $G$, then $H = G$ and $H' = G'$). Moreover, Condition
(b) follows from Theorem~\ref{FFVV}.

We consider Condition (c), following the same strategy as above. Let
$\bar H$ be an expansion of $H$.

If $q$ is void in $G$, then the transformation $\ren_{p\rightarrow q}$
is a reversible renaming, that is, $G = \ren_{q\rightarrow p}(H)$.
Moreover, if $t$ is the composition of the operations of the form
$\srcren_{s(p,i)\rightarrow s(q,i)}$, $\srcren_{in(p,i)\rightarrow
in(q,i)}$ and $\srcren_{out(p,i)\rightarrow out(q,i)}$, and if $t'$ is
the composition of the operations $\srcren_{s(q,i)\rightarrow s(p,i)}$,
$\srcren_{in(q,i)\rightarrow in(p,i)}$ and
$\srcren_{out(q,i)\rightarrow out(p,i)}$, then $\bar G = t'(\bar H)$ is
an expansion of $G$, $\bar H = t(\bar G)$. Moreover, if $\bar G'$ is an
expansion of $G'$, equivalent to $\bar G$, then $\bar H' = t(\bar G')$
is an expansion of $H'$.

We now assume that $q$ is not void in $G$. We need to consider several cases.

\medskip

\noindent\textbf{Case 1}. $p$ and $q$ are both large in $G$. Then $p$
is void and $q$ is large in $H$.

In order to build the desired $\bar{G}$, we split each $in(q,i)$-source
of $\bar{H}$\ into an $in(p,i)$-source and an $in(q,i)$-source. The
$in(p,i)$-source is linked by incoming edges to all $p$-ports of $G$,
and the $in(q,i)$-source is linked similarly to all $q$-ports. In the
same fashion, we split each $out(q,i)$-source of $\bar{H}$ into an
$out(p,i)$-source and an $out(q,i)$-source linked by ougoing edges to
all $p$-ports of $G$ and to all $q$-ports respectively. The term $t$
such that $\bar H = t(\bar G)$ is the composition of the operations
$\fus_{in(p,i)\rightarrow in(q,i)}$ and $\fus_{out(p,i)\rightarrow
out(q,i)}$.

The graph $\bar{G}$ does not contain $\overrightarrow{K}_{m+1,m+1}$,
since $\bar{H}$ does not (by Lemma~\ref{lemma YYY}). Hence $\bar{G}$ is
an expansion of $G$. Let now $\bar{G'}$ be an expansion of $G'$
equivalent to $\bar{G}$, and let $\bar{H}' = t(\bar{G}')$. It is easily
verified that $\bar H'$ is an expansion of $H'$, and as above, it
follows that $H \sim H'$.

\medskip

\noindent\textbf{Case 2}. $p$ is small and $q$ is large in $G$.

In order to build $\bar G$ from $\bar{H}$, we make the $p$-ports of $G$
into $s(p,i)$-sources, we delete the edges between the $in(q,i)$- and
the $out(q,i)$-sources and the $p$-ports of $G$. The term $t$ which
must do the opposite (that is, construct $\bar{H}$ from $\bar{G}$) is a
composition of source forgetting operations and of additions of new
edges. More precisely, for each $i,j$ such that $s(p,i)$ and $in(q,j)$
are source labels in $\bar G$, we use the operation $Z \longmapsto Z
\oplus (\alpha \longrightarrow \omega)$, where $(\alpha \longrightarrow
\omega)$ is the 2-vertex, 2-source, 1-edge graph, followed by the
operations $\fus_{\alpha\rightarrow s(p,i)}$ and
$\fus_{\omega\rightarrow in(q,j)}$. We then apply similar operations to
create edges from the $out(q,j)$- to the $s(p,i)$-sources. And we
finally apply the operations $\srcfg_{s(p,i)}$.

The graph $\bar G$ is a subgraph of $\bar H$ (up to source labels), so
$\bar G$ does not contain $\overrightarrow{K}_{m+1,m+1}$, and hence it
is an expansion of $G$. The proof continues as in the previous case.

\medskip

\noindent\textbf{Case 3}. $q$ is small and $p$ is large in $G$.

To build $\bar G$ from $\bar H$, we make the $q$-ports of $G$ into
$s(q,i)$-sources, we delete the edges between the $in(p,i)$-sources or
the $out(p,i)$-sources and the $q$-ports of $G$. In addition we rename
each $in(p,i)$-source to an $in(q,i)$-source, and each
$out(p,i)$-source to an $out(q,i)$-source. We can use the same
reasoning as in Case 2 to conclude in this case.

\medskip

\noindent\textbf{Case 4}. $p$ and $q$ are small in $G$, and
$\card(p_{G}) + \card(q_{G})\leq m$.

To build $\bar G$ from $\bar H$, we rename $s(q,i)$ into $s(p,i)$
whenever the $s(q,i)$-source of $\bar{H}$ is a $p$-port in $G$. The
term $t$ which does the opposite is a composition of source renamings.
The graph $\bar{G}$ does not contain $\overrightarrow{K}_{m+1,m+1}$,
otherwise $\bar{H}$ would do, since $\bar{G}$ is equal to $\bar{H}$ up
to source labels, and hence $\bar G$ is an expansion of $G$. The other
parts of the proof are the same.

\medskip

\noindent\textbf{Case 5}. $p$ and $q$ are small in $G$, and
$\card(p_{G}) + \card(q_{G})\geq m+1$.

\begin{figure}[htb]
    \centering
\begin{picture}(88,75)(0,-75)
 
\node[NLangle=0.0,Nw=32.0,Nh=12.0,Nmr=0.0](n0)(20.0,-16.0){$G$}

\node[NLangle=0.0,Nw=32.0,Nh=12.0,Nmr=0.0](n1)(66.0,-16.0){$H$}

\node[Nfill=y,fillcolor=Black,ExtNL=y,NLdist=1.5,Nw=0.8,Nh=0.8,Nmr=0.2](n2)(7.0,-10.0){$p$}

\node[Nfill=y,fillcolor=Black,ExtNL=y,NLdist=1.5,Nw=0.8,Nh=0.8,Nmr=0.2](n3)(16.0,-10.0){$p$}

\node[Nfill=y,fillcolor=Black,ExtNL=y,NLdist=1.5,Nw=0.8,Nh=0.8,Nmr=0.2](n4)(25.0,-10.0){$q$}

\node[Nfill=y,fillcolor=Black,ExtNL=y,NLdist=1.5,Nw=0.8,Nh=0.8,Nmr=0.2](n5)(34.0,-10.0){$q$}

\node[Nfill=y,fillcolor=Black,ExtNL=y,NLdist=1.5,Nw=0.8,Nh=0.8,Nmr=0.2](n6)(52.0,-10.0){$q$}

\node[Nfill=y,fillcolor=Black,ExtNL=y,NLdist=1.5,Nw=0.8,Nh=0.8,Nmr=0.2](n7)(61.0,-10.0){$q$}

\node[Nfill=y,fillcolor=Black,ExtNL=y,NLdist=1.5,Nw=0.8,Nh=0.8,Nmr=0.2](n8)(70.0,-10.0){$q$}

\node[Nfill=y,fillcolor=Black,ExtNL=y,NLdist=1.5,Nw=0.8,Nh=0.8,Nmr=0.2](n9)(79.0,-10.0){$q$}

\node[NLangle=0.0,Nw=32.0,Nh=12.0,Nmr=0.0](n10)(20.0,-48.0){$\bar G$}

\node[NLangle=0.0,Nw=32.0,Nh=12.0,Nmr=0.0](n11)(66.0,-48.0){$\bar H$}

\node[Nfill=y,fillcolor=Black,ExtNL=y,NLdist=1.5,Nw=0.8,Nh=0.8,Nmr=0.2](n12)(7.0,-42.0){$\scriptstyle
s(p,1)$}

\node[Nfill=y,fillcolor=Black,ExtNL=y,NLdist=1.5,Nw=0.8,Nh=0.8,Nmr=0.2](n13)(16.0,-42.0){$\scriptstyle
s(p,2)$}

\node[Nfill=y,fillcolor=Black,ExtNL=y,NLdist=1.5,Nw=0.8,Nh=0.8,Nmr=0.2](n14)(25.0,-42.0){$\scriptstyle
s(q,1)$}

\node[Nfill=y,fillcolor=Black,ExtNL=y,NLdist=1.5,Nw=0.8,Nh=0.8,Nmr=0.2](n15)(34.0,-42.0){$\scriptstyle
s(q,2)$}

\node[Nfill=y,fillcolor=Black,Nw=0.8,Nh=0.8,Nmr=0.2](n16)(52.0,-42.0){}

\node[Nfill=y,fillcolor=Black,Nw=0.8,Nh=0.8,Nmr=0.2](n17)(61.0,-42.0){}

\node[Nfill=y,fillcolor=Black,Nw=0.8,Nh=0.8,Nmr=0.2](n18)(70.0,-42.0){}

\node[Nfill=y,fillcolor=Black,Nw=0.8,Nh=0.8,Nmr=0.2](n19)(79.0,-42.0){}

\node[Nfill=y,fillcolor=Black,NLangle=-180.0,Nw=0.8,Nh=0.8,NLdist=7,Nmr=0.2](n20)(55.0,-28.0){$\scriptstyle
out(q,1)$}

\node[Nfill=y,fillcolor=Black,NLangle=0.0,Nw=0.8,Nh=0.8,NLdist=7,Nmr=0.2](n21)(73.0,-28.0){$\scriptstyle
in(q,2)$}

\drawedge(n20,n16){}

\drawedge(n20,n17){}

\drawedge(n20,n18){}

\drawedge(n20,n19){}

\drawedge(n16,n21){}

\drawedge(n17,n21){}

\drawedge(n18,n21){}

\drawedge(n19,n21){}

\node[NLangle=-90.0,Nfill=y,fillcolor=Black,Nw=0.8,Nh=0.8,NLdist=1.5,Nmr=0.2](n22)(52.0,-74.0){$\scriptstyle
s(p,1)$}

\node[NLangle=-90.0,Nfill=y,fillcolor=Black,Nw=0.8,Nh=0.8,NLdist=1.5,Nmr=0.2](n23)(61.0,-74.0){$\scriptstyle
s(p,2)$}

\node[NLangle=-90.0,Nfill=y,fillcolor=Black,Nw=0.8,Nh=0.8,NLdist=1.5,Nmr=0.2](n24)(70.0,-74.0){$\scriptstyle
s(q,1)$}

\node[NLangle=-90.0,Nfill=y,fillcolor=Black,Nw=0.8,Nh=0.8,NLdist=1.5,Nmr=0.2](n25)(79.0,-74.0){$\scriptstyle
s(q,2)$}

\node[NLangle=-180.0,Nfill=y,fillcolor=Black,Nw=0.8,Nh=0.8,NLdist=7,Nmr=0.2](n26)(55.0,-60.0){$\scriptstyle
out(q,1)$}

\node[Nfill=y,fillcolor=Black,NLangle=0.0,Nw=0.8,Nh=0.8,NLdist=7,Nmr=0.2](n27)(73.0,-60.0){$\scriptstyle
in(q,2)$}

\drawedge(n26,n22){}

\drawedge(n26,n23){}

\drawedge(n26,n24){}

\drawedge(n26,n25){}

\drawedge(n22,n27){}

\drawedge(n23,n27){}

\drawedge(n24,n27){}

\drawedge(n25,n27){}

\put(38,-67){\hbox{$E =$}}

\end{picture}
\caption{$m=2$ and $\bar H = t(\bar G) =
\srcfg_{s(p,1),s(p,2),s(q,1),s(q,2)}(\bar G \parallel E)$}
\label{fig: figure 2}
    \end{figure}
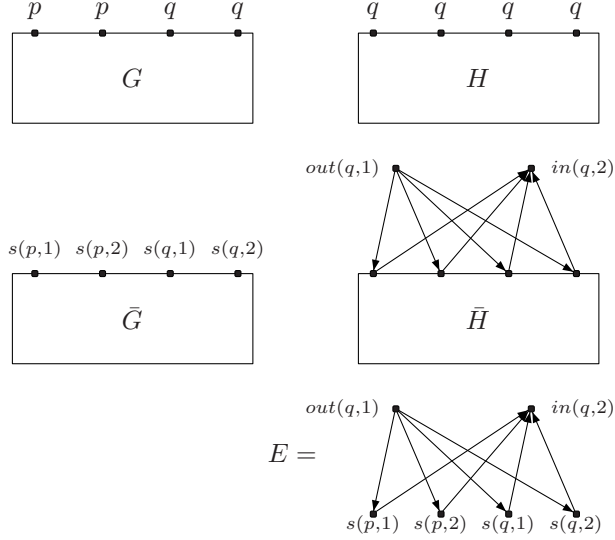

To build $\bar G$ from $\bar H$, we make the $p$-ports (resp.
$q$-ports) of $G$ into $s(p,i)$-sources (resp. $s(q,i)$-sources), we
delete the edges between the $in(q,i)$- and $out(q,i)$-sources and the
$p$- and $q$-ports of $G$, and we delete the $in(q,i)$- and
$out(q,i)$-sources. The term $t$ which does the opposite is a
composition of additions of new edges and of $\srcfg$ operations, as in
Case 2, see Figure~\ref{fig: figure 2}. The graph $\bar{G}$ does not contain $\overrightarrow
{K}_{m+1,m+1}$, otherwise $\bar{H}$ would too, since $\bar{G}$ is a
subgraph of $\bar{H}$ (up to source labels), and hence $\bar G$ is an
expansion of $G$. The proof continues as in the previous cases.

\medskip

This concludes the proof that $G\sim G'$ implies $\ren_{p\rightarrow
q}(G) \sim \ren_{p\rightarrow q}(G')$.

\paragraph{The operation $\otimes_{J}$}

We now consider the operation $\otimes_{J}$ where $J \subseteq (P\times
Q) \cup (Q\times P)$, $P$ and $Q$ are disjoint. Let $G \sim G'$ in
$\GP(P)$, $K \sim\ K'$ in $\GP(Q)$, $H = G\otimes_{J} K$ and $H' =
G' \otimes_{J} K'$. We want to prove that $H \sim H'$.

We first consider the very special case where $J = \emptyset$, and the
operation $\otimes_{J}$ is simply the disjoint union. Then $H$ contains
$\overrightarrow{K}_{m+1,m+1}$ if and only if $G$ or $K$ does, if and
only if $G'$ or $K'$ does, if and only if $H'$ does.

Asuming that $H$ does not contain $\overrightarrow{K}_{m+1,m+1}$, an
application of Theorem~\ref{FFVV} ensures, as for the operations of
port forgetting or renaming that $H$ and $H'$ satisfy the same small
first-order formulas.

We now consider an expansion $\bar H$ of $H$. It is necessarily of the
form $\bar{H} = \bar{G} \oplus \bar{K}$ where $\bar{G}$ and
$\bar{K}$ are expansions of $G$ and $K$ respectively. Then there
exist expansions $\bar{G}'$ and $\bar{K}'$ of $G'$ and $K'$
respectively, which are equivalent to $\bar{G}$ and $\bar{K}$. One then
verifies that $\bar{H}' = \bar{G}' \oplus \bar{K}'$ is an expansion of
$H'$, which is equivalent to $\bar{H}$.

\medskip

Next we assume that $J$ is a singleton, $J = \{(p,q)\}$, that is, $G
\otimes_{J} K = \add_{p,q}(G\oplus K)$ with $p\in P$ and $q\in Q$.

Since $G$ and $G'$ on one hand, and $K$ and $K'$ on the other satisfy
the same small first-order formulas, Theorem~\ref{FFVV} shows that $H =
\add_{p,q}(G \oplus K)$ contains $\overrightarrow {K}_{m+1,m+1}$ if and
only if $H' = \add_{p,q}(G' \oplus K')$ does. Assume now this is not
the case and consider an expansion $\bar{H}$ of $H$.

Again there are several cases. Note that $p$ and $q$ cannot both be
large in $G$ and $K$ respectively. We claim that $\bar{H}$ can defined
as $t(\bar{G},\bar{K})$ where $t$ is an $\HR_{\sep}$-term, $\bar{G}$ is
an expansion of $G$ and $\bar{K}$ is an expansion of $K$. As for the
other operations, we consider expansions $\bar G'$ and $\bar K'$ of
$G'$ and $K'$, equivalent to $\bar G$ and $\bar K$. Although it is a
bit tedious, we verify formally that $\bar H' = t(\bar G' \oplus \bar
K')$ is an expansion of $H'$. It follows that $\bar H'$ is equivalent
to $\bar H$, and hence $H \sim H'$.

\medskip

\noindent \textbf{Case 1}. $p$ is large in $G$ and $q$ is small in $K$.

Then $H$ has edges from all $p$-ports of $G$ to all $q$-ports of $K$,
which are actually $s(q,i)$-sources in $\bar{H}$. For each of these
$s(q,i)$-sources, say $x$, we create a new vertex $x'$, and each edge
coming from $G$ to $x$ is redirected towards $x'$. We make $x'$ into an
$in(p,j)$-source (for some appropriate $j$) of the expansion $\bar{G}$
of $G$ we are constructing. The desired expansion $\bar{K}$ of $K$ is
just the subgraph of $\bar{H}$ induced by the set of vertices of $K$.
And $\bar{G}$ consists of the subgraph of $\bar{H}$ induced by the
vertices of $G$ together with $x'$ and all these redirected edges. Then
the $\HR_{\sep}$-term $t$ needs only to fuse in $\bar{G} \oplus
\bar{K}$ the above described $in(p,j)$-sources with the corresponding
$s(q,i )$-sources. This can be done by a combination of the operation
$\oplus$ and those of the form $\fus_{in(p,j)\rightarrow s(q,i)}$. The
only point to check is that $\bar{G}$ does not contain
$\overrightarrow{K}_{m+1,m+1}$. We can apply Lemma~\ref{lemma YYY}
because $\bar{H}$ is obtained from $\bar{G} \oplus \bar{K}$ by fusions
of pairs of vertices which are not adjacent and have no incoming edges
with the same source (because $G$ and $K$ are disjoint) and no outgoing
edge at all.

Then there exist expansions $\bar{G}'$ and $\bar{K}'$ of $G'$ and $K'$
respectively, equivalent to $\bar{G}$ and $\bar{K}$. By letting
$\bar{H}' = t(\bar{G}',\bar{K}')$, we get the desired expansion of
$H'$, equivalent to $\bar{H}$.

\begin{figure}[htb]
    \centering
    \begin{picture}(90,80)(0,-80)
     
    \node[NLangle=0.0,Nw=20.0,Nh=12.0,Nmr=0.0](n0)(12.0,-24.0){$G$}

    \node[NLangle=0.0,Nw=12.0,Nh=28.0,Nmr=0.0](n1)(34.0,-16.0){$K$}

    \node[NLangle=0.0,Nw=20.0,Nh=12.0,Nmr=0.0](n2)(62.0,-24.0){$G$}

    \node[NLangle=0.0,Nw=12.0,Nh=28.0,Nmr=0.0](n3)(84.0,-16.0){$K$}

    \put(53,-35){$G\otimes_{J}K = \add_{p,q}(G\oplus K)$}

    \node[Nfill=y,fillcolor=Black,ExtNL=y,NLangle=-90.0,NLdist=1.2,Nw=0.8,Nh=0.8,Nmr=0.2](n4)(5.0,-18.0){$p$}

    \node[Nfill=y,fillcolor=Black,ExtNL=y,NLangle=-90.0,NLdist=1.2,Nw=0.8,Nh=0.8,Nmr=0.2](n5)(12.0,-18.0){$p$}

    \node[Nfill=y,fillcolor=Black,ExtNL=y,NLangle=-90.0,NLdist=1.2,Nw=0.8,Nh=0.8,Nmr=0.2](n6)(19.0,-18.0){$p$}

    \node[Nfill=y,fillcolor=Black,ExtNL=y,NLangle=-0.0,NLdist=1.2,Nw=0.8,Nh=0.8,Nmr=0.2](n7)(28.0,-6.0){$q$}

    \node[Nfill=y,fillcolor=Black,ExtNL=y,NLangle=-0.0,NLdist=1.2,Nw=0.8,Nh=0.8,Nmr=0.2](n8)(28.0,-12.0){$q$}

    \node[Nfill=y,fillcolor=Black,ExtNL=y,NLangle=-90.0,NLdist=1.2,Nw=0.8,Nh=0.8,Nmr=0.2](n9)(55.0,-18.0){$p$}

    \node[Nfill=y,fillcolor=Black,ExtNL=y,NLangle=-90.0,NLdist=1.2,Nw=0.8,Nh=0.8,Nmr=0.2](n10)(62.0,-18.0){$p$}

    \node[Nfill=y,fillcolor=Black,ExtNL=y,NLangle=-90.0,NLdist=1.2,Nw=0.8,Nh=0.8,Nmr=0.2](n11)(69.0,-18.0){$p$}

    \node[Nfill=y,fillcolor=Black,ExtNL=y,NLangle=-0.0,NLdist=1.2,Nw=0.8,Nh=0.8,Nmr=0.2](n12)(78.0,-6.0){$q$}

    \node[Nfill=y,fillcolor=Black,ExtNL=y,NLangle=-0.0,NLdist=1.2,Nw=0.8,Nh=0.8,Nmr=0.2](n13)(78.0,-12.0){$q$}

    \drawedge(n9,n12){}
    \drawedge(n9,n13){}
    \drawedge(n10,n12){}
    \drawedge(n10,n13){}
    \drawedge(n11,n12){}
    \drawedge(n11,n13){}

    \node[NLangle=0.0,Nw=20.0,Nh=12.0,Nmr=0.0](n14)(12.0,-72.0){$\bar G$}

    \node[NLangle=0.0,Nw=12.0,Nh=28.0,Nmr=0.0](n15)(34.0,-64.0){$\bar K$}

    \node[NLangle=0.0,Nw=20.0,Nh=12.0,Nmr=0.0](n16)(62.0,-72.0){}

    \node[NLangle=0.0,Nw=12.0,Nh=28.0,Nmr=0.0](n17)(84.0,-64.0){}

    \put(70,-83){$N$}

    \node[Nfill=y,fillcolor=Black,ExtNL=y,NLangle=-90.0,NLdist=1.2,Nw=0.8,Nh=0.8,Nmr=0.2](n18)(5.0,-66.0){}

    \node[Nfill=y,fillcolor=Black,ExtNL=y,NLangle=-90.0,NLdist=1.2,Nw=0.8,Nh=0.8,Nmr=0.2](n19)(12.0,-66.0){}

    \node[Nfill=y,fillcolor=Black,ExtNL=y,NLangle=-90.0,NLdist=1.2,Nw=0.8,Nh=0.8,Nmr=0.2](n20)(19.0,-66.0){}

    \node[Nfill=y,fillcolor=Black,ExtNL=y,NLangle=90.0,NLdist=1.2,Nw=0.8,Nh=0.8,Nmr=0.2](n21)(2.0,-50.0){$\scriptstyle
    in(p,1)$}

    \node[Nfill=y,fillcolor=Black,ExtNL=y,NLangle=90.0,NLdist=1.2,Nw=0.8,Nh=0.8,Nmr=0.2](n22)(12.0,-50.0){$\scriptstyle
    in(p,2)$}

    \node[Nfill=y,fillcolor=Black,ExtNL=y,NLangle=90.0,NLdist=1.2,Nw=0.8,Nh=0.8,Nmr=0.2](n23)(22.0,-50.0){$\scriptstyle
    in(p,3)$}

    \node[Nfill=y,fillcolor=Black,ExtNL=y,NLangle=-0.0,NLdist=1.2,Nw=0.8,Nh=0.8,Nmr=0.2](n24)(28.0,-54.0){$\scriptstyle
    s(q,1)$}

    \node[Nfill=y,fillcolor=Black,ExtNL=y,NLangle=-0.0,NLdist=1.2,Nw=0.8,Nh=0.8,Nmr=0.2](n25)(28.0,-60.0){$\scriptstyle
    s(q,2)$}
    
    \drawedge(n18,n21){}
    \drawedge(n18,n22){}
    \drawedge(n18,n23){}
    \drawedge(n19,n21){}
    \drawedge(n19,n22){}
    \drawedge(n19,n23){}
    \drawedge(n20,n21){}
    \drawedge(n20,n22){}
    \drawedge(n20,n23){}

    \node[Nfill=y,fillcolor=Black,ExtNL=y,NLangle=-90.0,NLdist=1.2,Nw=0.8,Nh=0.8,Nmr=0.2](n26)(55.0,-66.0){}

    \node[Nfill=y,fillcolor=Black,ExtNL=y,NLangle=-90.0,NLdist=1.2,Nw=0.8,Nh=0.8,Nmr=0.2](n27)(62.0,-66.0){}

    \node[Nfill=y,fillcolor=Black,ExtNL=y,NLangle=-90.0,NLdist=1.2,Nw=0.8,Nh=0.8,Nmr=0.2](n28)(69.0,-66.0){}

    \node[Nfill=y,fillcolor=Black,ExtNL=y,NLangle=90.0,NLdist=1.2,Nw=0.8,Nh=0.8,Nmr=0.2](n29)(52.0,-50.0){}

    \node[Nfill=y,fillcolor=Black,ExtNL=y,NLangle=-0.0,NLdist=1.2,Nw=0.8,Nh=0.8,Nmr=0.2](n30)(78.0,-54.0){}

    \node[Nfill=y,fillcolor=Black,ExtNL=y,NLangle=-0.0,NLdist=1.2,Nw=0.8,Nh=0.8,Nmr=0.2](n31)(78.0,-60.0){}

    \drawedge(n26,n30){}
    \drawedge(n26,n31){}
    \drawedge(n27,n30){}
    \drawedge(n27,n31){}
    \drawedge(n28,n30){}
    \drawedge(n28,n31){}

    \drawedge(n26,n29){}
    \drawedge(n27,n29){}
    \drawedge(n28,n29){}

\end{picture}
\caption{$N = t(\bar G,\bar K) = \srcfg_{all}(\bar G \parallel
\srcren_{s(q,1)\rightarrow in(p,2),s(q,2)\rightarrow in(p,3)}(\bar
K))$}
\label{fig: figure 3}
    \end{figure}
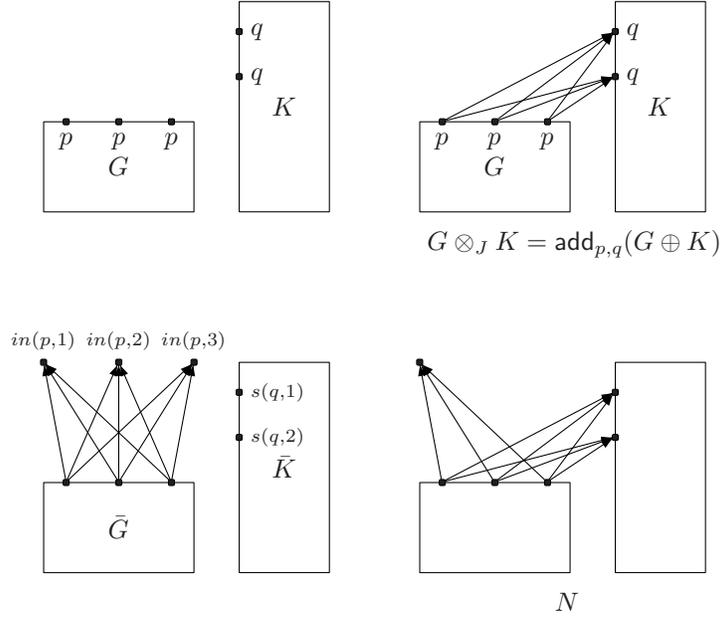

This case is illustrated in Figure~\ref{fig: figure 3}, where $m=3$
and $N$ is the constructed expansion of $G\otimes_{J}K$.

\medskip

\noindent \textbf{Case 2}. $p$ is small in $G$ and $q$ is large in $K$.

It is fully similar to the first case, creating new $out(q,j)$-sources
instead of $in(p,j)$-sources. We omit the details.

\medskip

\noindent \textbf{Case 3}. $p$ is small in $G$\ and $q$ is small in $K$.

Let $\bar G$ be the subgraph with sources of $\bar H$ consisting of the
vertices of $G$, and let $\bar K$ be defined similarly in terms of $K$.
Then $\bar{H}$ is obtained from $\bar{G} \oplus \bar{K}$ by the
addition of edges from each $s(p,i)$-source of $\bar{G}$ to each
$s(q,j)$-source of $\bar{K}$, which can be done by an $\HR_{\sep}$-term
(see Case 2 of the discussion of the renaming operation). Since
$\bar{G}$ and $\bar{K}$ are subgraphs of $\bar{H}$, they cannot contain
$\overrightarrow{K}_{m+1,m+1}$ and hence, they are in fact expansions
of $G$ and $K$ as desired. The proof continues as above.

\medskip

\noindent \textbf{Case 4}. $p$ is void in $G$ or $q$ is void in $K$.

Then $\add_{p,r}$ acts as the identity on $G\oplus K$, so $\otimes_{J}$
acts as $\oplus$ on $(G,K)$ and we are back to a previously studied
case. Recall that if $p$ (resp. $q$) is void in $G$ (resp. $K$), then
it is void in every $\sim$-equivalent graph with source.

\medskip

This concludes the study of the case where $J$ is a singleton in
$P\times Q$. The case where $J$ is a singleton in $Q\times P$ is of
course similar.

The proof is actually the same in the general case where $J$ is not a
singleton. We need only do the same constructions for all elements
$(p,q)$ in $J$. The only possible difficulty could arise from the use
of Lemma \ref{lemma YYY} to verify that the graphs $\bar{G}$ and
$\bar{K}$ obtained from $\bar{H}$ by the creation of vertices (like
$x'$ in Case 1 above) and the redirection of edges do not contain
$\overrightarrow{K}_{m+1,m+1}$, and hence are expansions. Thus let us
consider the transformation of $\bar G \oplus\bar{K}$ into $\bar{H}$.
It consists in a sequence of fusions of pairs of vertices. Whenever we
fuse an $in(p,i)$-source of $\bar{G}$, say $x$, with an $s(q,j)$-source
of $\bar{K}$, say $y$, we must verify that the fusions performed
previously keep the hypothesis of Lemma \ref{lemma YYY} valid. It is
clear that $x$ and $y$ are not adjacent, since $x$ is adjacent with
vertices of $G$ only. Because of previous fusions, there may exist an
edge from some $z$ in $G$ to $y$. However, this edge comes from a
previously applied operation $\add_{p',q}$ with $p' \neq p$. It follows
that there is no edge from $z$ to $x$. An analogous argument also
applies to fusions between an $out(p,i)$-source of $G$ and an
$s(q,j)$-source of $K$, and also when we exchange the roles of $G$ and
$K$. Hence, finally, we can apply Lemma~\ref{lemma YYY} to deduce that
$\bar{G}$ and $\bar{K}$ do not contain $\overrightarrow{K}_{m+1,m+1}$
because $\bar{H}$ does not. Hence, they are expansions of $G$ and $K$,
as we needed to check.

\medskip

This concludes the proof of Theorem~\ref{no Kmm}.

\subsection{Other finiteness conditions}\label{other finiteness}

We now consider some consequences of Theorem~\ref{no Kmm}. Let
$K_{n,n}$ be the undirected complete bipartite graph with $n+n$
vertices, that is, $K_{n,n}$ is the undirected graph underlying
$\overrightarrow K_{n,n}$. We say that a (directed) graph is
\emph{without $K_{n,n}$} if its undirected underlying graph
has no subgraph isomorphic to $K_{n,n}$.

We say that a graph $G$ is uniformly \emph{$k$-sparse} if $\card(E(H))
\leq k\ \card(V(H))$ for every finite subgraph $H$ of $G$, where $V(H)$
and $E(H)$ are the sets of vertices and edges of $H$. A set of graphs
is \emph{uniformly $k$-sparse} if each of its elements is.

\begin{proposition}\label{prop k-sparse}
    Let $L\subseteq \Grph$ be a set of graphs, satisfying one of the 
    following properties:
    \begin{itemize}
	\item[] $L$ is without $\overrightarrow K_{n,n}$ for some $n$
	\item[or] $L$ is without $K_{n,n}$ for some $n$
	\item[or] $L$ consists only of planar graphs
	\item[or] $L$ is uniformly $k$-sparse for some $k$
	\item[or] $L$ consists only of graphs of tree-width at most
	$k$ for some $k$.
    \end{itemize}
    Then $L$ is $\HR$-recognizable if and only if $L$ is
    $\VR$-recognizable.
\end{proposition}

\proof
By Corollary \ref{VR2HR}, it is always the case that a
$\VR$-recognizable set of graphs is $\HR$-recognizable.

If $L$ is without $\overrightarrow K_{n,n}$ for some $n$, the converse
implication was proved in Theorem~\ref{no Kmm}. Lemma \ref{no Kpp}
below shows that $L$ is without $K_{p,p}$ for some $p$ if and only if
it is without $\overrightarrow K_{n,n}$ for some $n$.

It is well-known that planar graphs are without $K_{3,3}$ (planarity is
a property of the underlying undirected graph, and $K_{3,3}$ is the
undirected graph underlying $\overrightarrow K_{3,3}$). It follows that
planar graphs are also without $\overrightarrow K_{3,3}$, and the
result follows from Theorem~\ref{no Kmm}.

It is easily seen that $\overrightarrow K_{2k+1,2k+1}$ is not
$k$-sparse. So if $L$ is uniformly $k$-sparse, then it is without
$\overrightarrow K_{2k+1,2k+1}$.

Finally, it is known that graphs of tree-width at most $k$ are
uniformly $(k+1)$-sparse (see for instance \cite{BCXIV}), which yields
the last assertion.
\eop

\begin{lemma}\label{no Kpp}
    Let $p$ be an integer. There exists an integer $n$ such that a
    directed graph without $\overrightarrow K_{p,p}$, is without
    $K_{n,n}$.
\end{lemma}

\proof
We use the particular case of Ramsey's Theorem for bipartite graphs,
given as Theorem 1 in \cite[p. 95]{Graham}. It states that for each
$p$, there exists an integer $n$ such that, if the edges of $K_{n,n}$
are partitioned into two sets $A$ and $B$, then either $A$ or $B$
contains the edges of a subgraph isomorphic to $K_{p,p}$.

So let us assume that $U,W\subseteq V(G)$, where $U$ and $W$ are
disjoint sets of $n$ elements and there is an edge between $u$ and $w$
(in one or both directions) for each $(u,w)\in U\times W$. Let $A$ be
the set of pairs $(u,w)\in U\times W$ such that the edge is from $u$ to
$w$, and let $B = (U\times W)\setminus A$. Then there exist sets
$U'\subseteq U$ and $W'\subseteq W$, with cardinality $p$, such that
$U'\times W'\subseteq A$ or $W'\times U'\subseteq B$. In either case,
we get a subgraph of $G$ isomorphic to $\overrightarrow K_{p,p}$.

Note hat a quick and direct proof can be given with $n = p2^{2p}$,
but we do not know the minimal $n$ yielding the result.
\eop

%
%
\begin{remark}
    The statement relative to bounded tree-width sets of graphs in
    Proposition~\ref{prop k-sparse} is also a consequence (in the case
    of finite graphs) of Lapoire's result \cite{Lapoire}, which states
    that, in a graph of tree-width at most $k$, one can construct a
    width-$k$ tree-decomposition by monadic second-order (MSO)
    formulas. This can be used to show that every $\HR$-recognizable
    set of graphs of bounded tree-width is definable in Counting
    Monadic Second-order (CMSO) logic, using edge set quantifications.
    Courcelle showed \cite{BCVI} that, for finite graphs of bounded
    tree-width, edge set quantifications can be replaced by vertex set
    quantifications. The considered set is therefore definable in CMSO
    logic with vertex set quantifications only, and hence is
    $\VR$-recognizable by another of Courcelle's results \cite{BCVII}.
\end{remark}

\begin{remark}
    It is proved in \cite{BCI} that every set of square grids is
    $\HR$-recog\-nizable. It follows from Theorem~\ref{no Kmm} that every
    such set is also $\VR$-recognizable. Hence, there are uncountably
    many $\VR$-recognizable sets of graphs, so we cannot hope for an
    automata-theoretic or a logical characterization of
    $\VR$-recognizability --- in contrast with the situation prevailing
    for words, trees and some special classes of graphs, see
    \cite{ThomasHdbook,LWTCS,LWIC,EsikNemeth,KuskeSP,KuskeMSC}.
\end{remark}

\subsection{$\HR$-recognizable sets which are not $\VR$-recognizable}\label{HR not VR}

The aim of this short section is to establish the existence of
$\HR$-recognizable sets which are not $\VR$-recognizable. We first
establish a lemma.

\begin{lemma}\label{cliques HR-rec}
    Every set of cliques (of the form $K_{n}$, $n\ge 1$) is
    $\HR$-recognizable.
\end{lemma}

\proof
Let $L$ be a set of undirected cliques (recall that an undirected graph
is a graph where the edge relation is symmetric). We provide a locally
finite $\calC\calS$-congruence on $\GS_{\sep}$ which saturates $L$ (see
Section~\ref{sec: other variants}). By Proposition~\ref{prop: other
variants}, this establishes that $L$ is $\HR$-recognizable.

For each finite set $C$ of source labels, let $G^i(C)$ be the set of
graphs in $\GS_{\sep}(C)$ having at least one internal vertex (i.e., a
vertex which is not a source), and let $G^{s}(C)$ be the set of graphs
in $\GS_{\sep}(C)$, in which every vertex is a source. In particular,
$G^{s}(C)$ is finite.

Let $\equiv$ be the following equivalence relation on $\GS_{\sep}$. We
use the operation $\sqbox_{C}$, as in Section~\ref{sec: other
variants}. If $G, G'\in\GS_{\sep}(C)$, we let $G\equiv\ G'$ if and only
if either $G = G'$, or $G,G'\in G^{i}(C)$ and for every $H \in
G^{s}(C)$, $G\sqbox_{C} H\in\ L$ iff $G'\sqbox_{C} H\in L$.

Note that for each $C$, there are only finitely many $\equiv$-classes
in $\GS_{\sep}(C)$, --- namely at most $p+2^{p}$, where $p$ is the
cardinality of $G^{s}(C)$.

Moreover, $\equiv$ saturates $L$. Indeed, suppose that $G, G' \in
\GS_{\sep}(C)$, $G\equiv G'$ and $G\in L$. Let $H$ be the graph in
$\GS_{\sep}(C)$ consisting of distinct $c$-sources ($c\in C$) and no
edges. Then we have $G = G \sqbox_{C} H$ and $G' = G'\sqbox_{C} H$. It
follows from the definition of $\equiv$ that $G'\in L$.

Finally, we check that $\equiv$ is a $\calC\calS$-congruence. Let $G,
G', H, H' \in \GS_{\sep}(C)$, with $G\equiv G'$ and $H \equiv H'$: we
want to show that $G\sqbox_{C} H \equiv G'\sqbox_{C} H'$. We observe
that if both $G$ and $H$ have internal vertices, then $G\sqbox_{C} H$
is not a clique (by definition of operation $\sqbox_{C}$), and hence
cannot be in $L$. The rest of the proof is a straightforward
verification.
\eop

We can now prove the following.

\begin{proposition}
    There is an $\HR$-recognizable set of graphs which is not
    $\VR$-recognizable.
\end{proposition}

\proof
Let $A$ be a set of integers which is not recognizable in
$\langle\mathbb{N},\form{succ},0\rangle$, for instance the set of prime
numbers, and let $L$ be the set of cliques $K_{n}$ for $n\in A$. Then
$L$ is $\HR$-recognizable by Lemma~\ref{cliques HR-rec}.

We now consider a set of $\VR$-terms describing $L$ and using exactly 2
port labels, $p$ and $q$. Recall that $\p$ denotes the $\VR$-constant
of type $\{p\}$, that is, the graph with a single vertex that is a
$p$-port and no edges. The constant $\q$ is defined similarly. Now let
$k_{1} = \p$, and $k_{n+1} = \ren_{q\rightarrow
p}\add_{p,q}\add_{q,p}(k_{n}\oplus \q)$. It is not difficult to verify
that $k_{n}$ denotes the clique $K_{n}$ where all the vertices are
$p$-ports, $K_{n}$ itself is denoted by the term
$\mdf_{\emptyset}k_{n}$, and the set $K$ of all $\VR$-terms of the form
$k_{n}$ is recognizable (as a set of terms, or trees). If $L$ is
$\VR$-recognizable, then the set of $\VR$-terms in $K$ that denote
graphs in $L$ is recognizable. This set consists of all the terms of
the form $\mdf_{\emptyset}k_{n}$ with $n\in A$, and it can be shown by
standard methods that it is not recognizable. It follows that $L$ is
not $\VR$-recognizable.
\eop

\subsection{Sparse graphs and monadic second-order logic}\label{sparse MSO}

Since graphs are relational structures, logical formulas can be used to
specify sets of graphs. Monadic second-order logic is especially interesting
because 

\smallskip

\textsl{every monadic second-order definable set of finite graphs is
$\VR$-recognizable (Courcelle \cite{BCVII,HbGraGraRozenberg97}).}

\smallskip

There is actually a version of monadic second-order logic allowing
quantifications on edges and sets of edges (one replaces the graph
under consideration by its incidence graph; we omit details). We say
that a set is $MS_{2}$-definable if it is definable by a monadic
second-order formula with edge and edge set quantifications, and that
we use the phrase $MS_{1}$-definable to refer to the first notion.
It is immediately verified (from the definition) that

\smallskip

\textsl{Every $MS_{1}$-definable set is $MS_{2}$-definable.}

\smallskip

\noindent The two following statements are more difficult.

\smallskip

\textsl{Every $MS_{2}$-definable set of simple graphs is $\HR$-recognizable
(Courcelle \cite{BCI}).}

\smallskip

\textsl{If a set of simple graphs is uniformly $k$-sparse for some $k$ and
$MS_{2}$-definable, then it is $MS_{1}$-definable (Courcelle
\cite{BCXIV}).}

\smallskip

This is somewhat analogous to the situation of Theorem~\ref{no Kmm}
(see Proposition~\ref{prop k-sparse}). However the combinatorial
conditions are different: if a set of graphs is uniformly $k$-sparse
for some $k$, it is without $K_{t,t}$ for some $t$, but the converse
does not hold. It is proved in the book by Bollobas \cite{Bol} that,
for each $t\geq2$, there is a number $a$ such that for each $n$, there
is a graph with $n$ vertices and $an^{b}$ edges that does not contain
$K_{t,t}$, where $b = 2t/(t+1)$. For these graphs, the number of edges
is not linearly bounded in terms of the number of vertices, so they are
not uniformly $k$-sparse for any $k$.

It is not clear how to extend Courcelle's proof in \cite{BCXIV}, to use
the condition \textit{without $K_{t,t}$} instead of \textit{uniformly
$k$-sparse}.

\section{Simple graphs vs multi-graphs}\label{sec simple multiple}

The formal setting of relational structures is very convenient to deal
with simple graphs, as we have seen already. It can also be used to
formalize multi-graphs (i.e., graphs with multiple edges), if we
consider two-sorted relational structures.

Formally, a \textit{multi-graph with sources} in $C$ is a structure of
the form $G = \langle V, E, \inc, (c_{G})_{c\in C}\rangle$ where $V$ is
the set of vertices, $E$ is the set of edges, each $c_{G}$ is an
element of $V$, and $\inc$ is a ternary relation of type $E\times
V\times V$. We interpret the relation $\inc(e,x,y)$ to mean that $e$ is
an edge from vertex $x$ to vertex $y$. We denote by $\GS_{m}(C)$ the
set of multi-graphs with sources in $C$. As in the study of $\StS$ or
$\GS$, we assume that the finite sets of source labels $C$ are taken in
a fixed countable set. We let $\GS_{m}$ be the union of the
$\GS_{m}(C)$ for all finite sets $C$ of source labels.

Graphs and hypergraphs with multiple edges and hyperedges are often
used, see the volume edited by Rozenberg \cite{Rozenberg}. In this
context, it is in fact frequent to consider operations on multi-graphs
that are very similar to the $\HR$-operations on $\GS$. More precisely,
the operations of disjoint union, source renaming, source forgetting
and source fusion can be defined naturally on multigraphs with sources:
thus $\GS_{m}$ can be seen naturally as an $\HR$-algebra.

It is clear that each simple graph in $\GS(C)$ can be considered as
an element in $\GS_{m}(C)$. It is important to note however that the 
$\HR$-operations on $GS_{m}$, when applied to such simple graphs, do
not necessarily yield the same result as in $\GS$. For instance, let 
$a,b$ be distinct elements of $C$, and let $G\in\GS(C)$ be a simple
graph. The action of fusing the $a$-source and the $b$-source of $G$ 
may now result in multiple edges: if there were arrows in both
directions between $a_{G}$ and $b_{G}$, or if there were arrows to
(resp. from) a vertex of $G$ from (resp. to) both $a_{G}$ and $b_{G}$.
In contrast, the same operation in $\GS(C)$ yields $\fus_{a,b}(G)$,
an element of $\GS(C)$ by definition. To avoid confusion, we will
denote by $\mfus_{a,b}$ this operation when used in $\GS_{m}$.

Fortunately, we do not have this sort of problem with the other
operations: applying the operations of disjoint union, source renaming
or source forgetting to simple graphs considered as elements of
$\GS_{m}$ yields the same result as applying the same operations
within the algebra $\GS$.

We let $\HR_{m}$ be the signature on $\GS_{m}$ consisting of the
operations of the form $\oplus$, $\srcfg_{a}$, $\srcren_{a\rightarrow
b}$ and $\mfus_{a,b}$. Thus, $\GS_{m}$ is an $\HR_{m}$-algebra. We
observe that, as a signature (that is, as a set of symbols denoting
operations), $\HR_{m}$ is in natural bijection with $\HR$. So we don't
really need to introduce the new notation $\HR_{m}$, and we could very
well say that $\GS_{m}$ is an $\HR$-algebra. We simply hope, by
introducing this notation, to clarify our comparative study of
recognizable subsets in the algebras $\GS$ and $\GS_{m}$. This
distinction will be useful in the proofs of Theorems \ref{L vs i(L) rec}
and \ref{L vs u(L) rec}.

To summarize and amplify the above remarks, let us introduce the
following notation. We denote by $\imath\colon \GS\rightarrow
\GS_{m}$ the natural injection. For each multi-graph $G$, we denote
by $u(G)$ the simple graph obtained from $G$ by fusing multiple edges
(with identical origin and end): that is, $u$ is a mapping from
$\GS_{m}$ onto $\GS$. Elementary properties of $\imath$ and $u$ are
listed in the next proposition.

\begin{proposition}\label{i, u morphisms}
    The mapping $u\colon\GS_{m}\rightarrow\GS$ is a homomorphism of
    $\HR$-algebras. The mapping $\imath\colon\GS\rightarrow\GS_{m}$ is
    not a homomorphism, but it commutes with the operations of the form
    $\oplus$, $\srcfg_{a}$ and $\srcren_{a\rightarrow b}$.
    
    $\imath$ does not commutes with the operations of the form
    $\fus_{a,b}$, but if $G\in\GS$, then $\imath(\fus_{a,b}(G)) =
    \imath(u(\mfus_{a,b}(\imath(G))))$.
    
    Finally, if $G\in\GS$, then $\imath(G) = u\inv(G) \cap \imath(\GS)$
    and $u(\imath(G)) = G$.
\end{proposition}

We now prove the following theorems, which describe the interaction
between $\HR_{m}$-recognizability of sets of multi-graphs and
$\HR$-recognizability of sets of simple graphs.

\begin{theorem}\label{simple graphs rec}
    The set of simple graphs is $\HR_{m}$-recognizable. More
    precisely, for each finite set of source labels $C$,
    $\imath(\GS(C))$ is $\HR_{m}$-recognizable.
\end{theorem}

\begin{theorem}\label{L vs i(L) rec}
    Let $C$ be a finite set of source labels and let $L\subseteq
    \GS(C)$. Then $L$ is $\HR$-recognizable if and only if $\imath(L)$
    is $\HR_{m}$-recognizable.
\end{theorem}

\begin{theorem}\label{L vs u(L) rec}
    Let $C$ be a finite set of source labels and let $L\subseteq
    \GS_{m}(C)$. If $L$ is $\HR_{m}$-recognizable, then $u(L)$ is
    $\HR$-recognizable.
\end{theorem}

\subsection{Proof of Theorem~\ref{simple graphs rec}}

We first introduce the notion of the type of a multi-graph: as for
the elements of $\StS$, if $G\in\GS_{m}(C)$, we let $\zeta(G)$ be the
restriction of $G$ to its $C$-sources and to the edges between them. 
We also denote by $\zeta$ the relation on $\GS_{m}$ induced by this
type mapping: two multi-graphs $G,H\in\GS_{m}(C)$ are
$\zeta$-equivalent if $\zeta(G) = \zeta(H)$.

\begin{lemma}\label{zeta for multi}
    The type relation $\zeta$ is an $\HR_{m}$-congruence on $\GS_{m}$.
    Moreover, for each finite set of source labels $C$, the elements of
    $\imath(\GS(C))$ can be found in only a finite number of
    $\zeta$-classes.
\end{lemma}

\proof
The result follows from the following, easily verifiable identities, where the
multi-graphs $G$, $H$ are assumed to have the appropriate sets of
sources.
\begin{eqnarray*}
    \zeta(G\oplus H) &=& \zeta(G) \oplus \zeta(H) \\
    \zeta(\srcren_{a\rightarrow b}(G)) &=& \srcren_{a\rightarrow
    b}(\zeta(G))\\
    \zeta(\mfus_{a,b}(G)) &=& \mfus_{a,b}(\zeta(G))\\
    \zeta(\srcfg_{a}(G)) &=& \zeta(\srcfg_{a}(\zeta(G))).
\end{eqnarray*}
The finiteness of the number of $\zeta$-classes containing elements of
$\imath(\GS(C))$ follows from the fact that there are only finitely
many source-only simple graphs with sources in $C$.
\eop

We also introduce the following finite invariant for a simple graph
$G\in\GS(C)$. We define $\eta(G)$ to be the set of all pairs
$\{a,b\}$ of elements of $C$ such that $a\ne b$, $a_{G}\ne b_{G}$ and
there exists a vertex $x$ of $G$ with either  edges from $x$ to both
$a_{G}$ and $b_{G}$, or edges to $x$ from both $a_G$ and $b_{G}$. The
set $\eta(G)$ can be viewed as a symmetric anti-reflexive relation on
$C$.

\begin{lemma}\label{charact multi}
    Let $G$ be a simple graph in $\GS(C)$ and let $a\ne b$ be elements
    of $C$. Then $\mfus_{a,b}(G)$ has multiple edges if and only if
    $\{a,b\} \in\eta(G)$ or $\mfus_{a,b}(\zeta(G))$ has multiple edges.
\end{lemma}

\proof
We first observe that $\mfus_{a,b}(G)$ has multiple edges if and only
if $a_{G}\ne b_{G}$ and at least one of the following situations
occurs: there are edges in both directions between $a_{G}$ and $b_{G}$,
or there is a vertex $x$ of $G$ with edges from (resp. to) both $a_{G}$
and $b_{G}$ (this includes the case where there is a loop at $a_{G}$ or
$b_{G}$ and an edge in either direction between $a_{G}$ and $b_{G}$).
That is, $\mfus_{a,b}(G)$ has multiple edges if and only
$\{a,b\}\in\eta(G)$ or there are edges in both directions between
$a_{G}$ and $b_{G}$.

We also observe that $\mfus_{a,b}(\zeta(G))$ is a subgraph of
$\mfus_{a,b}(G)$, so the former is simple if the latter is. Finally,
the existence of edges in both directions between $a_{G}$ and $b_{G}$
is sufficient to ensure that $\mfus_{a,b}(\zeta(G))$ has multiple edges.

These observations put together suffice to prove the lemma.
\eop

We are now ready to prove Theorem~\ref{simple graphs rec}. Let $\simeq$
be the following relation, defined on each $\GS_{m}(C)$. We let
$G\simeq G'$ if both $G$ and $G'$ have multiple edges, or both $G$ and
$G'$ are simple graphs, $\zeta(G) = \zeta(G')$ and $\eta(G) =
\eta(G')$.

It is immediate that $\simeq$ is an equivalence relation, saturating
$\imath(\GS(C))$. It follows from Lemma~\ref{zeta for multi} and from
the fact that $\eta(G)$ is a subset of the finite set $C\times C$, that
$\simeq$ is locally finite. So we only need to show that $\simeq$ is 
an $\HR_{m}$-congruence.

We need to describe the interaction between the mapping $\eta$ and
the $\HR_{m}$-operations. As observed in Proposition \ref{i, u
morphisms}, all $\HR_{m}$-operations preserve simple graphs except for
the operations of the form $\mfus_{a,b}$. Assuming that $G,H$ are
simple graphs with the appropriate sets of sources, we easily verify 
the following:

\begin{eqnarray*}
    \eta(G\oplus H) &=& \eta(G) \cup \eta(H) \\
    \eta(\srcfg_{a}(G)) &=& \eta(G) \setminus \{\{a,b\} \mid b\in C,\
    \{a,b\}\in\eta(G)\} \\
    \eta(\srcren_{a\rightarrow b}(G)) &=& \eta(G) \setminus \{\{a,c\}\mid
    c\in C,\ \{a,c\}\in\eta(G)\}\\
    &&\hskip1cm\cup \{\{b,c\}\mid c\in C,\ \{a,c\}\in\eta(G)\}
\end{eqnarray*}
Moreover, if $a_{G} \ne b_{G}$ and $\mfus_{a,b}(G)$ is simple (if it
isn't, its $\eta$-image is not defined), then $\eta(\mfus_{a,b}(G))$
consists of:

(1)\enspace  all pairs in $\eta(G)$,

(2)\enspace all pairs $\{c,d\}$ such that there are edges in $\zeta(G)$
from $a$ to $c$ and from $b$ to $d$, or from $c$ to $a$ and from $d$ to
$b$,

(3)\enspace all pairs $\{a,c\}$ (resp. $\{b,c\}$) such that
$\{b,c\}\in\eta(G)$ (resp. $\{a,c\}\in\eta(G)$),

(4)\enspace all pairs $\{a,c\}$ and $\{b,c\}$ such that there are edges
in $\zeta(G)$ between $a$ and $b$ (in either direction) and between $a$
or $b$ and $c$ (in any direction).

Let us justify this statement: it is easy to see that all these pairs
belong to $\eta(\mfus_{a,b}(G))$. In particular, $\eta(G) \subseteq
\eta(\mfus_{a,b}(G))$ since, as $\mfus_{a,b}(G)$ is assumed to be
simple, there is no $\{c,d\}\in \eta(G)$ such that $a_{G} = c_{G}$
and $b_{G} = d_{G}$.

Conversely, let us consider distinct edges in $G' = \mfus_{a,b}(G)$,
from $y$ to $x$ and from $z$ to $x$, as in Figure~\ref{fig: distinct
edges} (note that $x$ and $y$ may be equal), such that $y = e_{G'}$
and $z = f_{G'}$ for $e,f\in C$.
\begin{figure}[htb]
    \centering
    \begin{picture}(90,25)(0,-22)
    \node(n0)(0.0,0.0){$y$}

    \node(n1)(0.0,-20.0){$z$}

    \node(n2)(24.0,-8.0){$x$}

    \node(n3)(50.0,-8.0){$z$}

    \node(n4)(78.0,-8.0){$x{=}y$}

    \drawedge(n0,n2){}

    \drawedge(n1,n2){}

    \drawedge(n3,n4){}
    
    \drawloop[loopangle=-2.29](n4){}
    \end{picture}
    \caption{Distinct edges in $\mfus_{a,b}(G)$}
    \label{fig: distinct edges}
\end{figure}
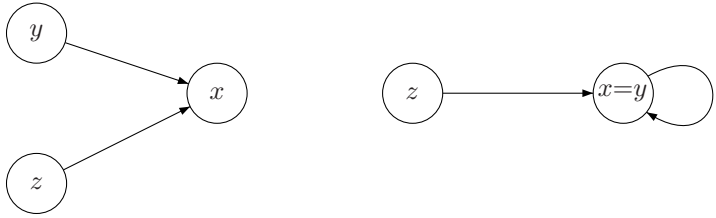
If neither $x$, nor $y$ nor $z$ is the $a$- and $b$-source in $G'$,
then we are in case (1), \textit{i.e.}, $\{e,f\}\in\eta(G)$. If $x$ is
the $a$- and $b$-source in $G'$ but neither $y$ nor $z$ is, then
$\{e,f\}$ satisfies case (1) or (2). If $y$ is the $a$- and $b$-source
in $G'$ but neither $x$ nor $z$ is, then $\{e,f\}$ satisfies case (3).
The same holds by symmetry if $z$ is the only one of these three
vertices to be the $a$- and $b$-source in $G'$. Finally if $x = y$
(resp. $x = z$) and is the $a$- and $b$-source,in $G'$ then there is an
edge between the $a$- and the $b$-source in $G$ and $\{e,f\}$ satisfies
case (4). The case of edges from $x$ to $y$ and to $z$ is symmetrical.

\medskip

In particular, $\eta(G\oplus H)$, $\eta(\srcfg_{a}(G))$,
$\eta(\srcren_{a\rightarrow b}(G))$ and $\eta(\mfus_{a,b}(G))$ are
entirely determined by $\eta(G)$, $\zeta(G)$ and $\eta(H)$.

Let us now consider $G, G', H, H'$ in $\GS_{m}$ (with the appropriate
sets of sources) such that $G\simeq G'$ and $H\simeq H'$. If $G$ is 
not simple, then neither are $G'$, $G\oplus H$, $\srcfg_{a}(G)$,
$\srcren_{a\rightarrow b}(G)$ and $\mfus_{a,b}(G)$. In particular, we
have $G\oplus H \simeq G'\oplus H'$, $\srcfg_{a}(G)\simeq
\srcfg_{a}(G')$, $\srcren_{a\rightarrow b}(G)\simeq
\srcren_{a\rightarrow b}(G')$ and $\mfus_{a,b}(G)\simeq
\mfus_{a,b}(G')$.

Assume now that $G$ and $H$ are simple. Then so are $G\oplus H$,
$\srcfg_{a}(G)$ and $\srcren_{a\rightarrow b}(G)$, and we have seen
that their $\eta$-images are determined by $\eta(G)$ and $\eta(H)$.
Since $\zeta$ is an $\HR_{m}$-congruence (Lemma~\ref{zeta for multi}),
it follows that $\simeq$ is preserved by the operations $\oplus$,
$\srcfg_{a}$ and $\srcren_{a\rightarrow b}$.

By Lemma \ref{charact multi}, whether $\mfus_{a,b}(G)$ is simple, is
determined by $\zeta(G)$ and $\eta(G)$, and hence $\mfus_{a,b}(G)$ and
$\mfus_{a,b}(G')$ are both non-simple (and then $\simeq$-equivalent) or
both simple. In the latter case, their $\eta$-images are equal since
they are both determined by $\eta(G) = \eta(G')$ and $\zeta(G) =
\zeta(G')$. Thus $\simeq$ is preserved by the operation $\mfus_{a,b}$.
This concludes the proof of Theorem~\ref{simple graphs rec}.

\subsection{Proof of Theorem~\ref{L vs i(L) rec}}

Recall that we want to show that for each $L\in\GS(C)$, $L$ is
$\HR$-recognizable if and only if $\imath(L)$ is $\HR_{m}$-recognizable.

One direction is quickly established: we know from Proposition~\ref{i,
u morphisms} that $\imath(L) = u\inv(L) \cap \imath(\GS(C))$. If $L$ is
$\HR$-recognizable, then $u\inv(L)$ is $\HR_{m}$-recognizable since $u$
is a homomorphism. In view of Theorem~\ref{simple graphs rec}, it
follows that $\imath(L)$ is $\HR_{m}$-recognizable as well.

Conversely, let us assume that $\imath(L)$ is $\HR_{m}$-recognizable
and let $\equiv$ be a locally finite $\HR_{m}$-congruence on $\GS_{m}$
saturating $\imath(L)$. We want to define a locally finite
$\HR$-congruence $\sim$ on $\GS$ saturating $L$.

For each symmetric anti-reflexive relation $A$ on a finite set of
source labels $D$ and for each graph $G\in\GS(D)$, let
$\del_{A}(G)\in\GS(D)$ be the graph obtained from $G$ by deleting the
edges between the $a$-source and the $b$-source for each pair $\{a,b\}$
in $D$. Let also $\fus_A$ be the composition of the operations
$\fus_{a,b}$ for all $\{a,b\} \in D$, in any order.

For $G,G'\in\GS(D)$, we let $G \sim G'$ if $\imath(G) \equiv
\imath(G')$, $\zeta(G) = \zeta(G')$ and, for each symmetric
anti-reflexive relation $A$ on $D$,
$$\imath\fus_{A}\del_{A}(G) \equiv \imath\fus_{A}\del_{A}(G').$$
The relation $\sim$ is clearly an equivalence relation, and it is
locally finite since $\equiv$ and $\zeta$ are. Moreover, it saturates
$L$ since $G\in L$ if and only if $\imath(G)\in\imath(L)$, and $\equiv$
saturates $\imath(L)$. The rest of the proof consists in showing that
$\sim$ is an $\HR$-congruence.

\paragraph{The source renaming operation}
Let $G\sim G'$ in $\GS(D)$. Then $\imath(G) \equiv \imath(G')$. Since
$\equiv$ is a congruence and in view of Proposition~\ref{i, u
morphisms}, $\imath(\srcren_{a\rightarrow b}(G)) =
\srcren_{a\rightarrow b}(\imath(G)) \equiv \srcren_{a\rightarrow
b}(\imath(G')) = \imath(\srcren_{a\rightarrow b}(G'))$. It also follows
from Lemma \ref{FFVV light} that $\zeta(\srcren_{a\rightarrow b}(G))
=\zeta(\srcren_{a\rightarrow b}(G'))$.

Let us now consider a symmetric anti-reflexive relation $A$ on the set
of source labels of $\srcren_{a\rightarrow b}(G)$. It is easily
verified that
$$\del_{A}\srcren_{a\rightarrow b} = \srcren_{a\rightarrow
b}\del_{B},$$
where $B = \{\{c,d\}\in A \mid \{c, d\} \cap\{a,b\} = \emptyset\} \cup
\{\{a,d\} \mid \{b,d\}\in A\}$. We also note that if $c,d\in
C\setminus\{a,b\}$, then $\fus_{c,d}$ and $\srcren_{a\rightarrow b}$
commute. Moreover $\fus_{b,d}\srcren_{a\rightarrow b} =
\srcren_{a\rightarrow b}\fus_{a,d}$ and
$\fus_{c,b}\srcren_{a\rightarrow b} = \srcren_{a\rightarrow
b}\fus_{c,a}$. Thus $\fus_{A}\srcren_{a\rightarrow b} =
\srcren_{a\rightarrow b}\fus_{B}$.

Now, using the fact that $\imath$ commutes with $\srcren_{a\rightarrow
b}$ we have
\begin{eqnarray*}
    \imath\fus_{A}\del_{A}\srcren_{a\rightarrow b}(G) &=&
    \imath\fus_{A}\srcren_{a\rightarrow b}\del_{B}(G) \\
    &=& \imath\srcren_{a\rightarrow b}\fus_{B}\del_{B}(G) \\
    &=& \srcren_{a\rightarrow b}\imath\fus_{B}\del_{B}(G).
\end{eqnarray*}
Since $\equiv$ is an $\HR_{m}$-congruence, it follows that
$$\imath\fus_{A}\del_{A}\srcren_{a\rightarrow b}(G) \equiv
\imath\fus_{A}\del_{A}\srcren_{a\rightarrow b}(G')$$
and, finally, that $\srcren_{a\rightarrow b}(G) \sim
\srcren_{a\rightarrow b}(G')$.

\paragraph{The source forgetting operation}
The proof is the same as for the source renaming operation, with this
simplifying circumstance that $\del_{A}\srcfg_{a} = \srcfg_{a}\del_{A}$
and $\fus_{A}\srcfg_{a} = \srcfg_{a}\fus_{A}$ (since $a$ is not a
source label of $\srcfg_{a}(G)$, and hence does not occur in $A$).

\paragraph{The source fusion operation}
Let $G\sim G'$ in $\GS(D)$. Here it is not immediate that
$\imath(\fus_{a,b}(G)) \equiv \imath(\fus_{a,b}(G'))$. However, if we
let $A = \{\{a,b\}\}$, we know that
$$\imath\fuse_{A}\del_{A}(G) \equiv \imath\fuse_{A}\del_{A}(G').$$
We note that $\fuse_{A}\del_{A}(G)$ is equal to $\fus_{a,b}(G)$ if $G$
has no edge between its $a$- or $b$-source, or if it has a loop at
either. Otherwise, $\fus_{a,b}(G)$ is equal to $\fuse_{A}\del_{A}(G)$
with a loop added to its $a$-source, that is:
$$\fus_{a,b}(G) = \srcfg_{\alpha}\srcfg_{\beta}
\fus_{a,\alpha}\fus_{b,\beta}(\fus_{A}\del_{A}(G) \oplus
E)\eqno{(*)}$$
where $\alpha$ and $\beta$ are source labels not in $D$ and $E$ is the
graph in $\GS(\{\alpha,\beta\})$ with 2 vertices and a single edge from
its $\alpha$-source to its $\beta$-source.

Observe also that the existence of loops at, or edges between the $a$-
and $b$-source of $G$ is a condition that depends only on $\zeta(G)$,
so it will be satisfied by both $G$ and $G'$ or by neither.

In the first case, where $\fuse_{A}\del_{A}(G) = \fus_{a,b}(G)$,
we find immediately that $\imath(\fuse_{a,b}(G)) \equiv
\imath(\fuse_{a,b}(G'))$. In the second case, the same
$\equiv$-equivalence is derived from Proposition~\ref{i, u morphisms}
and Equation ($*$) above.

By Lemma \ref{FFVV light}, $\zeta$-equivalence is preserved by the
operation $\fus_{a,b}$.

Now let $A$ be a symmetric anti-reflexive relation on $D$: we consider
the graph $\imath\fuse_{A}\del_{A}\fus_{a,b}(G)$. Our first observation
is that $\del_{A}\fus_{a,b} = \fus_{a,b}\del_{B}$ where
$$B = A \cup \{\{a,c\} \mid \{b,c\} \in A\} \cup \{\{b,c\} \mid \{a,c\}
\in A\}.$$
Next, we observe that $\fus_{A}\fus_{a,b} = \fus_{a,b}\fus_{B}$. Thus
we have
$$\imath\fuse_{A}\del_{A}\fus_{a,b}(G) = \imath\fuse_{A}\fus_{a,b}\del_{B}
= \imath\fuse_{a,b}\del_{B}\fus_{B}(G),$$
and hence $\imath\fuse_{A}\del_{A}\fus_{a,b}(G) \equiv
\imath\fuse_{A}\del_{A}\fus_{a,b}(G')$. It follows that $\fus_{a,b}(G)
\sim \fus_{a,b}(G')$.

\paragraph{The disjoint union operation}
Let $G\sim G'$ in $\GS(C)$ and $H\sim H'$ in $\GS(D)$ (where $C$ and
$D$ are disjoint). Since $\imath$ and $\zeta$ preserve $\oplus$, we
have $\imath(G \oplus H) \equiv \imath(G' \oplus H')$ and
$\zeta(G\oplus H) = \zeta(G' \oplus H')$.

Now let $A$ be a symmetric anti-reflexive relation on $C\cup D$. Let
$Q$ (resp. $R$) be the restriction of $A$ to $C$ (resp. $D$) and let $P
= A \cap ((C\times D)\cup(D\times C))$. It is easily verified that
\begin{eqnarray*}
    \del_{A}(G \oplus H) &=& \del_{Q}(G) \oplus \del_{R}(H) \\
    \fus_{A}\del_{A}(G \oplus H) &=& \fus_{P}(\fus_{Q}\del_{Q}(G) \oplus 
    \fus_{R}\del_{R}(H)).
\end{eqnarray*}
It now follows from Proposition~\ref{i, u morphisms} that
\begin{eqnarray*}
    \imath\fus_{A}\del_{A}(G \oplus H) &=&
    \imath\fus_{P}(\fus_{Q}\del_{Q}(G) \oplus \fus_{R}\del_{R}(H)) \\
    &=& \imath u \mfus_{P}\imath(\fus_{Q}\del_{Q}(G) \oplus
    \fus_{R}\del_{R}(H)) \\
    &=& \imath u \mfus_{P}(\imath\fus_{Q}\del_{Q}(G) \oplus
    \imath\fus_{R}\del_{R}(H)).
\end{eqnarray*}
Thus $\imath\fus_{A}\del_{A}(G \oplus H) \equiv
\imath\fus_{A}\del_{A}(G' \oplus H')$, and hence $G \oplus H \sim G'
\oplus H'$.

This concludes the proof of Theorem~\ref{L vs i(L) rec}.

\subsection{Proof of Theorem~\ref{L vs u(L) rec}}

Let $L\in\GS_{m}(C)$ be $\HR_{m}$-recognizable, and let $\equiv$ be a
locally finite $\HR_{m}$-congruence saturating $L$. We want to show that
$u(L)$ (a subset of $\GS(C)$) is $\HR$-recognizable.

Let $G, G'\in \GS(D)$. We let $GÊ\sim G'$ if, for each $H\in u\inv(G)$,
there exists $H'\in u\inv(G')$ such that $H\equiv H'$, and
symmetrically, for each $H'\in u\inv(G')$, there exists $H\in u\inv(G)$
such that $H\equiv H'$.

The relation $\sim$ is easily seen to be a locally finite equivalence
relation on $\GS$, saturating $u(L)$. There remains to see that
$\sim$ is an $\HR$-congruence.

We first establish the following lemma.

\begin{lemma}\label{for 6.4}
    Let $G\in\GS_{m}$ and let $H,K\in\GS$.
    \begin{itemize}
	\item $u(G) = H\oplus K$ if and only if there exist
	multi-graphs $H',K'$ such that $G = H'\oplus K'$, $u(H') = H$
	and $u(K') = K$.
	\item $u(G) = \srcfg_{a}(H)$ if and only if there exists a
	multi-graph $H'$ such that $G = \srcfg_{a}(H')$ and $u(H') = H$.
	\item $u(G) = \srcren_{a\rightarrow b}(H)$ if and only if there
	exists a multi-graph $H'$ such that $G = \srcren_{a\rightarrow
	b}(H')$ and $u(H') = H$.
	\item $u(G) = \fus_{a,b}(H)$ if and only if there exists a
	multi-graph $H'$ such that $G = \mfus_{a,b}(H')$ and $u(H') =
	H$.
    \end{itemize} 
\end{lemma}

\proof
Recall that $G$ and $u(G)$ have the same set of vertices, and each
edge $e$ of $u(G)$ arises from the identification $n(e)\ge 1$ edges
of $G$ between the same vertices.

If $u(G) = H\oplus K$, each edge of $u(G)$ is in exactly one of $H$ and
$K$. Let $H'$ (resp. $K'$) be the graph obtained from $H$ (resp. $K$)
by replacing each edge $e$ by $n(e)$ parallel edges. Then $G = H'\oplus
K'$, $u(H') = H$ and $u(K') = K$, as required.

The proof of the statements relative to the operations $\srcfg_{a}$ and
$\srcren_{a\rightarrow b}$ is done in the same fashion.

Let us finally consider the case where $u(G) = \fus_{a,b}(H)$. If
$a_{H} = b_{H}$, that is, $H = u(G)$, then $G = \mfus_{a,b}(G)$ and
we can let $H' = G$.

If $a_{H}\ne b_{H}$, we let $H'$ be obtained from $H$
be obtained from $H$ as follows: for each vertex $x$, each edge $e$ from 
$x$ to $y$ ($y\ne a,b$) is replaced by $n(e)$ parallel edges, and the
edges from $x$ to $a$ and $b$ are duplicated to a total of $n(e)$
edges.
\eop

We can now conclude the proof of Theorem~\ref{L vs u(L) rec}, by
proving that $\sim$ is an $\HR$-congruence. Let $G\sim G'$ and $H\sim
H'$. Let $K\in u\inv(G\oplus H)$. By Lemma~\ref{for 6.4}, $K = L \oplus
M$ for some $L\in u\inv(G)$ and $M\in u\inv(H)$. Since $G\sim G'$ and
$H\sim H'$, there exist $L'\in u\inv(G')$ and $M'\in u\inv(H')$ such
that $L'\equiv L$ and $M'\equiv M$. Let $K' = L'\oplus M'$. Then $K' =
L'\oplus M'\equiv L \oplus M = K$ and $K' \in u\inv(G' \oplus H')$. By
symmetry, this shows that $G \oplus H \sim G' \oplus H'$.

The verification that $\sim$ is preserved by the other $\HR$-operations
proceeds along the same lines. This concludes the proof of
Theorem~\ref{L vs u(L) rec}.

\section{Graph algebras based on graph substitutions}\label{sec:
modular}

The class $\Grph$, defined in Section~\ref{sec rel structures}, has
already been discussed in terms of the signatures $\calS$, $\VR$ and
$\HR$ since it is a domain in each of the three algebras $\StS$, $\GP$
and $\GS$. In this section, we consider a different set of operations
on $\Grph$, arising from the theory of the modular decomposition of
graphs, which makes $\Grph$ an algebra (one-sorted for a change!). This
algebraic framework was considered by the authors, in \cite{BCX} and
\cite{PWJALC}.

We first recall the definition of the composition operation on graphs.
Let $H$ be a graph with vertex set $[n] = \{1,\ldots,n\}$ ($n\ge 2$).
If $G_{1},\ldots, G_{n}$ are graphs, then the composite $H\langle
G_{1},\ldots,G_{n}\rangle$ is obtained by taking the disjoint union of
the graphs $G_{1},\ldots,G_{n}$, and by adding, for each edge $(i,j)$
of $H$ where $i\ne j$, an edge from every vertex of $G_{i}$ to every
vertex of $G_{j}$.

We say that a graph is \textit{indecomposable}, or \textit{prime}, if
it cannot be written non-trivially as a composition (a composition is
trivial if each of its arguments is a singleton). It is easily verified
that if $H$ and $H'$ are isomorphic graphs, then the corresponding
composition operations yield isomorphic graphs. So we fix a set
$\calF_{\infty}$ of representatives of the isomorphism classes of
indecomposable graphs. In particular, we may assume that every graph in
$\calF_{\infty}$ has a vertex set of the form $[n]$ for some $n\ge 2$.
We also denote by $\calF_{\infty}$ the resulting \textit{modular
signature}, consisting of the composition operations defined by these
graphs. The $\calF_{\infty}$-algebra of graphs is denoted by
$\Grph^{\calF_{\infty}}$.

It turns out that every finite graph admits a \textit{modular
decomposition}, that is, it can be expressed from the single-vertex
graph using only operations from $\calF_{\infty}$. This fact has been
rediscovered a number of times in the context of graph theory and of
other fields using graph-theoretic representations. We refer to
\cite{Mohring-Radermacher} for a historical survey, and to \cite{MS99}
for a concise presentation. In other words, $\Grph$ is generated by the
signature $\calF_{\infty}$ augmented with the constants $\vloop$ and
$\v$, which denote a single vertex graph, respectively with and without
a single loop edge.

\begin{remark}
    The modular decomposition of a graph is unique up to certain simple
    (equational) rules, see for instance \cite{PWJALC}. Moreover, the
    modular decomposition of a graph can be computed in linear time
    \cite{MS93,MS99,CournierHabib}.
\end{remark}

Our first results connect $\VR$-recognizability and
$\calF_{\infty}$-recognizability.

\begin{proposition}\label{VR-F}
    Every $\VR$-recognizable set of graphs is
    $\calF_{\infty}$-recognizable.
\end{proposition}

\proof
In view of Proposition~\ref{easy recog facts} and Theorem~\ref{thm VR
equivalence}, it suffices to show that every operation in
$\calF_{\infty}$ is $\VR^{+}$-derived.

For each integer $i$, let $\mark_{i}$ be the unary operation on $\GP$,
of type $\emptyset\rightarrow\{i\}$, defined as follows: given a graph
without ports, it simply marks every vertex with port label $i$
(leaving the set of vertices and the edge relation unchanged). Note that
$\mark_{i}$ is a qfd unary operation, and hence a
$\VR^{+}$-operation.

Let $H$ be an $n$-ary operation, that is, a graph in $\calF_{\infty}$
with vertex set $[n]$, and let $\edg_{H}$ be its edge relation. If
$G_{1},\ldots,G_{n}$ are finite graphs, the construction of $H\langle
G_{1},\ldots,G_{n}\rangle$ can be described as follows:

- construct the disjoint union, $\mark_{1}(G_{1}) \oplus
\cdots \oplus \mark_{n}(G_{n})$, an element of $\GP([n])$;

- apply (in any order) to this disjoint union the operations
$\add_{i,j}$ for all $i,j\in[n]$ such that $(i,j)$ is an edge of $H$
and $i\ne j$;

- forget all ports, that is, apply the operation $\mdf_{\emptyset}$.

\noindent This completes the verification that the operation defined by
$H$ can be expressed as a $\VR^{+}$-term, and hence the proof.
\eop

The following result shows that the converse of Proposition~\ref{VR-F}
does not hold.

\begin{proposition}
    Every set of prime graphs is $\calF_{\infty}$-recognizable, and
    there is a set of prime graphs which is not $\VR$-recognizable.
\end{proposition}

\proof
Let $L$ be a set of prime graphs, and let $\equiv$ be the relation on
$\Grph$ defined as follows. We let $G\equiv H$ if one of the
following holds:
\begin{itemize}
    \item neither $G$ nor $H$ is prime;
    
    \item $G$ and $H$ are both \textsf{1} (the graph with one vertex
    and no edge);
    
    \item $G$ and $H$ are both not \textsf{1}, prime and in $L$;
    
    \item $G$ and $H$ are both not \textsf{1}, prime and not in $L$.
\end{itemize}
This is clearly an equivalence relation with four classes, which
saturates $L$. Moreover, $\equiv$ is an $\calF_{\infty}$-congruence.
Indeed, let $K$ be a graph with $n$ vertices; for $i=1,\ldots,n$, let
$G_{i} \equiv H_{i}$ for each $i$. If for some $i$, $G_{i} \ne
\textsf{1}$, then $H_{i} \ne \textsf{1}$, and neither $K\langle
G_{1},...,G_{n}\rangle$ nor $K\langle H_{1},...,H_{n}\rangle$ is prime:
therefore they are equivalent. Otherwise, $G_{i} = H_{i} = \textsf{1}$
for each $i$, $K\langle G_{1},...,G_{n}\rangle$ and $K\langle
H_{1},...,H_{n}\rangle$ are both equal to $K$, and hence they are
equivalent.
This concludes the proof that every set of prime graphs is
$\calF_{\infty}$-recognizable.

Before we exhibit a set of prime graphs which is not $\VR$-recognizable,
we define inductively a sequence of $\VR$-terms written with three port
labels $a,b,c$. We let
$$t_{0} = \add_{a,b}(a\oplus\ b), \qquad t_{n+1} = \ren_{c\rightarrow
b}(\ren_{b\rightarrow a}(\add_{b,c}(t_{n}\oplus c))).$$
The term $\mdf_{\emptyset}(t_{n})$ (forgetting all port labels in
$t_{n}$) denotes the string graph $P_{n+2}$, with $n+2$ vertices, say
$1,\ldots,n+2$ and edges from $i$ to $i+1$ for each $1\le i\le n+1$.
Each of these graphs is prime.

Now let $A$ be a set of positive integers that is not recognizable in
$\langle\mathbb{N},\form{succ},0\rangle$ and let $L$ be the set of
all terms $P_{n}$ with $n\in A$. From the above discussion, we know
that $L$ is $\calF_{\infty}$-recognizable. If $L$ was
$\VR$-recognizable, standard arguments would show that the set of
$\VR$-terms $t_{n}$ ($n\in A$) would be recognizable as well, and it
would follow that $A$ is recognizable, contradicting its choice.
\eop

Now let $\calF$ be a finite subsignature of the modular signature
$\calF_{\infty}$. A graph which can be constructed from one-vertex
graphs using only operations from $\calF$ is called an $\calF$-graph.
The next result deals with sets of $\calF$-graphs. This finiteness
condition (the elements of $L$ are built by repeated composition of a
finite number of graph-based operations) is non-trivial. In fact, for
many natural classes of graphs such as rectangular grids, it is not
satisfied: since grids are indecomposable, a set of graphs containing
infinitely many grids cannot satisfy our finiteness condition. But that
condition is satisfied by other classical classes (e.g.
cographs, series-parallel posets), see \cite{BCX,PWJALC}.

Using results of Courcelle \cite{BCX}, we can show the following
result, which yields in particular a weak converse of
Proposition~\ref{VR-F}.

\begin{theorem}\label{FSVR}
    Let $\calF$ be a finite subsignature of $\calF_{\infty}$ and let
    $L$ be a set of $\calF$-graphs.  The following properties are
    equivalent:
    \begin{enumerate}
	\item $L$ is $\calS$-recognizable;
	\item $L$ is $\VR$-recognizable.
	\item $L$ is $\calF_{\infty}$-recognizable.
	\item $L$ is $\calF$-recognizable.
    \end{enumerate}
\end{theorem}

\proof
The equivalence of (1) and (2) can be found in Theorem~\ref{thm VR
equivalence}. Proposition~\ref{VR-F} shows that (2) implies (3). And
(3) implies (4) as an immediate consequence of Proposition~\ref{easy
recog facts} since $\calF$ is a subsignature of $\calF_{\infty}$. The
fact that (4) implies (1) is a consequence of two results of Courcelle:
\cite[Theorem 4.1]{BCX}, which states that if a set of $\calF$-graphs
is $\calF$-recognizable, then it is definable in a certain extension of
$MS$-logic; and \cite[Theorem 6.11]{BCX}, which states that all sets
definable in this logical language are $\calS$-recognizable.
\eop

\begin{remark}
    Theorem~\ref{FSVR} states that for sets of graphs with only
    finitely many prime subgraphs, all four notions of
    recognizability are equivalent. Presented in this fashion, the
    statement is somewhat similar to that of Theorem~\ref{no Kmm}.
\end{remark}

\section{Conclusion}

In this article, we have investigated the recognizability of sets of
graphs quite in detail, focusing on the robustness of the notion, which
was not immediate since many signatures on graphs can be defined.
Although we had in mind sets of graphs, we have proved that embedding
graphs in the more general class of relational structures does not
alter recognizability. We have proved that the very same structural
conditions that equate $\VR$-equational and $\HR$-equational sets of
graphs, also equates $\HR$-recognizability and $\VR$-recognizability.

Summing up, we have defined a number of tools for handling
recognizability.
Some questions remain to investigate.

\medskip

\noindent$\bullet$\enspace When is it true that a quantifier-free
operation preserves recognizability?

\smallskip

Results in this direction have been established in Courcelle
\cite{BC-MSCS}. Are they applicable to quantifier-free definable
operations? In particular, is it true that the set of disjoint unions
of two graphs, one from each of two $\VR$-recognizable sets is
$\VR$-recognizable ?

\medskip

\noindent$\bullet$\enspace Which quantifier-free definable operations
can be added to the signature $\HR$, in such a way that the class of
$\HR$-recognizable sets is preserved (as is the case when we extend
$\VR$ to $\VR^{+}$)? The paper by Blumensath and Courcelle
\cite{BluCour}, which continues the present research, considers unary
non qfd operations that can be added to $\VR^+$ and to $\StS$ while
preserving the classes of equational and recognizable sets.

\medskip

\noindent$\bullet$\enspace Our example of an $\HR$-recognizable, not
$\VR$-recognizable set of cliques, is based on the weakness of the
parallel composition of graphs with sources, i.e., the fact that this
operation is not able to split large cliques. Can one find another
example, based on a different argument? If one cannot, what does this
mean?

\medskip

We conclude with an observation concerning the finiteness of
signatures. Whereas all finite words on a finite alphabet can be
generated by this alphabet and only one operation, dealing with finite
graphs (by means of grammars, automata and related tools) requires
infinite signatures. More precisely, one needs infinitely many
operations to generate all finite unlabelled graphs (see
Remark~\ref{economical} below). On the other hand, applications to
testing graph properties require the consideration of algebras
generated by a finite signature. Here is the reason.

\medskip

Let $M$ be an $\calF$-algebra of graphs. If the unique valuation
homomorphism $val_{M}\colon T(\calF)\rightarrow M$ (which evaluates a
term into an element of $M$) is surjective, i.e., if $\calF$ generates
$M$, then a subset $L$ of $M$ is recognizable if and only if
$val_{M}^{-1}(L)$ is a recognizable set of terms (see
Proposition~\ref{easy recog facts} and Section~\ref{significance}). And
the membership of a term in a recognizable set can be verified in
linear time by a finite deterministic (tree) automaton. Hence the
membership of a graph $G$ in $L$ can be checked as follows:

(1) One must first find some term $t$ such that $val_{M}(t)=G$, 

(2) then one checks whether $t$ belongs to $val_{M}^{-1}(L)$. 

\medskip

The latter step can be done in time proportional to the size of $t$,
usually no larger than the number of vertices of $G$. Although any term
$t$ with value $G$ gives the correct answer, it may be difficult to
find at least one (graph parsing problems may be $NP$-complete).

Because of this fact many hard problems (in particular if they are
expressed in Monadic Second-order logic) can be solved in linear time
on sets of graphs of bounded tree-width, and also on sets of graphs of
bounded clique-width, provided the graphs are given with appropriate
decompositions, see Courcelle \cite{HbGraGraRozenberg97}, Courcelle and
Olariu \cite{BC-Olariu} or Downey and Fellows \cite{DF}. If the
decompositions are not given, one can achieve linear time for graphs 
of bounded tree-width and $MS_{2}$ problems using a result by
Bodlaender \cite{BodSIAM}, and polynomial time for graphs of bounded
clique-width and $MS_{1}$ problems using a result by Oum and Seymour 
\cite{OumSeymour}.

However, even if $\calF$ is infinite or is finite without generating
the set $M$, recognizability remains interesting as an algebraic
concept, and for every restriction to a finitely generated subset of
$M$, we are back to the ``good'' case of a finitely generated algebra.

Finally, we think that infinite signatures can be used for checking
graph properties defining recognizable sets. This will not be possible
by \textit{finite} tree-automata if the graph algebra is not finitely
generated, but it can perhaps be done with automata using ``oracles''.
An oracle would be a subroutine handling some verifications for big
subgraphs that cannot be decomposed by the operations under
consideration. This idea needs of course further elaboration.

\begin{remark}\label{economical}
    We asserted above that finite unlabelled graphs cannot be generated
    with a finite signature. This is not entirely correct, and we
    briefly describe here a signature with 6 operations on a 2-sorted
    algebra which generates, somewhat artificially, all finite graphs
    (undirected and without loops). These operations have no good
    behaviour with respect to automata and verification questions, and
    such an ``economical'' generation of graphs is useless.
    
    The 2 sorts are $\form o$, the set of finite graphs equipped with a
    linear order of their vertex set, and $\form u$, the set of
    ordinary, unordered graphs. There is one unary operation of type
    $\form o\rightarrow\form u$, which forgets the order on the vertex
    set. All other operations are unary, of type $\form
    o\rightarrow\form o$: one consists in adding one new vertex, to be
    the new least element; one adds an (undirected) edge between the
    two least vertices; one performs a circular shift of the vertices;
    and one swaps the two least vertices. The three last operations
    leave the graph unchanged if it has less than 2 vertices. Finally,
    one adds a 6th, nullary operation, of type $\form o$: the constant
    $0$, standing for the empty graph with no vertices.
\end{remark}

\appendix
\section{Equivalences of logical formulas}\label{appendix}

In this appendix, we discuss some equivalences and transformations of
logical formulas which can be used to give upper bounds for the index
of congruences considered in this paper, and to complete the proof of
the effectiveness of certain notions (e.g. quantifier-free definition
schemes).

More specifically, we make precise in what sense we can state, as we do
in the body of the paper, that the set of first-order (resp. monadic
second-order) formulas over finite sets of relations, constants and
free variables, and with a bounded quantification depth, can be
considered as finite. Moreover, explicit upper bounds on the size of
these finite sets are derived, which can be used to justify the
termination of some of our algorithms, and in evaluating their
complexity. That these upper bounds have unbounded levels of
exponentiation is not unexpected, and even unavoidable by Frick and
Grohe \cite{FrickGrohe}.

\subsection{Boolean formulas}

Let $p_{1},\ldots,p_{n}$ be Boolean variables and let $B_{n}$ be the 
set of Boolean formulas written with these variables. It is well
known that $B_{n}$ is finite up to logical equivalence. For further
reference, we record the following more precise statement.

\begin{proposition}\label{red bool}
    There exists a subset $B_{n}^{red}$ of $B_{n}$, of cardinality
    $2^{2^n}$ such that every formula in $B_{n}$ can be effectively
    transformed into an equivalent formula in $B_{n}^{red}$.
\end{proposition}

\proof
We let $B_{n}^{red}$ be the set of Boolean formulas in disjunctive
normal form, where in each disjunct, variables occur at most
once and in increasing order, no two disjuncts are equal, and
disjuncts are ordered lexicographically. These constraints guarantee 
the announced cardinality of $B_{n}^{red}$; the rest of the proof is 
classical.
\eop

Of course, the formula in $B_{n}^{red}$ equivalent to a given formula,
is not always the shortest possible.

\subsection{First-order formulas, semantic equivalence}

Let us consider finite sets $R$ and $C$, of relational symbols and of
constants (nullary relations, source labels) as in Section~\ref{sec rel
structures}. Recall that, if $X$ is a finite set, $FO(R,C,X)$ denotes
the set of first-order formulas in the language of $(R,C)$-structures,
with free variables in $X$. For unproved results in this section, we
refer the reader to \cite{borger}.

Several notions of semantic equivalence of formulas can be defined. If
$\phi,\psi\in FO(R,C,X)$, say that $\phi\equiv \psi$ if for every
$(R,C)$-structure $S$ and for every assignment of values in $S$ to the
elements of $X$, $\phi$ and $\psi$ are both true or both false. Say
also that $\phi\equiv_{\omega}\psi$ if the same holds for every finite
or countable $(R,C)$-structure $S$, and $\phi\equiv_{f}\psi$ if $S$ is
restricted to being finite.

The equivalences $\equiv$ and $\equiv_{\omega}$ coincide by the
L\"owenheim-Skolem theorem. Indeed this theorem states that if a closed
formula has an infinite model, then it has one of each infinite
cardinality: to prove our claim, it suffices to apply it to the formula
$\exists \vec x\ \neg(\phi(\vec x) \Leftrightarrow \psi(\vec x))$. We
note that this equivalence cannot be extended to monadic second-order
formulas: there exists an MS formula with a unique model, isomorphic to
the set of integers $\mathbb N$ with its order.

Each of these three equivalences is known to be undecidable.

The equivalence $\equiv$ (or $\equiv_{\omega}$ since we consider only
first-order formulas) is semi-decidable: by G\"odel's completeness
theorem, $\phi\equiv\psi$ if and only if the formula $\forall \vec x\
(\phi(\vec x) \Leftrightarrow \psi(\vec x))$ has a proof, which is a
recursively enumerable property.

Trakhtenbrot proved that one cannot decide whether a first-order
formula is true in every finite structure, thus proving that
$\equiv_{f}$ is not decidable. However, the negation of $\equiv_{f}$ is
semi-decidable: if $\phi\not\equiv_{f}\psi$, a counter-example can be
produced by exploring systematically all finite $(R,C)$-structures.
This is a proof also that $\equiv$ and $\equiv_{f}$ do not coincide.

\subsection{First-order formulas, a syntactic equivalence}

We now describe a syntactic equivalence $\approx$ on formulas, which
refines the semantic equivalences $\equiv$ and $\equiv_{f}$: that is,
if $\phi \approx \psi$, then $\phi\equiv \psi$ and
$\phi\equiv_{f}\psi$.

If $b\in B_{n}$, and if $\phi_{1},\ldots,\phi_{n} \in FO(R,C,X)$, we
denote by $b(\phi_{1},\ldots,\phi_{n})$ the formula in $FO(R,C,X)$
obtained by replacing each occurrence of $p_{i}$ in $b$ by $\phi_{i}$.
It is clear that if $b$ and $b'$ are equivalent Boolean formulas, then
$b(\phi_{1},\ldots,\phi_{n}) \equiv b'(\phi_{1},\ldots,\phi_{n})$.

A \textit{Boolean transformation step} consists in replacing in a
first-order formula, a sub-formula of the form
$b(\phi_{1},\ldots,\phi_{n})$ by the equivalent formula
$b'(\phi_{1},\ldots,\phi_{n})$, where $b,b'\in B_{n}$ are equivalent.
Then we let $\phi \approx \psi$ if $\phi$ can be transformed into
$\psi$ by a sequence of Boolean transformation steps and of renaming 
of bound variables.

It is clear that if $\phi \approx \psi$, then $\phi \equiv \psi$. We
want to show that each first-order formula is effectively equivalent to
an $\approx$-equivalent formula of the same quantifier height, and to
give an upper bound on the number of $\approx$-equivalence classes of
formulas of a given height.

\subsubsection{Quantifier-free formulas}

Let $QF(R,C,X)$ be the set of quantifier-free formulas in $FO(R,C,X)$.
Such formulas are Boolean combinations of atomic formulas. Let
$Atom(R,C,X)$ be the set of these atomic formulas. Note that each
atomic formula is either of the form $x = y$, where $x$ and $y$ are in
$X \cup C$, or $r(x_{1},\ldots,x_{\rho(r)})$ where $r$ is a
$\rho(r)$-ary relation in $R$ and the $x_{i}$ are in $X\cup C$. Letting $n =
\card(X)$ and $c = \card(C)$, it is easily verified that

$$\card(Atom(R,C,X)) = (n+c)^2 + \sum_{r\in R}(n+c)^{\rho(r)}.$$
We let $f(R,c,n)$ be this function. Note that if we allow for the
(effective) syntactic simplifications of identifying the formulas of
the form $x = x$ with the constant $\true$, and of identifying the
formulas $x = y$ and $y = x$, we can lower the value of $f(R,c,n)$ to
$1 + \frac 12(n+c)(n+c-1) + \sum_{r\in R}(n+c)^{\rho(r)}$.

We then have the following.

\begin{proposition}\label{red qf}
    There exists a subset $QF^{red}(R,C,X)$ of $QF(R,C,X)$, of
    cardinality $2^{2^{f(R,c,n)}}$, such that every formula in
    $QF(R,C,X)$ can be effectively transformed to an
    $\approx$-equivalent formula in $QF^{red}(R,C,X)$.
\end{proposition}

\proof
By definition of quantifier-free formulas, $QF(R,C,X)$ is the set of
all formulas of the form $b(\phi_{1},\ldots,\phi_{n})$, where $b$ is a
Boolean formula and the $\phi_{i}$ are atomic formulas. Now let
$QF^{red}(R,C,X)$ be the set of all formulas of the form
$b(\phi_{1},\ldots,\phi_{n})$, where $b\in B_{n}^{red}$ and the
$\phi_{i}$ are pairwise distinct atomic formulas. The proof of the
precise statement is now immediate, using Proposition~\ref{red bool}.
\eop

\begin{example}
    Let us consider graphs with sources, so that $R$ consists of a
    single, binary edge relation. Then $f(R,c,0) = 2c^2$ and
    $\card(QF^{red}(R,C,\emptyset)) = 2^{2^{2c^2}} = q(c)$. Thus the
    type equivalence $\zeta$ (see Section~\ref{sec elementary
    properties} and Lemma \ref{type equivalence}) has at most
    $2^{q(c)}$ classes in $\GS(C)$.
\end{example}

\begin{remark}\label{remark A4}
    Again, we are not claiming that the set $QF^{red}(R,C,X)$ is as
    small as possible. On quantifier-free formulas, the equivalence
    $\equiv$ is decidable, because $\phi \equiv \psi$ is false if and
    only if the closed formula $\exists \vec x (\phi(\vec x)
    \not\Leftrightarrow \psi(\vec x))$ is satisfiable, and the
    satisfiability problem for existential formulas in prenex normal
    form is decidable (see \cite{borger}). Thus one can modify
    Proposition~\ref{red qf} by letting $QF^{red}(R,C,X)$ be the set of
    lexicographically minimal formulas in each $\equiv$-class: the same
    statement of Proposition~\ref{red qf} would then hold with $\equiv$
    instead of $\approx$. In particular, the transformation would still
    be effective, although very inefficient. It is not clear whether
    the cardinality of the new set of reduced quantifier-free formulas
    would be significantly smaller.
\end{remark}

\subsubsection{Quantifier depth of first-order formulas}

Recall that the quantifier depth of a first-order formula is the
maximal number of nested quantifiers. If we let $FO_{k}(R,C,X)$ be
the set of formulas in $FO(R,C,X)$ of quantifier depth at most $k$, a
formal definition is as follows: $FO_{0}(R,C,X) = QF(R,C,X)$ and, for
each $k \ge 0$, $FO_{k+1}(R,C,X)$ is the set of Boolean combinations of
formulas in
\begin{eqnarray*}
    \widehat{FO}_{k}(R,C,X) &=& FO_{k}(R,C,X) \\
    && \cup\enspace \{\exists y\ \phi\mid \phi\in FO_{k}(R,C,X\cup\{y\})\} \\
    && \cup\enspace \{\forall y\ \phi\mid \phi\in FO_{k}(R,C,X\cup\{y\})\}.
\end{eqnarray*}
Using the same recursion, let us define sets of ``reduced'' formulas of
every quantifier depth. First we fix an enumeration of the countable
set of variables. Next, we let $FO_{0}^{red}(R,C,X) = QF^{red}(R,C,X)$.
For each $k\ge 0$, we then let $FO_{k+1}^{red}(R,C,X)$ be the set of
formulas of the form $b(\phi_{1},\ldots,\phi_{n})$ where $b\in
B_{n}^{red}$ and the $\phi_{i}$'s are in
\begin{eqnarray*}
    \widehat{FO}^{red}_{k}(R,C,X) &=& FO^{red}_{k}(R,C,X) \\
    &\cup&  \{\exists y\ \phi\mid \phi\in
    FO^{red}_{k}(R,C,X\cup\{y\}),\hbox{ $y$ minimal not in $X$}\} \\
    &\cup& \{\forall y\ \phi\mid \phi\in
    FO^{red}_{k}(R,C,X\cup\{y\}),\hbox{ $y$ minimal not in $X$}\}.
\end{eqnarray*}

\begin{proposition}\label{red fo}
    For each $k\ge 0$, the set $FO_{k}^{red}(R,C,X)$ is finite.
    Moreover, every formula in $FO_{k}(R,C,X)$ can be effectively
    transformed to an $\approx$-equivalent formula in
    $FO_{k}^{red}(R,C,X)$.
\end{proposition}

\proof
Let $n = \card(X)$ and $c = \card(C)$, let $g(k,R,c,n)$ be the
cardinality of $FO_{k}^{red}(R,C,X)$, and let $h(k,R,c,n)$ be the
cardinality of $\widehat{FO}_{k}^{red}(R,C,X)$. It is elementary to
verify that these functions can be bounded as follows:
\begin{eqnarray*}
    g(0,R,c,n) &\le& 2^{f(R,c,n)} \hbox{ and for $k > 0$} \\
    g(k,R,c,n) &\le& 2^{2^{h(k,R,c,n)}} \\
    h(k,R,c,n) &\le& 3g(k-1,R,c,n+1).
\end{eqnarray*}

The rest of the proof is immediate, from the recursive definitions. 
\eop

\begin{remark}
    Since there is a procedure to transform each first-order formula
    into an $\approx$-equivalent formula in ``reduced form'', we can
    consider a new equivalence relation on first-order formulas: to
    yield the same reduced formula. This equivalence is decidable and
    it refines $\approx$ (and hence $\equiv$).
\end{remark}

\begin{remark}
    In Proposition \ref{red fo}, we can still consider replacing each
    formula by the lexicographically least equivalent formula, but this
    method is not effective, since the equivalence of first-order
    formulas is not decidable.
%
%
\end{remark}

\subsection{Monadic second-order formulas}

A very similar analysis can be conducted for monadic second-order
formulas of bounded quantifier depth. One difference is that the
L\"owenheim-Skolem theorem does not hold for these formulas, so the
semantic equivalence of formulas based on coincidence on all finite or
countable models does not imply coincidence on all models. Moreover,
since there is no complete proof systems for such formulas, the
equivalences $\equiv$ and $\equiv_{\omega}$ are not semi-decidable.

For the rest, one can follow the same techniques as above, to prove the
following result. We denote by $MS_{k}(R,C,W)$ the set of monadic
second-order formulas of quantification depth $k$ in the language of
$(R,C)$-structures, with their first- and second-order free variables
in $W$.

\begin{proposition}
    For every finite $R,C,W,k$, one can construct a finite subset
    $MS_{k}^{red}(R,C,W)$ of $MS_{k}(R,C,W)$ such that, for every
    formula in $MS_{k}(R,C,W)$, one can construct effectively an
    $\equiv$-equivalent formula in $MS_{k}^{red}(R,C,W)$.
\end{proposition}


\end{document}